\def\mearth{{\rm\,M_\oplus}}
\def\rearth{{\rm\,R_\oplus}}

\documentclass[preprint2]{proto}
\usepackage{times}
\usepackage{xcolor}
\usepackage{graphicx}
\usepackage{wasysym}
\usepackage{float}
\usepackage{dblfloatfix}

\begin{document}

\title{\textbf{\LARGE ORIGIN OF EARTH'S WATER: SOURCES AND CONSTRAINTS}}

\author {\textbf{\large Karen Meech}}
\affil{\small\em Institute for Astronomy, 2680 Woodlawn Drive, Honolulu HI 96822}

\author {\textbf{\large Sean N. Raymond}}
\affil{\small\em Laboratoire d'Astrophysique de Bordeaux, CNRS and Universit{\'e} de Bordeaux, All{\'e}e Geoffroy St. Hilaire, 33165 Pessac, France}

\begin{abstract}
\begin{list}{ } {\rightmargin 1in}
\baselineskip = 11pt
\parindent=1pc
{\small 
The origin of Earth's water is a longstanding question in the fields of planetary science, planet formation and astrobiology. In this chapter we review our current state of knowledge. Empirical constraints on the origin of Earth's water come from chemical and isotopic measurements of solar system bodies and of Earth itself. Dynamical models have revealed potential pathways for the delivery of water to Earth during its formation, most of which are anchored to specific models for terrestrial planet formation. Meanwhile, disk chemical models are focused on determining how the isotopic ratios of the building blocks of planets varied as a function of radial distance and time, defining markers of material transported along those pathways.  Carbonaceous chondrite meteorites -- representative of the outer asteroid belt -- provide a good match to Earth's bulk water content (although mantle plumes have been measured at a lower D/H).  What remains to be understood is how this relationship was established.  Did Earth's water originate among the asteroids (as in the classical model of terrestrial planet formation)? Or, more likely, was Earth's water delivered from the same parent population as the hydrated asteroids (e.g., external pollution, as in the Grand Tack model)?   We argue that the outer asteroid belt -- at the boundary between the inner and outer solar system -- is the next frontier for new discoveries.  The outer asteroid belt is icy, as shown by its population of icy bodies and volatile-driven activity seen on twelve main belt comets (MBCs); seven of which exhibit sublimation-driven activity on repeated perihelion passages.  Measurements of the isotopic characteristics of MBCs would provide essential missing links in the chain between disk models and dynamical models. Finally, we extrapolate to water delivery to rocky exoplanets. Migration is the only mechanism likely to produce very water-rich planets with more than a few percent water by mass (and even with migration, some planets are purely rocky). While water loss mechanisms remain to be studied in more detail, we expect that water should be delivered to the vast majority of rocky exoplanets.
\\~\\~\\~}

\end{list}
\end{abstract}  

\section{\textbf{INTRODUCTION}}
\label{sec:intro}

We have only one example of an inhabited world, namely Earth, with its thin veneer of water, the solvent essential to known life.  Is a terrestrial planet like Earth that lies in the habitable zone and has the ingredients of habitability a common outcome of planet formation or an oddity that relied on a unique set of stochastic processes during the growth and subsequent evolution of our solar system?

Ultimately, water originates in space, likely formed on dust grain surfaces via reactions with atoms inside cold molecular clouds \citep{Tielens1982,Jing2011}. Grain surface water ice then provides a medium for the chemistry that forms molecules of biogenic relevance \citep{herb2009}. Both water and organics are abundant ingredients in protoplanetary disks out of which solar systems form. In the disk, water is present both as a gas and a solid and is likely processed during planet formation. The Earth formed in our solar system's inner protoplanetary disk. \textcolor{black}{The sequence of events that led to the growth of a rocky Earth with a thin veil of water remain to be fully understood. Both thermal processing -- in particular, desiccation of the earliest generations of planetesimals due to $^{26}$Al heating~\citep{grimm93,monteux18,alexander19a} -- and Earth's orbital location within the planet-forming disk likely played important roles. Water only exists as ice past the disk's {\em snow line}\footnote{the radial distance from the star beyond which the temperatures are low enough for gases to condense as ice} which itself shifts inward in time as the disk evolves, and which may have been located interior to Earth's orbit during part of Earth's formation. Reconciling Earth's water content with meteoritic constraints and models for Solar System formation remain a challenge.
}

\begin{figure*}[ht!]
\begin{center}
\includegraphics[width=14cm]{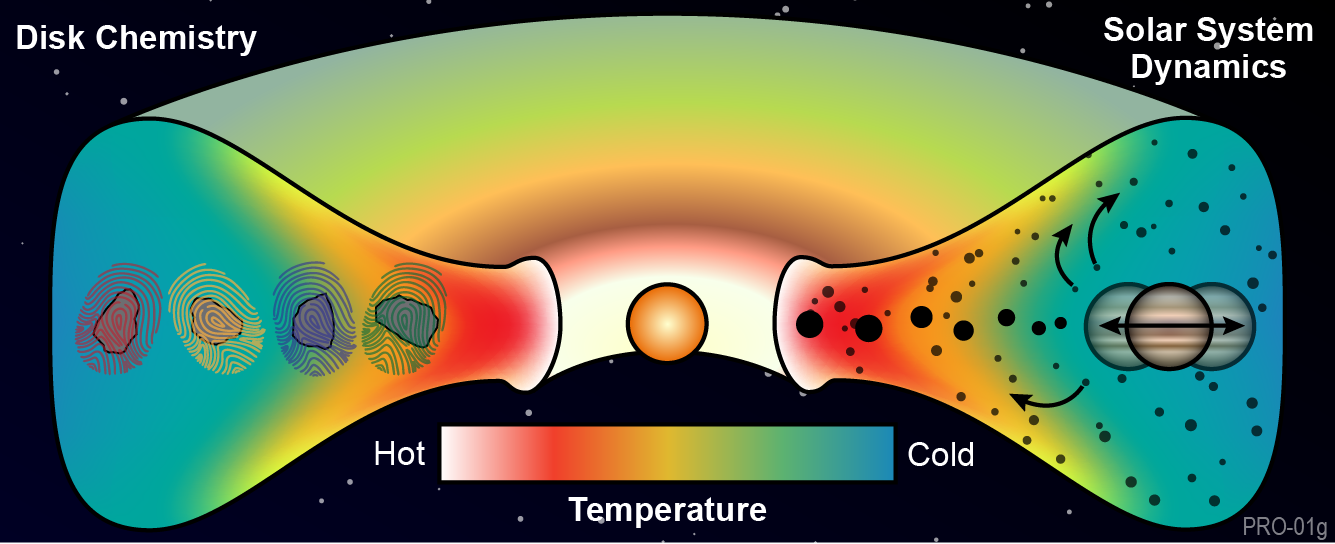}
\caption{Protoplanetary disk chemical signatures are implanted on planetesimals as ices freeze near the midplane.  This fingerprint is a sensitive measure of the planetesimal formation location. Subsequent planetesimal scattering as the giant planets grew will scramble this signature.  The key to understanding where inner solar system volatiles originated is to measure the isotopes in a primitive reservoir of material that records the dynamical history during formation.}
\label{fig:Disk}
\end{center}
\end{figure*}

While there has been significant work to investigate the origin of Earth's water--the wellspring of life--we still do not know if it came mostly from the inner disk or if it was delivered from the outer disk. No one knows if our solar system, with a planet possessing the necessary ingredients for life within the habitable zone, e.g. sufficient water and organics, is a cosmic rarity. Nor do we know whether the gas giants in our solar system aided or impeded the delivery of essential materials to the habitable zone. The answers to these questions are contained in volatiles\textcolor{black}{\footnote{In planetary science volatiles are compounds that are typically found as gases or as ices in the outer solar system.}} unaltered since the formation of the giant planets. To access this record, we need: (1) a population of bodies that faithfully records the history of volatile migration in the early solar system; (2) a source of volatiles that we can access; (3) knowledge that the volatiles were not altered by aqueous interaction with their parent body; and (4) measurements from multiple chemical markers with sufficient precision to distinguish between original volatile reservoirs. 

Comets were long thought to be the most likely ``delivery service'' of Earth's water based on a comparison of their D/H ratio's with that of Earth's ocean \citep{delsemme92,Owen1995}. But new models and data, including the {\it Rosetta} mission's survey of comet 67P/Churyumov-Gerasimenko, have cast doubt on this source. The relative abundances of 67P's volatile isotopes (D/H, N, C and noble gases) do not match those of Earth \citep{Altwegg2015,Marty2016}. Furthermore, comets will not provide an answer to the original source of Earth's water because they are dynamically unstable and can not be traced back to where they formed \citep{levison97,brasser13}. The only way to learn where Earth's water came from is to match the chemical fingerprints of inner solar system volatiles to a location in the protoplanetary disk. Such data can distinguish between competing models of solar system formation to specify where the water came from and how it was delivered.

Our solar system is quantifiably unusual compared to the thousands of known exoplanet systems \citep{martin15,mulders18,Raymond2018}. \textcolor{black}{After taking observational biases into account,} only $\sim$10\% of Sun-like stars have gas giants~\citep{cumming08,mayor11,fernandes19}, and only $\sim$10\% of outer gas giants have low-eccentricity orbits like Jupiter~\citep{butler06,udry07}, putting the solar system's orbital architecture in a $\sim$1\% minority~\citep[for a discussion, see][]{Raymond2018}. Many exoplanetary systems have Earth-sized planets in or near their host star's habitable zones~\citep[e.g., in the \textcolor{black}{TRAPPIST}-1 system;][]{gillon17}, and a handful of systems have outer gas giants and inner small planets~\citep[e.g., the Kepler-90 8-planet system;][]{cabrera14}. It remains unclear whether these planets are actually habitable. Liquid water is certainly a key ingredient, and it remains to be seen whether the solar system's formation and the planetary arrangement itself were critical in setting the conditions needed for life, in terms of the quantity and timing of volatile delivery. \textcolor{black}{However, it is important to ascertain the mechanisms for water delivery to the inner planets, as this is directly relevant to the habitability of those planets, and thus bears on the age-old question, ``Are we alone?''}

\begin{figure*}[ht!]
\begin{center}
\includegraphics[width=16cm]{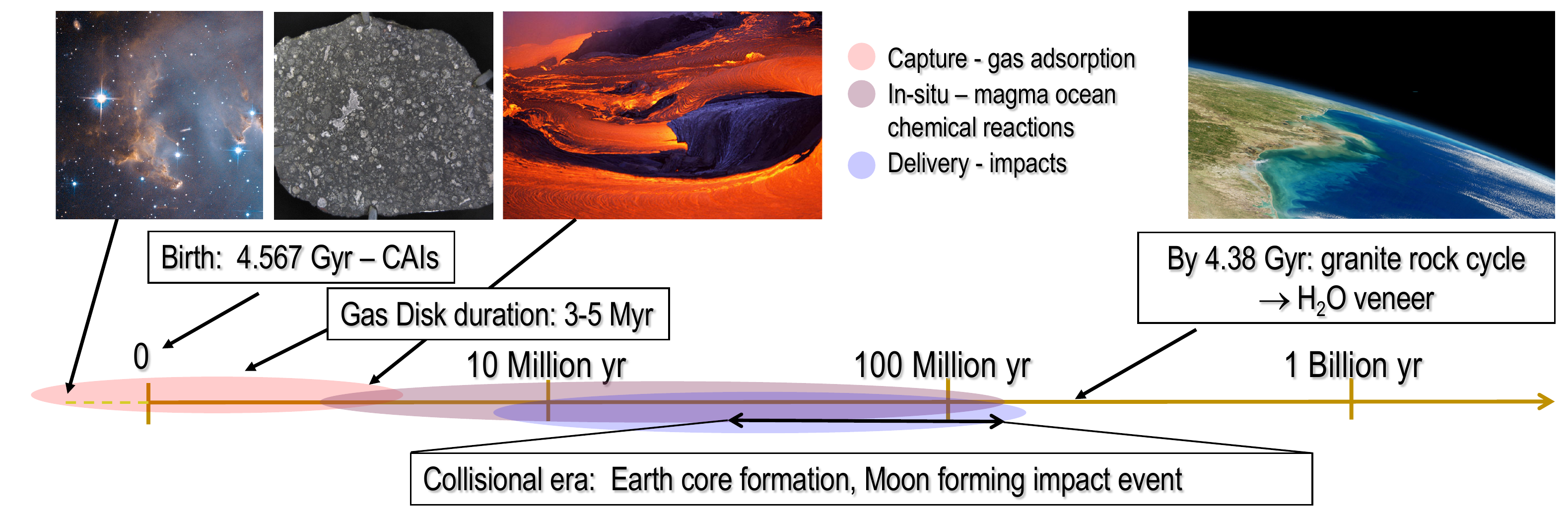}
\caption{\small 
Earth received its water early; by 4.38 Gy Earth had oceans. Images from:ESA and the Hubble Heritage team (STScI/auRA)-ESA/Hubble Collaboration; AMNH-Creative commons license, 2.0 (https://creativecommons.org/licenses/by/2.0/deed.en); NASA. 
}  
\label{fig:Timeline}
\end{center}
\end{figure*}

\subsection{What we Know about Earth's Water}
\label{sec:earth}

The inner solar system is relatively dry \citep{abe00,vanDischoeck2014}. Mercury is dry except for surface ice seen at the poles, likely deposited from recent exogenous impacts \citep{Deutsch2019}. Venus was once wet \citep{Donahue1982}, but experienced major water loss through hydrodynamic escape or impact-driven desiccation \citep{Kasting1983,Kurosawa2015}.  Likewise, Mars may have once had significant water, but has also lost a large fraction of its atmosphere to space via sputtering \citep{Jakosky2017}. While images of Earth from space suggest that it is a black oasis in space, rich in water, the \textcolor{black}{bulk Earth is in fact relatively dry.  However, its precise water content is uncertain. Assessments of the amount of water stored as hydrated silicates within the mantle vary between roughly one and ten ``oceans''~\citep{hirschmann06,Mottl2007,Marty2012,halliday13}, where one ocean is defined as the amount of water on Earth's surface (by mass, roughly $2.5 \times 10^{-4} \mearth$). \textcolor{black}{The amount of water in Earth's core is also uncertain~\citep{Badro2014,nomura14}, although recent laboratory experiments suggest that it is quite dry and that the bulk of Earth's water resides in the mantle and on the surface~\citep{clesi18}.}}

Earth's water was incorporated during its formation. Planets form around young stars in disks of gas and dust, whose characteristics may vary~\citep{bate18}. Disks are $\sim$99\% gas by mass. The solid component of the disk starts as sub micron-sized dust grains inherited from the interstellar medium. Water is provided to the disk as ice coatings on grains.  What is not clear is how much recycling of water there is in the disk.  The disk is flared, i.e., the thickness of the disk increases with radius, so the disk top and bottom surfaces are exposed to radiation from the star. This heats the surfaces, producing a vertical variation in temperature (in addition to a general decrease in temperature with distance from the star), as shown schematically in Fig.~\ref{fig:Disk}. \textcolor{black}{Matter is accreted from the disk onto the star, adding to the star's UV and x-ray flux. This can} alter water chemistry and isotopic composition as a function of position and time. Isotopes can preserve the signature of these kinds of disk processes and will be the key to tracing the volatiles in the early solar system.  

Cosmochemical and geochemical evidence provides clues that Earth's water arrived early (Fig.~\ref{fig:Timeline}; \textcolor{black}{see also the discussion in the chapter by Zahnle and Carlson}). The condensation of the first solar system solids (the Calcium and Aluminum Inclusions; CAIs) sets the time ``zero'' \textcolor{black}{for the birth of the Solar System} at 4.567 Gyr \citep{Amelin2002,Krot2009,Bouvier2010}.  As discussed in Section~\ref{sec:watermodel}, it is believed that water could have arrived at Earth (1) locally as gas that was adsorbed onto the surface of dust grains\textcolor{black}{~\citep{King2010,asaduzzaman14}}, or (2) it could have formed on Earth's surface if a magma ocean was in contact with a primordial hydrogen atmosphere\textcolor{black}{~\citep{Ikoma2006}}, or (3) water could have been dynamically delivered from the ice-rich outer solar system \textcolor{black}{~\citep{walsh11,raymond17}}. Both observations and models of protoplanetary disks show that the \textcolor{black}{lifetimes of gaseous protoplanetary disks} are very short, a few Myr \citep{haisch01,Ercolano2017}, so local processes would have had to have occurred early. Likewise Earth's magma ocean probably could have lasted anywhere from a \textcolor{black}{hundred thousand years} to more than $\sim$100 Myr \citep{Elkins2012,Hamano2013,monteux16}. Estimates of the final formation stages of the Earth (the core formation and the moon forming impact event) show that this did not extend beyond 140 Myr after the formation of the first solids \citep{Rubie2007,Fischer2018}.  \textcolor{black}{Evidence from Hadean zircons show that oceans \textcolor{black}{likely} existed on Earth within $\sim$150-250 Myr of the start of Solar System formation \citep{Mojzsis2001,wilde01}. This indicates that water (and accompanying organics and other volatiles) must have arrived as Earth was forming, consistent with models for Earth's chemical evolution during its accretion~\citep{wood10,dauphas17}.}

\begin{figure*}[hb]
\begin{center}
\includegraphics[width=16cm]{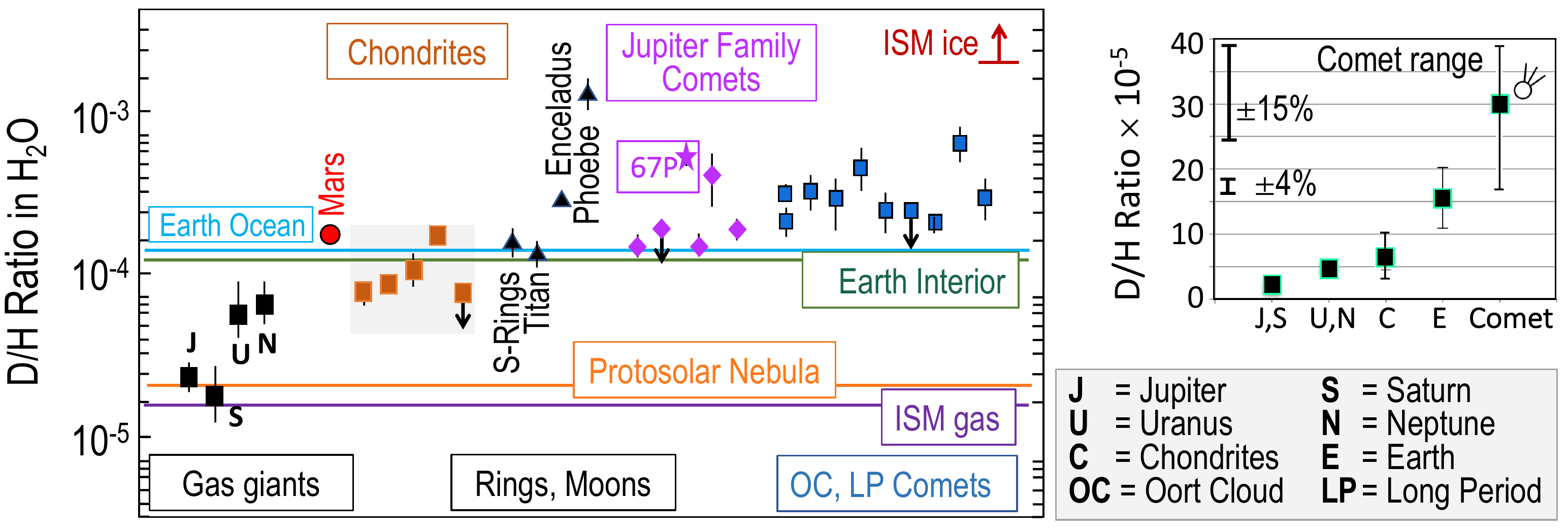}
\caption{\small 
D/H measurements from a variety of solar system reservoirs. The values and references are shown in Table~1. While high precision measurements of D/H can discriminate between source reservoirs, some reservoirs have similar D/H values. D/H alone can not identify the source of inner solar system water.
}  
\label{fig:DH}
\end{center}
\end{figure*}

\textcolor{black}{This chapter reviews current paradigms for the origins of Earth's water.  As water delivery is inherently linked with planet formation, we summarize the geochemical, cosmochemical, astrochemical, astronomical, and dynamical evidence for how habitable worlds form, addressing in particular the origin of water on rocky planets.} Isotopes and noble gases are powerful tracers of the early solar system chemical processes in the protoplanetary disk. Interpreting these fingerprints requires an understanding of how the planetesimals grew, moved around and were eventually incorporated into habitable planets (see Fig.~\ref{fig:Disk}). \textcolor{black}{The remainder of} section~\ref{sec:intro} introduces the history of exploring the origin of water through the isotopic record.  Section~\ref{sec:Constraints} summarizes the dynamical and chemical constraints and processes for solar system formation, Section~\ref{sec:watermodel} summarizes models for the delivery of water to the inner solar system and Section~\ref{sec:discuss} discusses the implications.

\subsection{D/H As a Tracer of Earth's Water}
\label{sec:DH}

Deuterated molecules have a highly temperature sensitive chemistry that can provide information about the physical conditions at the time of their formation. The original D/H ratio set in the big bang \citep{Spergel2003} and altered by stellar nucleosynthesis is measured from interstellar absorption lines as starlight passes through diffuse gases.  Deuterium is enriched in interstellar ices in cold dense regions in molecular clouds via complex gas phase and gas-grain chemistry reaction networks \citep{Millar1989}.

\begin{table}[h]
  \begin{center}
    {\label{tab:DH} Astronomical and solar system D/H values}
    \begin{tabular}{lccc}
    \hline
    {\bf Source} & \multicolumn{2}{c}{\bf D/H Value}  & {\bf Note} \\
    & [$\times$ 10$^{-5}$] & $\delta$D [$\permil$]$^{\dag}$ \\
    \hline
    \small
    Big Bang      & 2.62$\pm$0.19 & --832$\pm$12                & 1\\ 
    Local ISM gas & 2.3$\pm$0.24  & --852$\pm$15                & 2\\ 
    ISM/YSO ice   & 10-100 & -350 $\rightarrow$ +5419  & 3\\ 
    Protosolar    & 2.0$\pm$0.35  & --872$\pm$22                & 4\\ 
    SMOW          & 15.57         &     0                       & 5\\ 
    Earth Mantle  & $<$12.2       & $<$--217                    & 6\\ 
    Giant planets & 1.7-4.4       & -890 $\rightarrow$ --717    & 7\\ 
    Mars Mantle   & $<$19.9       & $<$278                      & 8\\ 
    C-Chondrites  & 8.3-18        & --467 $\rightarrow$ +156    & 9\\ 
    Rings, moons  & 14.3-130      & --82 $\rightarrow$ +7346    & 10\\
    Comets        & 16-53         & +27 $\rightarrow$ +2402     & 11\\
    \hline
    \end{tabular}
    \end{center}
    \caption{Notes:
    $^{\dag}$ $\delta$D in per mil [$\permil$] = ([D/H]{\textsubscript{sample}} / [D/H]{\textsubscript{SMOW}} -- 1) $\times$ 1000. 
    [1] \citet{Spergel2003} 
    [2] \citet{Linsky2007} 
    [3] \textcolor{black}{\citet{Cazaux2011,Coutens2014}} 
    [4] \citet{Geiss1998} 
    [5] \citet{Lecuyer1998} 
    [6] \citet{Hallis2015} 
    [7] \textcolor{black}{As measured in H$_2$ \citet{Lellouch2001,Feuchtgruber2013,Pierel2017}} 
    [8] \citet{Hallis2017} 
    [9] \citet{Alexander2012} 
    [10] \citet{Waite2009,Clark2019} 
    [11] \citet{Bockelee2015,Altwegg2015}. 
    }
    \normalsize
\end{table}

It had long been known that the D/H ratio of Earth's oceans \citep[e.g., Standard Mean Ocean Water, D/H\textsubscript{SMOW} = 15.576 $\times$ 10$^{-5}$,][]{Lecuyer1998} was significantly elevated (by a factor of 6.4) above that of the expected protosolar value \citep[][see Table~1]{Geiss1998}.  The measurement of a similarly elevated D/H value in water (2$\times$ SMOW) for Comet Halley by the {\it Giotto} mission \citep{Eberhardt1987,Eberhardt1995} led to the idea that comets could have been the source of Earth's water \citep{Owen1995} and that D/H could serve as the ``fingerprint'' to identify the origin of inner solar system water.

However, as more measurements of D/H in comets were obtained (mostly of long period comets, LPCs; see Fig.~\ref{fig:DH}) showing that the values were all elevated above that of SMOW \citep{Mumma2011}, it was clear that comets could not be the only source of Earth's water.

\begin{figure*}[ht!]
\begin{center}
\includegraphics[width=14cm]{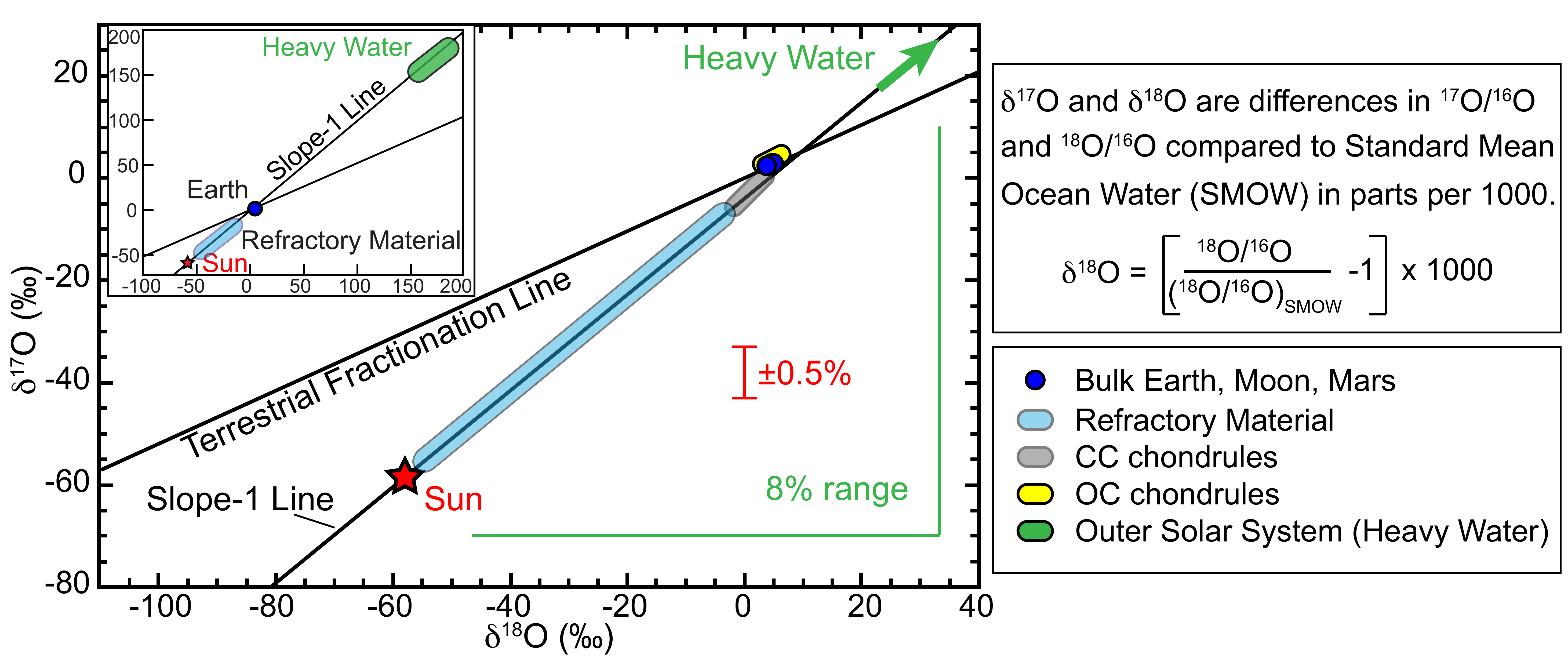}
\caption{The range of oxygen isotopic variation in the solar system is small, so distinguishing reservoirs requires very high precision measurements.}
\label{fig:O}
\end{center}
\end{figure*}

Initially, D/H measurements from astronomical sources were compared to Earth's oceans. However, since the oceans likely do not represent the bulk of Earth's water inventory, the comparison should be made to Earth's bulk D/H. The Earth's oceans are not a closed system; they interact with the atmosphere, and water is mixed into the mantle via subduction. \textcolor{black}{One scenario invoking large-scale loss of an early H-rich atmosphere proposes that} the atmospheric D/H may have increased over the age of the solar system by a factor of 2-9$\times$~\citep{Genda2008}. Determining the Earth's \textcolor{black}{primordial} D/H ratio thus requires sampling a primitive undegassed mantle source. Magma ocean crystallization models \citep{Elkins-Tanton2008} suggest that there may have been small volumes of late-solidifying material in the first 30-75 Myr of Earth's history that still exist near the core mantle boundary. Mantle plumes in Hawai`i, Iceland,  and Baffin Island appear to have tapped into undegassed, deep mantle sources, based on their He isotope ratios \citep{starkey2009,stuart2003,Jackson2010}. Measurements from the Icelandic plume show \textcolor{black}{a measurement for Earth's mantle that has a D/H lower than that of the oceans \citep[][see Section~\ref{sec:oxidation}]{Hallis2015}, although this may not be representative of the whole mantle}.

Among solar system water reservoirs, only carbonaceous chondrites have D/H ratios that span the estimates for Earth's ocean and interior \citep{marty06,Marty2012,Hallis2015,Alexander2012}. \textcolor{black}{While chondritic water is isotopically lighter than Earth's water, chondritic organics are heavier~\citep{alexander12}. These two reservoirs (water and organics) are expected to have equilibrated during accretion, and it is the bulk D/H isotopic composition that should be considered.}

\subsection{Oxygen Isotopic Variations}
\label{sec:O}

Isotopic measurements of D/H alone are insufficient to uniquely ascertain a formation location in the disk because of the complexity of disk chemical models (see $\S$~\ref{sec:fingerprints}). Oxygen has three isotopes whose variation in the disk depends on different physical processes, thus the variation can be used as an independent tracer.  
The standard isotope notation is defined as:
\begin{equation}
    \delta^{17} \mathrm{O} = \left(\frac{^{17}\mathrm{O}/^{16}\mathrm{O}_{sample}} {^{17}\mathrm{O}/^{16}\mathrm{O}_{SMOW}} - 1 \right)~\times 1000,
\end{equation}
\noindent (similarly for $\delta^{18}$O).

The oxygen isotopic composition of the Sun inferred from samples of the solar wind returned by the Genesis spacecraft is $^{16}\mathrm{O}$-rich 
\citep{McKeegan2011}, whereas nearly all solar system solids are $^{16}\mathrm{O}$-depleted relative to the Sun's value \textcolor{black}{(see Fig.~\ref{fig:O})}. Exceptions are refractory inclusions that formed from a gas of approximately solar composition and are $^{16}\mathrm{O}$-rich \citep{Scott2014}. Most physical and chemical fractionation processes depend on mass, and on an oxygen three-isotope plot of $\delta$$^{17}$O versus $\delta$$^{18}$O samples should fall on a line with a slope of $\sim$0.52. \textcolor{black}{Most samples from Earth (excluding the atmosphere)} fall along this line, called the Terrestrial Fractionation line (TFL). However, the compositions of chondrules and refractory inclusions in the primitive (unmetamorphosed and unaltered) carbonaceous chondrites fall along a mass-independent fractionation line with a slope of $\sim$1 (see Section~\ref{sec:oxygen}).
The physical processes controlling the distribution of oxygen isotopes in primitive solar system bodies are different from those for D/H, thus the oxygen isotopic composition of water can also provide clues about its origins (see Section~\ref{sec:oxygen}).

\begin{figure*}[ht]
\begin{center}
\includegraphics[width=13.0cm]{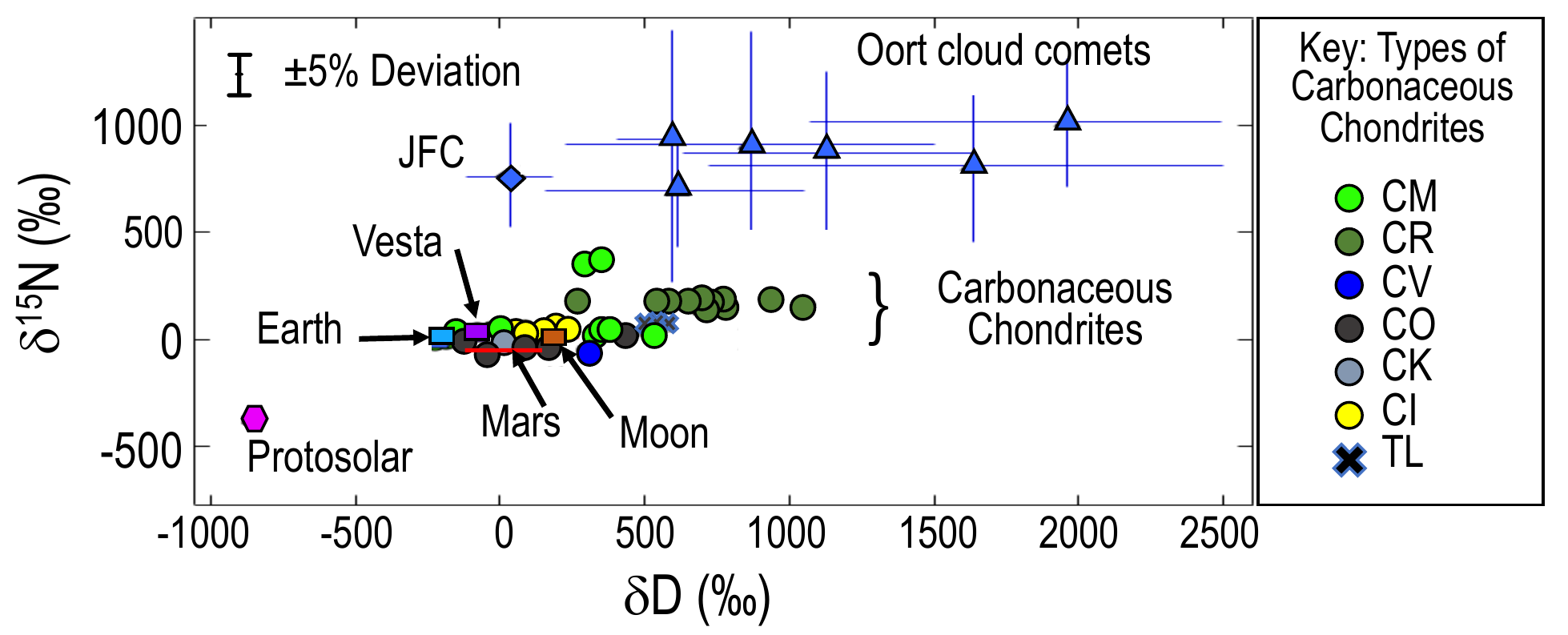}
\caption{Combining nitrogen and hydrogen isotope ratios helps discriminate between reservoirs. The N-isotope difference between the group defined by Earth, Vesta, the Moon, Mars, chondrites, the Oort cloud comets, and the protosolar value is large. The difference between the CR chondrites and the other chondrites is 15\%. \textcolor{black}{The cometary values are measured in a variety of different molecules including water, HCN and NH$_3$. It is challenging to infer the bulk isotopic compositions. }}
\label{fig:N}
\end{center}
\end{figure*}

\subsection{Nitrogen Isotopic Variations}
\label{sec:N}

Nitrogen isotopes can provide additional constraints for planetesimal formation distances \citep{Alexander2018}. There is a general trend of increasing $\delta^{15}$N with increasing distance of formation from the Sun, although again, we only have these measurements for a small number of comets. While many carbonaceous chondrites have nitrogen isotopic ratios that match that of the Earth, cometary values are very different from the telluric value.  The large cometary $^{15}$N excess relative to Earth's atmosphere indicates that their volatiles underwent isotopic fractionation at some point in the early solar system. Within the uncertainties on estimates of volatile budgets, \citet{Marty2012} and \citet{Alexander2012} have proposed that Earth's inventory of H, C and noble gases may be matched with a few percent CM/CI chondritic material (see Fig.~\ref{fig:N}). However, Earth's bulk nitrogen content appears to be depleted by an order of magnitude relative to those elements. It remains uncertain whether Earth preferentially lost much of its nitrogen during giant impacts or whether there remains an as-yet-unidentified reservoir of nitrogen in Earth's interior. 

\subsection{Noble Gases}
\label{sec:NG}

While Earth's D/H may be a match for carbonaceous chondrites, the noble gases are not a match \citep{Owen1995,Owen2000} \textcolor{black}{and an additional solar component is required. Ideas for how this solar component was acquired include: that it was accreted directly from the nebula, was solar wind implanted into the surfaces of small objects accreted by the Earth, or was delivered by cometary ices} \textcolor{black}{(see also the discussion in the Chapter by Zahnle \& Carlson)}. The noble gas fractionation patterns for the Martian and terrestrial atmospheres are similar, both depleted in Xe relative to the other noble gases compared to carbonaceous chondrites. The first detection of noble gases in comets was from the {\it Rosetta} mission and showed that Ar and Kr were solar \citep{Rubin2018}, but there were deficits in the heavy Xe isotopes \citep{Marty2017}. This suggests that there was at least a small contribution of cometary volatiles \textcolor{black}{that delivered little of Earth's water but a significant fraction of its (atmospheric) noble gases \citep{Marty2016,Marty2017}.}

\subsection{Multiple Fingerprints are Needed}
\label{sec:fingerprints}

\textcolor{black}{Generally, volatiles that have been heated and re-equilibrated with inner solar system gas will have a low (protosolar) D/H value, and bodies formed in the distant solar system will be high.} Most of the early comet D/H measurements were from long period comets (LPCs), which were thought to form closer to the sun than the Jupiter family comets (JFCs). Based on the expectation that the JFCs likely formed in the colder outer disk and the LPCs formed in the giant planet region closer to the sun \citep{Meech2004}, it was expected that the JFCs would have an even higher D/H ratio than the LPCs. However, the D/H measurement in the {\it EPOXI} mission target, JFC 103P/Hartley 2 matched that of Earth's oceans leading to claims that the JFC comet reservoir could have delivered Earth's water \citep{Hartogh2011}. 

More recently, the high D/H measurements from the {\it Rosetta} mission target (also a JFC; see Fig.~\ref{fig:DH}) have been interpreted to imply that comets did not bring Earth's water \citep{Altwegg2015}. \textcolor{black}{However, there are two key uncertainties. First, cometary isotopic ratios were only measured in certain molecules and may not represent the bulk isotopic compositions~\citep[recall that, while carbonaceous chondrites match Earth's bulk D/H, chondritic water is isotopically lighter whereas chondritic organics are isotopically heavier;][]{alexander12}.  Second, there may be a correlation between measured D/H values and the level of cometary activity, with more active comets having lower D/H~\citep{Lis2019}. This effect is likely to be small, however, and models show that this should not be a factor if comets are observed at perihelion \citep{Podolak2002}}. 

\begin{figure*}
\begin{center}
\includegraphics[width=15cm]{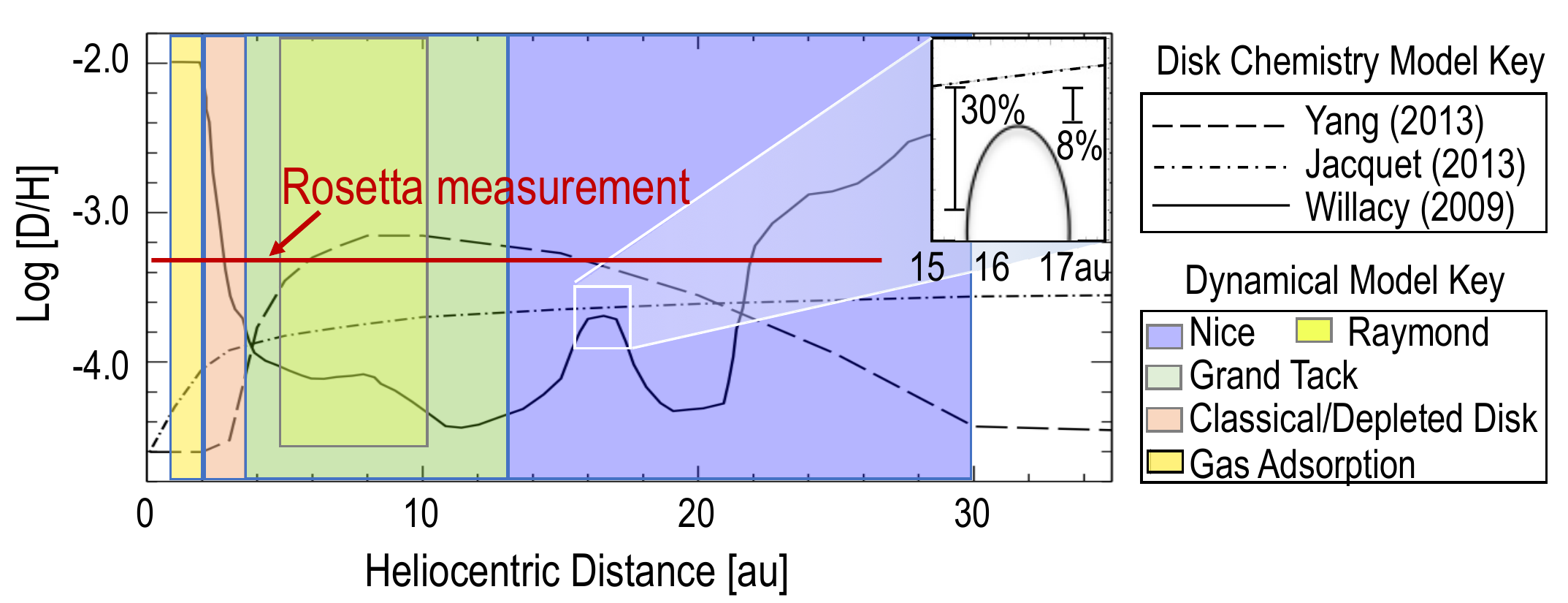}
\caption{Measuring only D/H cannot uniquely determine an origin location. The {\it Rosetta} measurement from comet 67P/Churyumov-Gerasimenko is consistent with at least two disk chemical models predicting formation at several heliocentric distances--compatible with several dynamical models. Furthermore, disk chemical models can have similar D/H at different distances, and different dynamical models can scatter from overlapping regions.}
\label{fig:Models}
\end{center}
\end{figure*}

Many of these D/H measurements have stimulated the development of new disk chemical models when data did not match predicted trends \citep{Aikawa2002,Willacy2009,Jacquet2013,Yang2013}. The real issue is that the predicted D/H variation along the mid-plane from chemical models of protosolar disks is complex, and D/H alone is not sufficient to determine a formation distance to compare to dynamical models.  \textcolor{black}{This was seen with the {\it Rosetta} D/H measurement \citep{Altwegg2015}, which was consistent with several disk chemical and dynamical models} (see Fig.~\ref{fig:Models}). 

\textcolor{black}{There is an additional complication. The D/H of water in ordinary and R chondrites is higher than in carbonaceous chondrites \citep{Alexander2012}, possibly due to either higher presolar water abundances in the inner solar system, or due to oxidation of metal by water and subsequent isotopic fractionation as the generated H$_2$ is lost \citep{alexander19a,alexander19b}.}

Finally, because of the complex chemistry and dynamical mixing as the planets grew, the chemical fingerprints are ``smeared''.  Multiple reservoirs can also have the same fingerprint. Thus, to understand the origins of inner solar system volatiles, we need to measure multiple isotopes, comparing these to the dynamical models. 


\section{Solar System Formation Constraints and Processes}
\label{sec:Constraints}

Given that the origin of Earth's water cannot be decoupled from its formation, we must consider the large-scale constraints on solar system formation. In this section, we briefly summarize the empirical constraints and the key processes of planet formation. For more detail we refer the reader to recent in-depth reviews focused on dynamical modeling and solar system formation~\citep[][\textcolor{black}{also chapters by Zahnle \& Carlson and by Raymond et al.}]{morbyraymond16,obrien18}.

\subsection{Planet Formation Model Empirical Constraints}
\label{sec:Empirical}

The following observations represent the fossil evidence of our solar system's formation, which successful planet formation scenarios must reproduce~\citep[see][]{chambers01,raymond09c}.

\subsubsection{The Planets' Orbits}\label{sec:orbit} 

Most of the mass in terrestrial planets is concentrated in a narrow ring between the orbits of Venus and Earth. The terrestrial planets themselves have near-circular, co-planar orbits which for years posed a problem for models of terrestrial accretion.  Meanwhile, the giant planets' low-eccentricity but spread-out orbits may indicate an orbital instability in our solar system's distant past~\citep{tsiganis05}, albeit one that was far weaker than those inferred for exoplanet systems~\citep{raymond10}.

\subsubsection{Small Body Populations} \label{sec:smbody}  
Solar system small bodies are the leftovers of planet formation, although there is no reason to think that they are a representative sample. The asteroid belt is very low in mass but has an excited orbital distribution. In broad strokes, the inner belt is dominated by dry objects (such as the S-types) and the outer belt by hydrated asteroids (e.g., the C-types) \citep{gradie82,demeo13}. \textcolor{black}{In addition to the asteroids,} volatile-rich bodies include comets, small satellites, and Kuiper belt objects. The Kuiper belt contains a total of $\sim $0.1 Earth masses, $\mearth$, \citep{gladman01} and the Oort cloud up to a few $\mearth$ \citep[][and references therein]{Boe2019}.

For a long time it was believed that comets formed in distinct regions, e.g., that the LPCs formed in the giant planet region and were scattered to the Oort cloud during formation, and that the JFCs were formed further out in the region of the Kuiper belt--eventually getting perturbed inward, during which time they became Centaurs, until their orbits were influenced by Jupiter \citep{Meech2004}. However, while there are clear trends in comet chemistry \citep{AHearn1995,Mumma2011}, they have not been clearly tied to dynamical class.  Comets likely formed over a range of distances outside the solar system's snow line and have experienced significant dynamical scattering.  \textcolor{black}{In considering possible sources for Earth's water we must therefore consider these objects as a population because individual comets cannot be traced back in time; rather, they sample the entire disk outside the snow line.}

\subsubsection {Meteorite Constraints on Growth Timescales} \label{sec:meteorite}  

Isotopic analyses of different types of meteorites provide vital constraints on formation timescales of different types of objects.  CAIs, the oldest dated solids to have formed in the solar system, are generally used as ``time zero'' for planet formation.  Age estimates of iron meteorites provide upper limits to the formation timescales of differentiated bodies~\citep{kruijer14}.  The existence of two types of chondritic meteorites with different isotopic anomalies (non-carbonaceous and carbonaceous) but similar ages has been interpreted as evidence for the rapid growth of Jupiter's core, which would have provided a barrier between these populations~\citep[][\textcolor{black}{see also chapter by Zahnle \& Carlson}]{kruijer17,desch18}.  Finally, the Hf/W system provides estimates of the timing of core formation, and suggest that Mars' growth was rapid~\citep{nimmo07,dauphas11}, whereas Earth's was prolonged~\citep{kleine09,jacobson14}.

These constraints are inherently tied to the conditions of planet formation.  For example, numerical experiments have shown that a smooth disk of solids extending from Mercury's orbit out to Jupiter's generally fails to reproduce the solar system because (1) the terrestrial planets' orbits are overly excited~\citep{chambers01,obrien06,raymond06b}; (2) Mars is too massive and grows too slowly~\citep{raymond09c,fischer14,izidoro15c}; and (3) the asteroids' orbits are under-excited~\citep{izidoro15c}.  Although we note that these problems may be solved if the giant planet instability happens {\it during} terrestrial planet formation; \citep[][see Section~\ref{sec:watermodel} and Figure~\ref{fig:3models} for a comparison between models]{clement18}.

\subsection{Key Planet Formation Processes}
\label{sec:Processes}

Planet formation models are built of processes.  Each process can be thought of as a puzzle piece, which must be assembled into a global picture of planetary growth (\textcolor{black}{see the chapter by Raymond et al. for a review dedicated entirely to this endeavor for the solar system and exoplanet systems}). We now very briefly summarize the key planet formation processes from the ground up.

\subsubsection{Disk Structure and Evolution}\label{sec:structure}

The underlying structure and dynamics of protoplanetary disks remain poorly understood~\citep{morbyraymond16}. An essential piece of the story is how angular momentum is transported within disks~\citep{turner14}. The radial surface density of gas and dust sets the stage for planet formation. Within a disk, the gas is subject to hydrodynamic pressure forces as well as gravity, and its motion deviates from pure Keplerian motion, with radial velocities that are generally slightly slower than the Keplerian velocity. Dust grows and drifts within the disk~\citep{birnstiel12}. Dust accumulates at pressure bumps, narrow rings at which the gas velocity matches the Keplerian velocity such that the drag force disappears~\citep{haghighipour03}. Exterior to a pressure bump the gas velocity is sub-Keplerian such that dust particles feel a headwind, lose orbital energy and drift radially inward.  Just interior to pressure bumps the gas velocity is super-Keplerian so particles feel a \textcolor{black}{tailwind}, driving them back outward. Very small dust particles remain strongly coupled to the gas, and large bodies have enough inertia to drift slowly; the fastest-drifting particles are ``pebbles''~\citep[e.g.,][]{ormel10,lambrechts12}.  
As disks evolve they cool down, and locations associated with specific temperatures -- e.g., the {\em snow line}, the radial distance beyond which a volatile (such as water) may condense as ice -- move inward. Gaseous disks are observed to dissipate on a characteristic timescale of a few Myr~\citep{haisch01,hillenbrand08}.

\subsubsection{Planetesimal Formation}\label{sec:planetesimal}

 Planetesimals are the smallest macroscopic bodies for which gravity dominates over hydrodynamical forces. Their origin has long been difficult to reproduce with formation models because any growth model must traverse the size scale at which particles start to decouple from the gas motion. These intermediate-sized particles then experience the headwind that causes them to rapidly spiral inward, preventing aggregation into larger bodies. This occurs at cm- to m-sizes, so this is sometimes called the ``meter barrier''~\citep{weidenschilling77b}. New models have demonstrated that mm-sized particles can be concentrated and clump directly into planetesimals via processes such as the streaming instability~\citep{youdin05,johansen09,simon16,yang17}, thus jumping over the meter barrier. The conditions for triggering the streaming instability vary in time and position within a given disk~\citep{drazkowska17,carrera17}.

\subsubsection{Pebble Accretion}\label{sec:pebble}

Once planetesimals form, they may grow by accreting other planetesimals as well as pebbles drifting inward through the disk~\citep{johansen17}. Here, ``pebbles'' are taken to be particles that drift rapidly through the gas and are typically mm- to cm-sized for typical disk parameters~\citep{ormel10,lambrechts12}.  Pebble accretion can be extremely fast under some conditions and objects can quickly grow to many Earth-masses in the giant planet region if there is a sufficient reservoir of pebbles~\citep{lambrechts14,morby15}.  Pebble accretion is self-limiting, as above a given mass~\citep[typically $10-20 \mearth$ at Jupiter's orbit][]{bitsch18} a core generates a pressure bump exterior to its orbit that holds back the inward-drifting pebbles.

\subsubsection{Gas Accretion}\label{sec:gas}

Cores that grow large enough and fast enough accrete gas from the disk. The {\em core-accretion} scenario for giant planet formation~\citep{pollack96} envisions the growth of $\sim$10 $\mearth$ cores followed by a slow phase of gas accretion.  When the mass in the gaseous envelope is comparable to the core mass, gas accretion can accelerate and quickly form Saturn- to Jupiter-mass gas giants. During this rapid accretion phase, the orbits of nearby planetesimals are destabilized and many are scattered inward, contaminating the inner planetary system~\citep{raymond17b}. Given \textcolor{black}{that most cores are not expected to grow into gas giants,} this model predicts a much higher abundance of ice giant-mass planets relative to gas giants, which has been confirmed by exoplanet statistics~\citep{gould10,mayor11,petigura13}. \textcolor{black}{On the other hand, some giant exoplanets -- in particular those at large orbital radii -- may form \textcolor{black}{rapidly} by direct gravitational collapse~\citep{boss97,mayer02,boley09}}.

\subsubsection{Orbital Migration}\label{sec:migration}

Gravitational interactions between a growing planet and its nascent gaseous disk generate density perturbations which torque the planet's orbit and cause it to shrink or grow~\citep[i.e., to {\it migrate} inward or outward,][]{kley12,baruteau14}. Migration matters for planets more massive than $\sim 0.1-1 \mearth$. In most cases migration is directed inward but the corotation torque, which depends on the local disk conditions, can in some instances be positive and strong enough to drive outward migration~\citep{kley08,paardekooper11}. In the context of the entire disk, in some regions planets migrate towards a common location, although these convergence zones themselves shift as disks evolve~\citep{lyra10,bitsch15}. Above a critical mass, a planet clears an annular gap in the gaseous disk and migration transitions to ``type 2''~\citep[as opposed to ``type 1'' for planets which do not open gaps][]{lin86,ward97,crida06}. Type 2 migration is generally slower than type 1 migration and is again directed inward in most instances.

\subsubsection{Giant Impacts}\label{sec:impacts}

Impacts between similar-sized massive objects are thought to be common in planet formation. The late phases of terrestrial planet growth are attributed to a small number of ever-larger giant impacts between growing rocky bodies~\citep{wetherill91,agnor99}. Giant impacts among large ice-rich cores have been invoked to explain the large obliquities of Uranus and Neptune~\citep{benz89,izidoro15b}. The final giant impact on Earth is believed to be the one that led to the formation of the Moon~\citep{benz86,canup01}.

\subsubsection{Late Accretion}\label{sec:late}

Giant impacts are thought to be energetic enough to trigger core formation events, which sequester siderophile (``iron-loving'') elements in the planet's core~\citep{harper96}. Highly-siderophile elements (HSEs) in a planet's mantle and crust are, therefore, considered to have been delivered by impacts with planetesimals {\em after} the giant impact phase~\citep{kimura74,day07,walker09}. This is called late accretion or the late veneer. Earth's HSE budget implies that roughly 0.5\% of an Earth mass was delivered in late accretion \textcolor{black}{\citep{walker15,morbywood15}}. 

\subsubsection{Dynamical Instability}\label{sec:instability}

After the disappearance of the gaseous disk, systems of planets may become unstable. This applies both to systems of low-mass planets (i.e., super-Earths) and gas giants~\citep{Raymond2018}. Our solar system's giant planets are thought to have undergone such an instability~\citep{tsiganis05,morby07,levison11}.  This instability can explain the giant planets' orbits and a multitude of characteristics of small body populations~\citep[for a review, see][]{nesvorny18}.  \textcolor{black}{This} ``Nice model'' was originally conceived to explain the Late Heavy Bombardment~\citep{gomes05}, a perceived spike in the impact rate on the Moon starting roughly 500 Myr after planet formation~\citep{tera74}.  However, a new interpretation of the evidence has led to the conclusion that there was probably no delayed bombardment but rather a smooth decline in the impact rate in the inner solar system~\citep{boehnke16,zellner17,morby18,hartmann19}. The instability is still thought to have occurred but may have taken place anytime in the first $\sim$100 Myr of solar system history~\citep{nesvorny18b}. The broad eccentricity distribution of exoplanets implies that instabilities are common~\citep[][the median eccentricity of giant exoplanets is $\sim 0.25$]{chatterjee08,ford08,juric08}. Instabilities in most giant planet systems are likely to have been far more violent than in our own solar system~\citep{raymond10,ida13}, and to have often disrupted growing terrestrial exoplanets and outer planetesimal disks~\citep{veras06,raymond11,raymond12}.

\subsection{Protoplanetary Disk Chemistry}
\label{sec:PPDiskChem}

Disk chemistry is responsible for the key chemical markers that can be used to trace the transport of water in the disk and this is transferred to the planetesimals as ice freezes on dust grains. \textcolor{black}{The isotopic composition is imprinted on planetesimals when they form. However, given the widespread planetesimal scattering and dynamical re-arrangement during and after planetary formation~\citep[e.g.,][]{levison08,walsh11,raymond17}, it is important to recognize that the present-day orbits of Solar System bodies may not reflect their formation locations.  For example, while Jupiter is often used as the boundary between the inner and outer Solar System, recent models suggest that the parent bodies of the carbonaceous chondrites likely originated beyond Jupiter~\citep{walsh11,kruijer17,raymond17}.}

Unique chemical signatures are thus imprinted on the icy material that is incorporated in the planetesimals that grow to form planets. Planet formation models incorporate volatile condensation onto grains \citep{Grossman1972} and water transport and condensation beyond the snow line \citep{Stevenson1988,Garaud2007}, inside which the temperatures are too warm for water ice to remain stable. 

Modern protoplanetary disk evolution models are based on chemical networks developed for the interstellar medium (ISM), and include disk structure, isotopic fractionation, and gas transport and incorporate the physics of grain growth, settling and radial migration. \textcolor{black}{Although models are always incomplete and remain crude representations of reality, recent observations have provided key constraints to help refine the models.} Models and observations \citep{Pontoppidan2014,vanDischoeck2014} reveal radial and vertical variations in thermal and chemical disk structure (Fig.~\ref{fig:Disk}). Infrared observations from {\it Spitzer} and {\it Herschel} \citep{Zhang2013,Du2014} and new spatially-resolved Atacama Large Millimeter Array (ALMA) telescope observations constrain the models \citep{Qi2013}. ALMA observations also set limits on ionization in protoplanetary disks \citep{Cleeves2014a,Cleeves2015}, providing constraints on models of deuterium-enrichment in water \citep{Cleeves2014b} due to ion-molecule reactions. These state-of-the-art disk-chemistry models \citep{Willacy2009,Jacquet2013,Yang2013} make very different testable predictions of radial isotope distributions in protoplanetary disks.

\subsubsection{Deuterium Chemistry} \label{sec:deuterium}

Observations and detailed models provide a good understanding of the complex chemistry of cold interstellar clouds in which new stars form \citep{Bergin2007b}. For example, low-temperature ion-molecule reactions drive deuterium fractionation \citep{Millar1989}. Water ice becomes enriched in deuterium, with a D/H ratio of 0.001 to 0.02 compared to a cosmic abundance of $2.6\times10^{-5}$. Physical processes in the disk control the temperature and radiation-field dependent chemistry. In the hot inner region of the disk, isotopic exchange reactions between water vapor and hydrogen gas reduce the D/H ratio to $\sim2\times10^{-5}$, the ``protosolar'' value (see Table~1). In the nebula's outer disk, D/H evolved from an initial supply of water from molecular cloud (``ISM ice'' in Fig.~\ref{fig:DH}) that was highly enriched in deuterium via ion-molecule and gas-grain reactions \citep{Herbst2003,vandishoeck13} at temperatures $<$30 K. \textcolor{black}{Indeed, high D/H water probably cannot be produced within disks themselves and must be inherited~\citep{Cleeves2014a,cleeves16}. Sharp gradients in the disk's D/H isotopic composition arise from mixing between inherited water and water that had re-equilibrated by isotopic exchange with hydrogen in the hot inner disk \citep[sometimes called the {\em protosolar nebula,}][]{Geiss1998,Lellouch2001,Yang2013,Jacquet2013}. \textcolor{black}{The radial extent of equilibration is uncertain, as stellar outbursts (also called FU Orionis outbursts) can strongly heat the disk for short periods (years to decades) and drive the snow line out to tens of au~\citep{cieza16}.}}

\subsubsection{Oxygen Isotope Fractionation} \label{sec:oxygen}

The most widely accepted mechanism for explaining the oxygen isotopic diversity among solar system materials is CO self-shielding \citep{Lyons2005}. The three oxygen isotopes ($^{16}$O,$^{17}$O,$^{18}$O) have dramatically different abundances ($\sim$2500, 1, 5, respectively). The wavelengths necessary to dissociate $^{12}$C$^{16}$O, $^{12}$C$^{17}$O, and $^{12}$C$^{18}$O are distinct, and the number of photons at each wavelength is similar in the UV continuum. At the edge of a dense molecular cloud or accretion disk, UV light dissociates the same fraction of all the three isotopologues. But as the light penetrates into the cloud or disk, the photons that dissociate the $^{12}$C$^{16}$O are depleted by absorption, so deeper in the cloud only $^{12}$C$^{17}$O and $^{12}$C$^{18}$O are dissociated. The resulting oxygen ions can either recombine into CO or combine with H$_2$ to form H$_2$O. Deeper in the cloud, the H$_2$O will be enriched in $^{17}$O and $^{18}$O. If the solar system started out with the composition of the Sun, self-shielding would have produced isotopically heavy water \textcolor{black}{in the outer parts of the disk}.  \textcolor{black}{\citet{Yurimoto2004} also suggest that this self-shielding could occur in the pre-solar molecular cloud, and this material was transported into the solar nebula by icy dust grains during the cloud collapse.  As they drifted in toward the sun and sublimated this enriched the inner disk gas.}

An alternative mechanism to explain the oxygen isotopic diversity is the Galactic Chemical Evolution (GCE) model \citep{Krot2010}, \textcolor{black}{although there are reasons why this model may not work \citep{Alexander2017}}. According to the GCE model, the solids and gas in the protosolar molecular cloud had different ages and average compositions; the solids were younger and $^{16}$O-depleted relative to the gas. According to the CO self-shielding model, O-isotope compositions of the primordial and thermally processed solids must follow \textcolor{black}{a slope 1.0 line, whereas there is no a-priori reason to believe that the GCE model results in the formation of solid and gaseous reservoirs falling on a slope 1.0 line \citep{Lugaro2012}. Patterns of oxygen isotope fractionation can be compared against these two models.} 

Models combined with {\it Genesis} observations \citep{McKeegan2011} indicate that primordial dust and gas had the same $^{16}$O-rich composition as the Sun. The result is an array of points with a slope $\sim$1 on an oxygen three-isotope plot (Fig.~\ref{fig:O}), with the initial CO plotted at the lower left (marked ``Sun'' on the diagram) and the $^{17,18}$O-rich water plotted at the upper right (marked ``Heavy Water'') \citep{Clayton1973,Yurimoto2008}. Outside the snow line, the heavy water froze on the surface of dust and settled to the disk mid-plane. Other compositions in the diagram can be produced by combining isotopically ``heavy'' water with isotopically ``light'' condensates with compositions similar to the Sun \citep{Lyons2005,Yurimoto2004}.
The total range in oxygen isotope variation seen in the solar system is \textcolor{black}{small (see range marked in Fig.~\ref{fig:O}), so distinguishing different reservoirs and formation distances requires high precision oxygen isotope information}. Self-shielding depends on UV intensity and gas densities, and therefore relates to solar distance, and time.

\subsubsection{Nitrogen Chemistry} \label{sec:nitrogen}

Nitrogen fractionation in the disk is dominated by different physical mechanisms than for D/H and oxygen, thus providing a third independent tracer. The primordial nebula and the Sun are significantly $^{15}$N-depleted relative to Earth's atmosphere \citep{Owen2001,Meibom2007,Marty2010,Anders1989}. Other reservoirs (CN and HCN in comets, some carbonaceous chondrite organics and Titan) are $^{15}$N-enriched (Fig.~\ref{fig:N}). This large fractionation may be inherited from the protosolar molecular cloud, where it is attributed to low-temperature ($<$10 K) ion-molecule reactions \citep{Anders1989} or it could have resulted from photochemical self-shielding effects in the protoplanetary disk \citep{Heays2014,Lyons2012}; the radial dependence on isotopic composition will differ significantly between these mechanisms. We expect the radial dependence of nitrogen-isotopes will be similar to those of oxygen. However, the radial dependence is additionally subject to low-temperature effects, active in the outer disk that is exposed to X-rays, making it distinguishable from oxygen.

\subsubsection{Noble Gases as a Thermometer} \label{sec:nobleT}

When water ice condenses from a gas at temperatures less than 100K, it condenses in the amorphous form and can trap other volatiles.  The amount of the trapped volatiles and their isotopic fractionation is a sensitive function of trapping temperature and pressure \citep{BarNun1985,Yokochi2012,Rubin2018}. Over the age of the solar system, \textcolor{black}{solar insolation and impacts can heat many small bodies above 137K}, the amporphous ice crystallization temperature. When the ice crystallizes, a fraction of the trapped gases is retained in the water ice \citep[possibly in the form of clathrates][]{Rubin2018,Laufer2017} and is released only when the ice sublimates. This is seen consistently in laboratory experiments \citep{BarNun1988}.  Measurements of the $^{84}$Kr/$^{36}$Ar ratio can provide a sensitive indication of temperature at which the gases were trapped \citep{Mousis2018}, and this can be linked through disk chemistry models to location in the protoplanetary disk.

\section{Models for the Origin of Earth's Water and Their Context in Planet Formation}
\label{sec:watermodel}

We now delve into the depths of models that describe exactly how Earth may have acquired its water.  There are two categories of models: those that propose that Earth's water could have been sourced locally, and those that require the {\em delivery} of water from more distant regions.  

Below we explain how each model works and its inherent assumptions. For some models that requires going into the dynamics they invoke to sculpt the solar system.  We then confront each model with the empirical constraints laid out in Section 2.1.

\subsection{Local accretion}
\label{sec:local}

To date, two models have been proposed that advocate for a local source of water on the terrestrial planets.  Here we describe those models and their challenges.

\subsubsection{Adsorption Onto Grains}\label{sec:adsorption}

It has been suggested that water could be ``in-gassed'' into growing planets~\citep[e.g.,][]{sharp17}. \citet{Drake2004} and \citet{Muralidharan2008} proposed that water vapor could be adsorbed onto silicate grains.  This would allow for a {\em local} source of water at 1 au within the Sun's planet forming disk.  Density functional theory calculations have shown that water vapor can indeed be adsorbed onto forsterite~\citep{muralidharan08} or olivine~\citep{stimpfl06,asaduzzaman14} grains.  In principle, this mechanism can explain the accretion of multiple oceans of water onto Earth. 

There are \textcolor{black}{four} apparent issues with this model. \textcolor{black}{First, it cannot account for the delivery of other volatiles to Earth including carbon, nitrogen and the noble gases. In fact, this model would require an additional, H-depleted source for these species.} Second, while the mechanism may accrete a water budget of perhaps a few oceans, it does not account for collisional loss of water as planetesimals grow, which can be substantial \citep{genda05}.  Second, it begs the question of why the enstatite chondrite meteorites -- which offer a close match to Earth's composition~\citep[albeit with some important differences, such that Earth cannot be made entirely of enstatite condrites; e.g., ][]{dauphas17} -- failed to incorporate any water and are extremely dry.  Third, this mechanism implies that Earth's water should have the same isotopic signature as the nebular gas.  However, the D/H ratio of the disk gas is likely to have been the same as the Sun's, which is a factor of 6.4$\times$ lower than Earth's oceans \citep[see][and Section~\ref{sec:intro}]{Geiss1998}. Finally, disk evolution models find that an individual parcel of gas in the vicinity of the snow line moves radially much faster than the snow line does \citep{morby16}.  This means that the water vapor in the inner disk likely does not remain but rather moves inward and is accreted onto the star much more quickly than it can be replenished. The inner disk's gas is therefore dry for a large fraction of the disk lifetime. 

\subsubsection{Oxidation of a Primordial H-rich Atmosphere} \label{sec:oxidation}

\citet{Ikoma2006} also proposed that Earth's water could have been acquired locally, but as a result of reactions between Earth's primordial atmosphere and surface.  During the magma ocean phase, atmospheric hydrogen can be oxidized by gas-rock interactions and produce water.  A magma ocean phase that meets the criteria for this mechanism is expected for planets more massive than $
\sim 0.3 \mearth$~\citep{Ikoma2006}, although the duration of a magma ocean is also a strong function of orbital distance~\citep{Hamano2013}.  

This mechanism produces terrestrial water directly from nebular gas. Yet the D/H ratio of nebular gas is a factor of $\sim$6.4$\times$ lower than Earth's~\citep{Geiss1998}.  However, most of this early hydrogen-rich atmosphere had to escape (Jean's escape), and the escape process leads to fractionation and an increase in the D/H ratio.  \cite{Genda2008} showed that the D/H of Earth's water can increase to its present-day value for certain values of the efficiency and timescale of hydrogen escape.  \textcolor{black}{However, the collateral effects of this loss on the isotopic fractionation of other species such as the noble gases remains unaddressed. Given that nitrogen should not be fractionated by this process, it appears to imply that Earth should have a Solar nitrogen isotopic composition, \textcolor{black}{which is not the case~\citep{Marty2012}.} }

Measurements of water thought to be sourced from an isolated deep mantle reservoir yield D/H values closer to the nebular one~\citep{Hallis2015}.  This indicates that at least a fraction of Earth's water \textcolor{black}{may have} a nebular origin.  There is also circumstantial evidence from noble gases to support the idea that Earth had a primordial nebular atmosphere~\citep{dauphas03,williams19}.  

It remains to be seen whether nebular gas can explain the origin of the bulk of Earth's water.  This mechanism clearly requires Earth to have accreted and later lost a thick hydrogen envelope.  While such a process is a likely outcome of growth within the gaseous disk: \textcolor{black}{and there is evidence that many $\sim$Earth-sized exoplanets have thick Hydrogen atmospheres~\citep[e.g.][]{wolfgang16}}, it seems a cosmic coincidence for the parameters to have been right for Earth to end up with the same D/H ratio (and $^{15}$N/$^{14}$N ratio) as carbonaceous chondrites~\citep{Marty2012}. \textcolor{black}{Carbonaceous chondrite-like objects thus seem a more likely source for Earth's water.}

\subsection{Water Delivery}
\label{sec:delivery}

\textcolor{black}{An alternate explanation for the origin of Earth's water is delivery from the outer Solar System.} The following models invoke sources of water that are separated from Earth's orbit.  Hence water is ``delivered'' from these regions to the Earth.  These models are driven by a number of processes, including a combination of the moving snow line and drifting pebbles (Section~\ref{sec:snow}), widening of terrestrial planets' feeding zones (Section~\ref{sec:feedingzone}), gravitational scattering of planetesimals during the growth and migration of the giant planets (Section~\ref{sec:pollution}), and inward migration of large planetary embryos (Section~ \ref{sec:migration}).

\subsubsection{Pebble ``Snow''}\label{sec:snow}

\begin{figure}[t]
\begin{center}
\includegraphics[width=8cm]{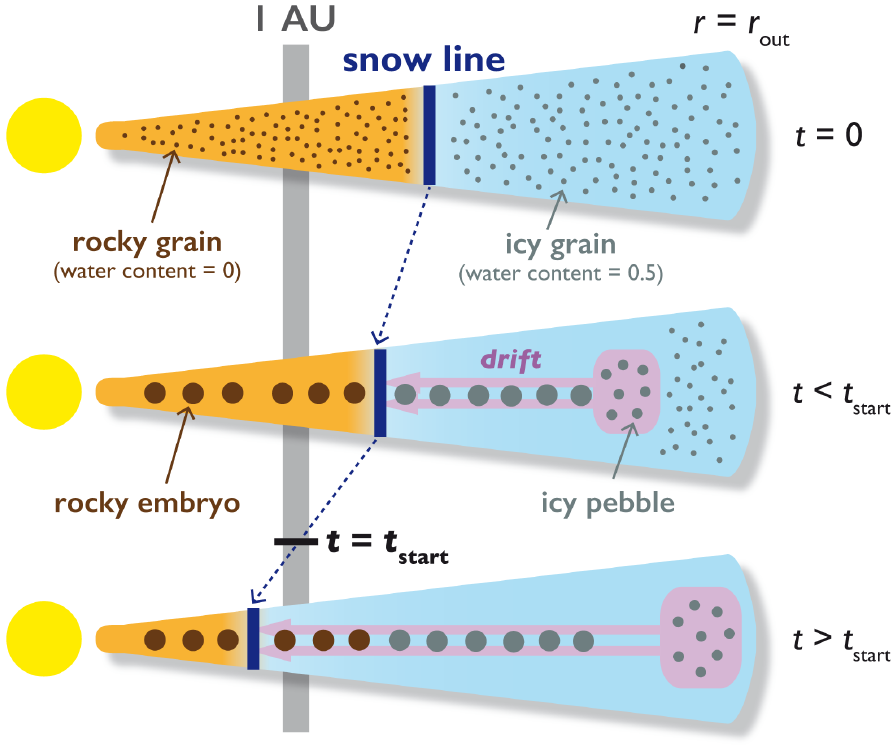}
\caption{Snapshots in time of the evolution of a disk, showing how the snow line sweeps inward as the disk cools.  Water may thus be delivered to rocky planets at 1 au by pebbles as they drift inward. We refer to this mechanism in the text as {\em pebble snow}. The pink region represents the ``pebble production front'', the outward-moving radial location in the disk where pebbles grow from dust and start to drift inward. From \citet{sato16}.}
\label{fig:Sato}
\end{center}
\end{figure}

As disks evolve and thin out, they cool down.  Condensation fronts move inward~\citep{dodson-robinson10}, including the water-ice snow line, located at $T\sim 150$~K. The exact evolution of the snow line depends on the thermal evolution of the disk and therefore on the assumed heating mechanisms, the most important of which are stellar irradiation and viscous heating. For alpha-viscosity disk models, the snow line starts in the Jupiter-Saturn region and moves inside 1 au within 1 Myr or so~\citep{sasselov00,lecar06,kennedy08}. The snow line also moves inward and generally ends up inside 1 au in more complex disk models~\citep{oka11,martin12,bitsch15,bitsch19b}. However, it is worth noting that radial variations in viscosity may under certain assumptions keep the snow line outside 1 au for the disk lifetime~\citep{kalyaan19}. Figure~\ref{fig:Sato} shows how a planet forming in a close-in, rocky part of the disk can become hydrated as the snow line sweeps inward~\citep{sato16}.  

\begin{figure*}[!t]
\begin{center}
\includegraphics[width=14cm]{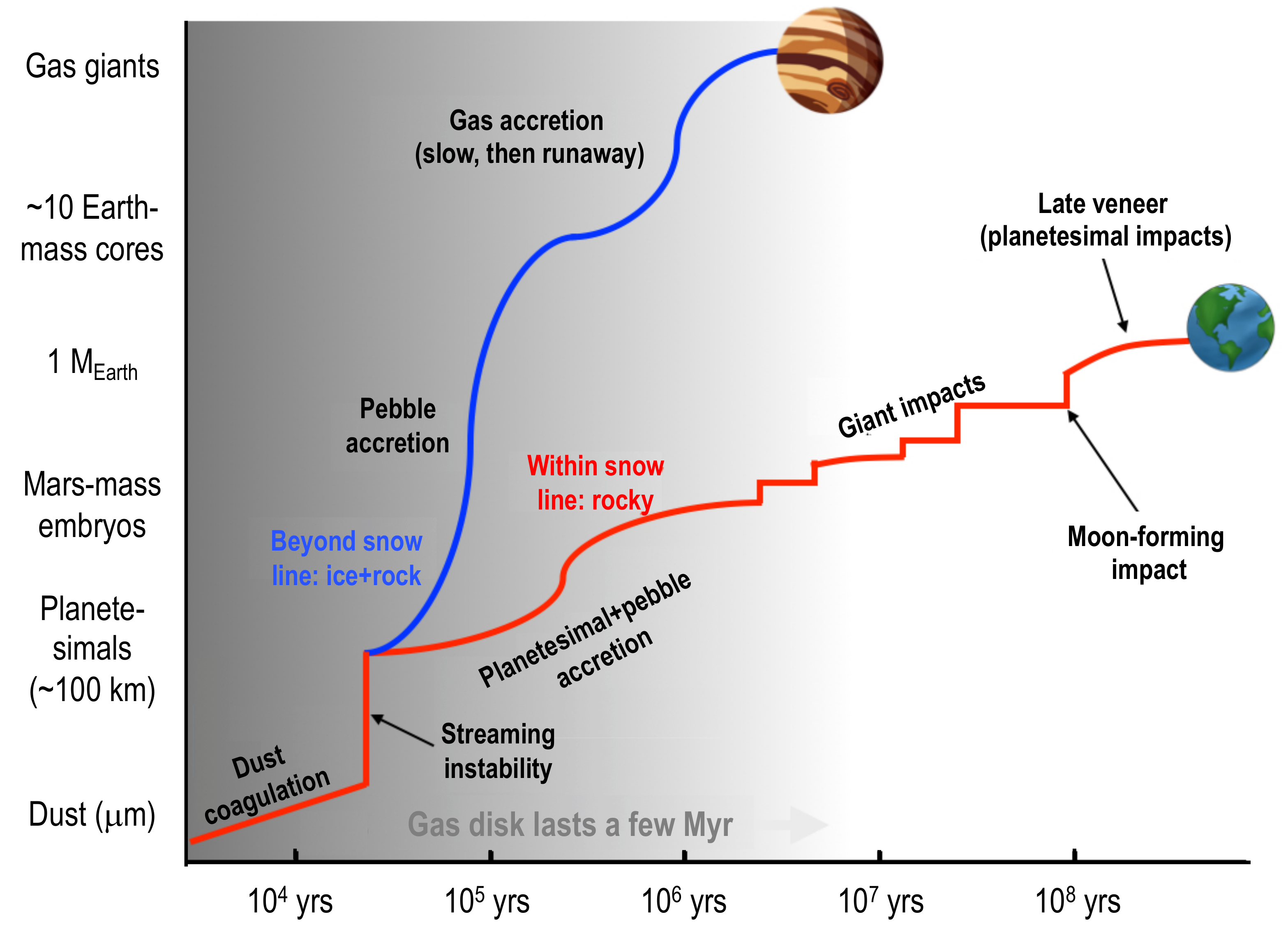}
\caption{A rough summary of the current understanding of the growth history of rocky and gas giant planets, illustrating the various planet formation processes. The cartoon images of Earth and Jupiter are from www.kissclipart.com.}
\label{fig:cartoon}
\end{center}
\end{figure*}

\textcolor{black}{Planetesimals are thought to form from dust and pebbles that are locally concentrated by drifting within the gas disk, followed by a phase of further concentration (e.g., by the streaming instability) to produce gravitationally bound objects~\citep[see, e.g., the review by][]{johansen14}. When planetesimals form, their compositions ``lock in'' the local conditions at that time. Yet most of the mass in solid bodies remains in dust and pebbles, which are small enough that their compositions likely change as they drift inward, in particular by losing their volatiles. Planetesimals continue to grow, in part by accreting pebbles (see Fig.~\ref{fig:cartoon} for a cartoon representation of the different phases of growth).  The planetesimal and pebble phases overlap for the entire gas disk lifetime.}

The evolution of the pebble flux controls the water distribution within the solids in the disk. When the snow line sweeps inward, the source of water is not condensing gas but inward-drifting particles~\citep[i.e., pebbles][]{morby16}. This is because the gas's radial motion is faster than the speed at which the snow line moves.  As the snow line moves inward, it does not sweep over water vapor that can condense.  Rather, the gas interior to the snow line is dry (or mostly dry) simply because it moves more quickly than the snow line itself.  The source of water at the snow line is instead in the form of ice-rich pebbles that drift inward from farther out in the disk.

There are two ways in which the flux of pebbles drifting inward through the disk can drop significantly: the pebble supply can be exhausted or the pebble flow can be blocked.

Dust within the disk coagulates to sizes large enough to drift inward rapidly~\citep[e.g.,][]{birnstiel12,birnstiel16}.  However, dust grows into pebbles faster closer-in to the star, where accretion timescales are short and densities high. So the dust is consumed faster in the disk interior, resulting in an outward-moving front at which pebbles are produced, after which the pebbles drift inward~\citep{lambrechts14,ida16}. When this {\it pebble production front} reaches the outer edge of the disk, the pebble flux drops drastically, as the source of pebbles is exhausted.  Taking this into account limits the degree to which an inward-sweeping snow line can hydrate rocky planets \textcolor{black}{because the mass in water-bearing pebbles drops off drastically in time}. Nonetheless, in many cases this mechanism can deliver Earth-like water budgets~\citep{ida19}.

The pebble flux can also be blocked, either by growing planets or by structures within the disk itself. Drifting particles follow the local pressure gradient~\citep{haghighipour03}, which drives pebbles monotonically inward in a perfectly smooth disk.  However, pressure ``bumps'' act as traps for inward-drifting particles. These are radially-confined regions in which the gas pressure gradient becomes high enough that the gas orbits at the Keplerian speed, thus eliminating the headwind and associated drag forces on pebbles. Such traps may exist naturally within the disk, or they can be produced by growing planets. Once a planet reaches a critical mass~\citep[of roughly $20 \mearth$ at Jupiter's orbit for typical disk parameters][]{lambrechts14} it generates a pressure bump exterior to its orbit, which acts as a trap for inward-drifting pebbles~\citep{morby12,lambrechts14,bitsch18}.  This not only starves the planet itself but also all other planets interior to its orbit.  

When the pebble flux is blocked by a growing planet, it renders the concept of the snow line ambiguous.  Given that inward-drifting pebbles are the source of water, the location at which the temperature drops below 150~K continues to move inward in the disk but does not bring any water along with it \textcolor{black}{(recall that the gas is dry because it moves much faster than the snow line)}.  In this way, the water distribution within a disk is ``fossilized'' at the time when an outer planet first grew large enough to block the pebble flux~\citep{morby16}. This fossilization is analogous to the snow line on a mountain, which marks the location at which the temperature reached zero Celsius while it was snowing (i.e., while pebbles were drifting).  Once it stops snowing, the snow line on a mountain is no longer linked with the local temperature \textcolor{black}{(as long as it does not warm up past the freezing point)}. After the inward drift of pebbles is cut off, redistribution of water within the disk requires dynamical processes that transport objects at larger size scales (e.g., via migration or gravitational scattering as invoked by the other water delivery mechanisms).

Can the pebble snow mechanism explain the origin of Earth's water?  The {\em pebble snow} mechanism can produce planets with water contents similar to Earth's~\citep{ida19}. However, understanding whether Earth could have accreted a large enough contribution from inward-drifting pebbles requires an understanding of the chemical properties of Earth's building blocks.  

Nucleosynthetic isotope differences are seen in a number of elements between the two main classes of meteorites: carbonaceous and non-carbonaceous \citep{warren11,kruijer17}.  These two populations appear to have been sourced from different reservoirs within the planet-forming disk, whose origins remain debated \citep{nanne19}. The rapid growth of Jupiter's core has been invoked as a mechanism to keep the two populations separate, by preventing the drift of outer, carbonaceous pebbles into the inner solar system, which is thought to have been dominated by non-carbonaceous material \textcolor{black}{\citep{budde16,kruijer17}}.  

Earth's water's D/H ratio is well-matched by carbonaceous chondrite meteorites (see Section~\ref{sec:DH} and Fig.~\ref{fig:DH}). The pebble snow model would thus invoke carbonaceous pebbles as the source of Earth's water, delivered late enough that the disk had cooled to the point that the snow line was located inside Earth's orbit. 

However, the pebble snow model does not appear to match empirical constraints. \textcolor{black}{The latest-forming chondrites have ages that extend to roughly 4 Myr after CAIs~\citep{Sugiura2014,Alexander2018,Desch2018},} likely the full length of the disk lifetime. This seems to indicate that the two reservoirs remained spatially separated throughout.  In other words, there are no known classes of chondritic meteorites with nucleosynthetic anomalies that lie in between, which would be a signature of the mixing.

This early separation between carbonaceous and non-carbonaceous pebble reservoirs seems to preclude drifting pebbles as the source of Earth's water.  If water-rich carbonaceous pebbles drifted inward to deliver water to Earth, then it should follow that the two reservoirs should have mixed and some meteorites should exist with intermediate compositions, which is not the case.


Dynamical models naturally implant carbonaceous planetesimals from Jupiter's orbit and beyond into the asteroid belt and terrestrial planet region~\citep[discussed in detail in Section~\ref{sec:pollution}; ][]{walsh11,raymond17}. This may indicate that the non-carbonaceous S-type asteroids formed in the inner solar system whereas the carbonaceous C-types were implanted as planetesimals. 

\subsubsection{Wide Feeding Zones} \label{sec:feedingzone}

In the well-studied framework of the {\em classical} model of terrestrial planet formation, the late phases of terrestrial accretion occur after the formation of the giant planets~\citep{wetherill91,wetherill96,chambers01,raymond06b,raymond09c,raymond14,obrien06,morishima10,fischer14,izidoro15c,kaib15}. \textcolor{black}{In Fig.~\ref{fig:cartoon}, late-stage accretion starts after dispersal of the gaseous disk, during the giant impact phase.}

\begin{figure}[t]
\begin{center}
\includegraphics[width=8cm]{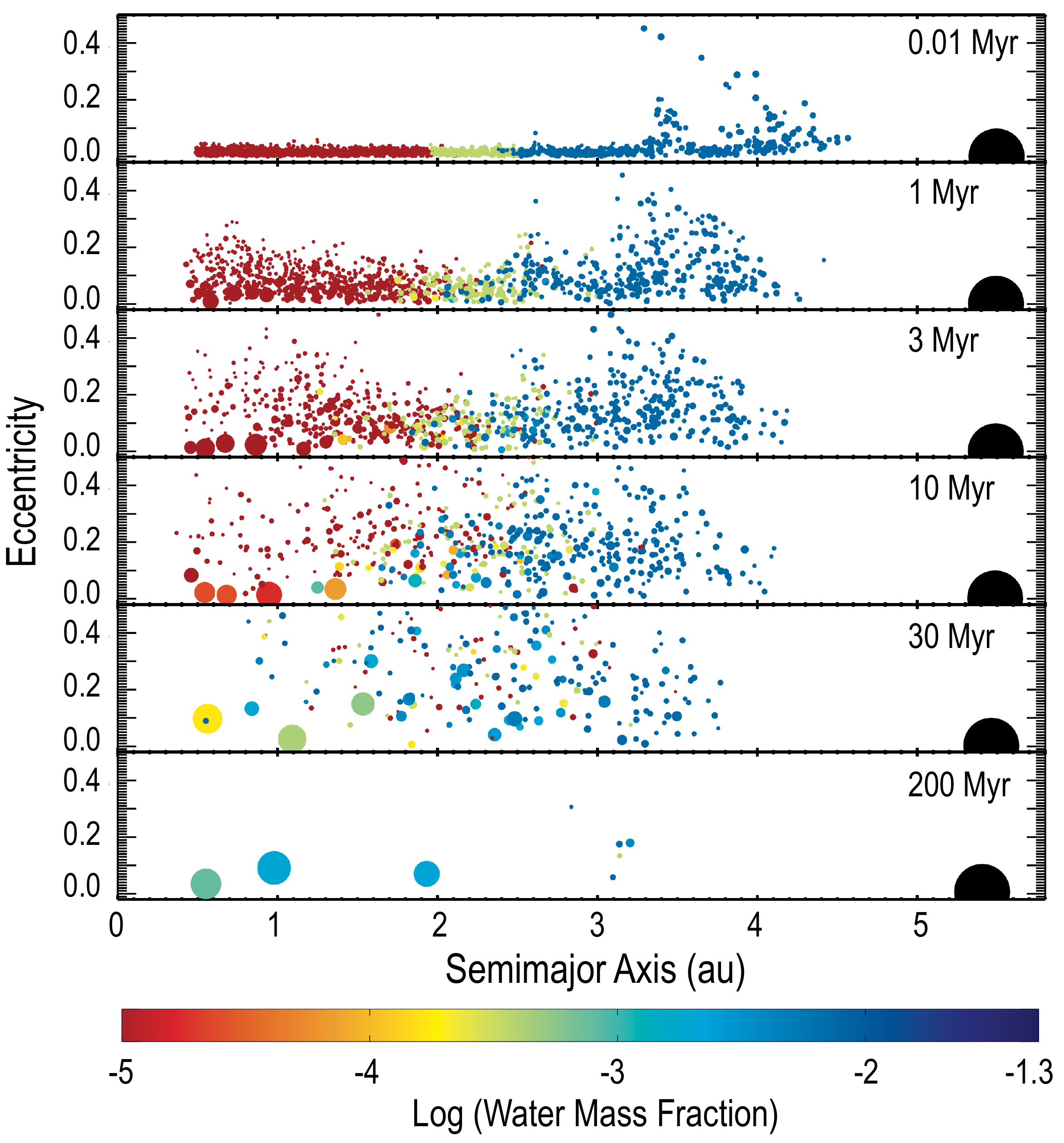}
\caption{Snapshots from a simulation of the classical model of terrestrial planet formation. Jupiter was included from the start, on a low-eccentricity orbit at 5.5 au (large black circle). Almost 2000 planetary embryos are represented by their relative sizes (proportional to their masses$^{1/3}$) and their water contents, which were imposed at the start of the simulation (red = dry, black = 5\% water by mass; see color bar). 
Three terrestrial planets formed, and each acquired material from beyond 2.5 au that delivered its water. Adapted from \cite{raymond06b}.}
\label{fig:classical}
\end{center}
\end{figure}

The late stage of terrestrial accretion (essentially starting from the ``giant impacts'' phase shown in Fig.~\ref{fig:cartoon}) thus starts from a population of roughly Mars-mass planetary embryos embedded in a sea of planetesimals, under the dynamical influence of the already-formed giant planets.  

Figure~\ref{fig:classical} shows an example simulation of the classical model. The compositional gradient in the present-day solar system is assumed to represent the initial conditions for planet formation: inside roughly 2.5 au objects are dry and between 2.5 au and Jupiter's orbit they have water contents of 5-10\% similar to carbonaceous chondrites~\citep{morby00,raymond04}.

In the inner disk, there is an effective wave of growth that sweeps outward in time, driven by self-gravity among the growing embryos~\citep{kokubo00}.  In the asteroid region, eccentricities are excited by Jupiter via secular and resonant forcing.  Dynamical friction keeps the most massive planetary embryos on near-circular orbits while planetesimals and smaller embryos often have eccentric and inclined orbits~\citep{obrien06,raymond06b}.  

Collisional growth continues for 10-100 Myr as the planets grow by giant impacts.  At the end of the simulation, three terrestrial planets have formed~\citep[see][for details]{raymond06b}. These include reasonable analogs to Venus and Earth and a planet close to Mars' orbit that is roughly ten times more massive than the actual planet.

The feeding zones of all three planets included a tail that extended into the outer asteroid belt. The water content of each planet in this simulation was thus sourced from the outer asteroid belt.

The wide feeding zones of the planets in Fig.~\ref{fig:classical} are a generic feature of late-stage accretion.  Eccentricity excitation implies that any planet's building blocks sample a wide region.  Water delivery is, therefore, a robust outcome of classical model-type accretion.

But the classical model has an Achilles heel: Mars.  Classical model simulations systematically form Mars ``analogs'' that are 5-10 times more massive than the real planet~\citep[quantified by][]{raymond09c,morishima10,fischer14,izidoro15c,kaib15}.  The problem is not Mars' absolute mass but the fact that it is so much less massive than Earth.  Accretion tends to form systems in which neighboring planets have comparable masses~\citep[e.g.,][]{lissauer87}. Models that succeed in reproducing the inner solar system invoke mechanisms to deplete Mars' feeding zone relative to Earth's (summarized in Fig.~\ref{fig:3models}). 

The {\em Early Instability} model~\citep{clement18,clement19,clement19b} matches the terrestrial planets (including the Earth/Mars mass ratio) while preserving many of the assumptions of the classical model. It assumes that the giant planet instability~\citep[sometimes referred to as the Nice model instability \textcolor{black}{because it was developed in the French town of Nice};][]{tsiganis05,morby07} took place shortly after the dissipation of the gaseous disk. 

Perturbations during the giant planets' instability act to strongly excite and deplete the asteroid belt and Mars region without strongly affecting the region within 1 au. Simulations match the terrestrial planets' mass distribution and the rate of success in matching the inner solar system correlates with that in matching the outer solar system~\citep{clement18}.  

In the Early Instability model, water is delivered to the growing terrestrial planets from the outer asteroid region in the same way as in Fig.~\ref{fig:classical}. That water would presumably have the same chemical fingerprint as today's C-type asteroids, represented by carbonaceous chondrites, and thus match the Earth.

The possibility that Earth's water could be a result of its wide feeding zone therefore rests on the viability of the Early Instability model itself.  To date there are three successful models that can explain the early evolution of the inner solar system~\citep[see][and also Fig.~\ref{fig:3models}]{Raymond2018}. Future studies will use empirical and theoretical arguments to evaluate these models.

\subsubsection{External Pollution} \label{sec:pollution}

Water may be delivered by a relatively low-mass population of volatile-rich planetesimals that ``rain down'' onto the terrestrial planet-forming region. In this scenario, the terrestrial planets would have formed predominantly from local rocky material but with a small amount of {\em external water-bearing pollution}. \textcolor{black}{The difference between this model and the classical model is that the polluting planetesimals are not simply an extension of the planets' feeding zones but rather were dynamically injected from more distant regions of the planet-forming disk.} These water-bearing planetesimals would have been scattered on high-eccentricity orbits by the growth and/or migration of the giant planets during the late parts of the gaseous disk phase. 

\begin{figure}[t]
\begin{center}
\includegraphics[width=8cm]{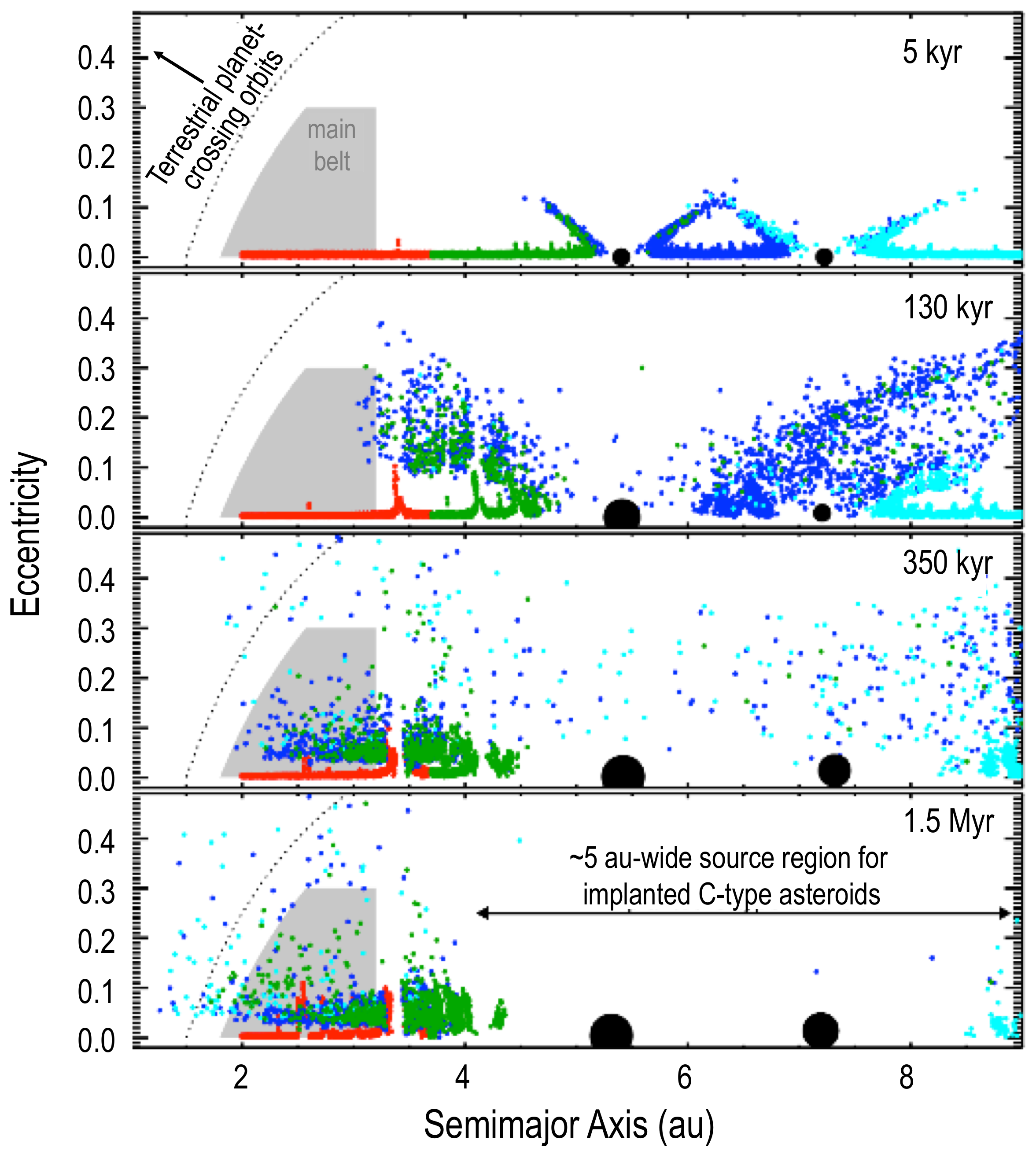}
\caption{Dynamical injection of planetesimals into the inner solar system during Jupiter and Saturn's growth. This figure shows snapshots of a simulation in which 100-km planetesimals interact gravitationally with the growing gas giants and by gas drag with the disk. Jupiter's mass was increased from a core to its final mass from 100-200 kyr and Saturn from 300-400 kyr. The colors of planetesimals serve to indicate their starting location.  A large number of planetesimals were captured on stable orbits in the outer asteroid belt, providing a good match to the C-type asteroids. Many planetesimals were scattered onto high-eccentricity orbits that cross the growing terrestrial planets' and may have delivered water to Earth. The source region for implanted planetesimals was from 4 to 9 au in this example, but can extend out to 20 au when migration and the ice giants are considered. From \cite{raymond17}.
}
\label{fig:injection}
\end{center}
\end{figure}
To date, two mechanisms have been proposed to produce a population of high-eccentricity planetesimals. The first is a general mechanism that applies to every instance of giant planet growth~\citep{raymond17}.  The second is inherently tied to the Grand Tack model~\citep{walsh11}.

In the core-accretion model~\citep{pollack96}, gas giant planets grow in two steps (see Fig.~\ref{fig:cartoon}).  First they accrete large solid cores of $5-20\mearth$~\citep[likely by pebble accretion; e.g.,][]{lambrechts12}.  Then they accrete gas from the disk.  Gas accretion proceeds slowly until a critical threshold is reached (likely the point at which the mass in the gaseous envelope is comparable to the solid core mass), after which accretion accelerates and the planet rapidly grows into a Saturn- to Jupiter-mass planet~\citep[e.g.,][]{lissauer09} and carves an annular gap in the disk~\citep{crida06}.

Figure~\ref{fig:injection} shows how the growth of gas giant planets affects nearby planetesimals~\citep[from][]{raymond17}.  When a planet such as Jupiter undergoes a phase of rapid gas accretion it destabilizes the orbits of nearby objects. Planetesimals undergo close encounters with the growing Jupiter and are scattered onto eccentric orbits across the solar system. Meanwhile, gas drag acts to damp the planetesimals' eccentricities. Planetesimals scattered inward can thus become decoupled from Jupiter as their aphelia decrease, and be trapped onto stable orbits in the inner solar system~\citep{raymond17,ronnet18}.  Saturn's growth has a similar effect, although planetesimals must be scattered first by Saturn and then by Jupiter to reach the inner solar system. Because Saturn is thought to have grown later than Jupiter when the planet-forming disk had evolved and was lower in mass than it was during Jupiter's formation, many planetesimals are scattered inward {\it past} the asteroid belt to the terrestrial planet region.  Many planetesimals destabilized by Saturn may also be captured in Jupiter's circumplanetary disk~\citep{ronnet18}.

The same population of scattered planetesimals end up crossing the terrestrial planets' orbits and populating the outer asteroid belt~\citep{raymond17}.  This mechanism naturally explains why carbonaceous chondrites (from C-type asteroids) are a chemical match to Earth's water.

The balance between planetesimals implanted into the belt and scattering toward the terrestrial zone depends on unknown parameters. The most important parameter is simply the strength of gas drag, which depends on a combination of the planetesimal size and the gas surface density in the inner disk at the time of planetesimal scattering. It is likely that there were many generations of planetesimal scattering into the inner solar system: during Jupiter and Saturn's growth and possible migration and the ice giants' growth and migration. In the example simulation from Fig.~\ref{fig:injection}, planetesimals are implanted from a 5 au-wide swath of the disk.  However, this is a {\em minimum} width, as taking migration and the ice giants into account can extend the source region to past 20 au~\citep{raymond17}.

\begin{figure*}
\begin{center}
\includegraphics[width=15cm]{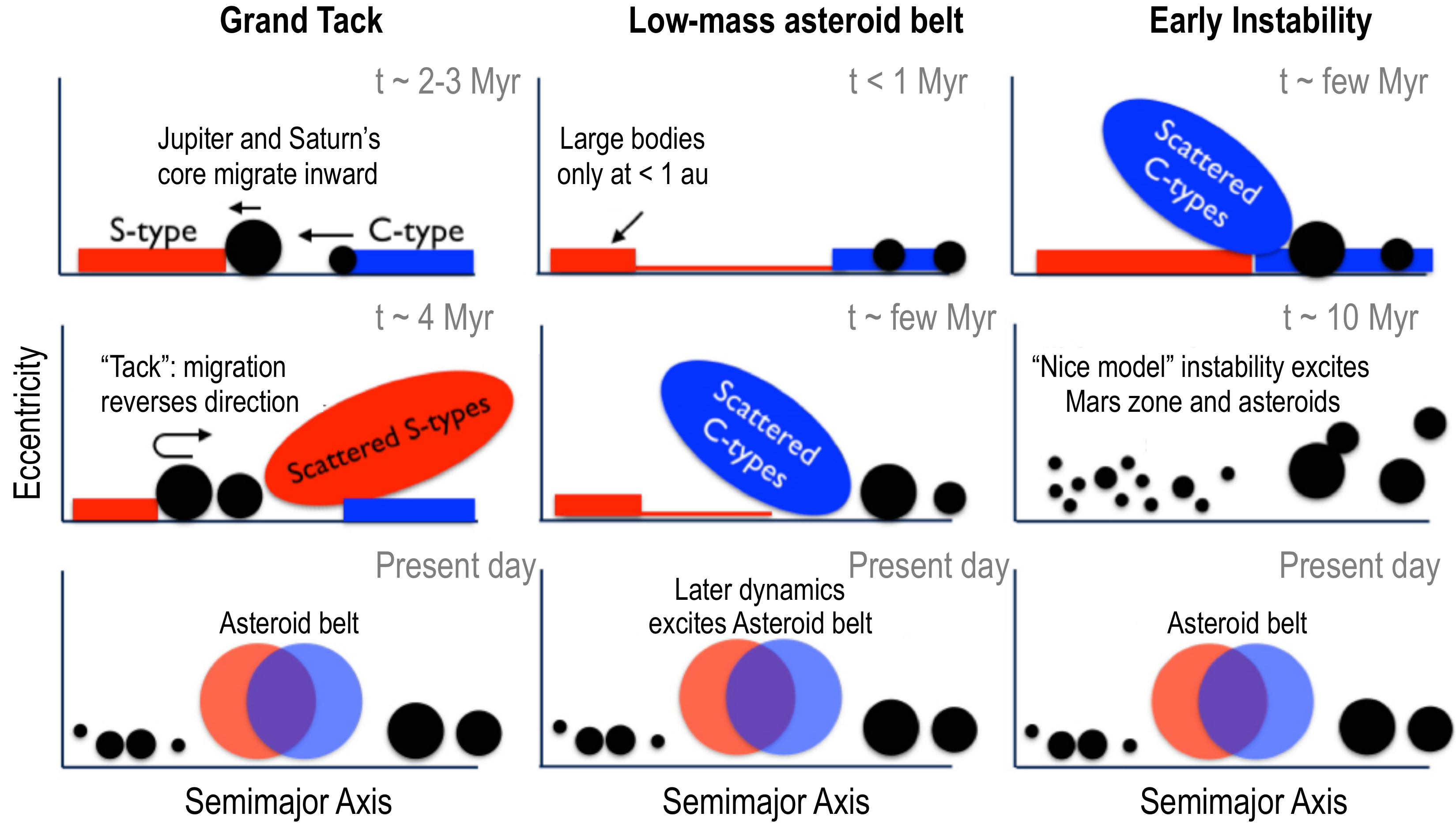}
\caption{Cartoon of the evolution of three models that can match the inner solar system.  From \cite{Raymond2018}.}
\label{fig:3models}
\end{center}
\end{figure*}

The mechanism illustrated in Fig.~\ref{fig:injection} is generic and applies to any instance of giant planet formation~\citep{raymond17}. It may thus explain the initial conditions of the classical model. This mechanism has also been shown to be robust to a number of migration histories for the giant planets~\citep{raymond17,ronnet18,pirani19}.

This mechanism also naturally provides a source of C-types and terrestrial water for the {\em Low-mass Asteroid Belt} model, another viable model for terrestrial planet formation.  The Low-mass Asteroid Belt model proposes that the terrestrial planets did not form from a broad disk of rocky material but rather from a narrow annulus~\citep[for details see][and Fig.~\ref{fig:3models}]{Raymond2018}. In this model, the large Earth/Mars mass ratio is a simple consequence of a primordial mass deficit in the Mars region~\citep[][this would also explain the large Venus/Mercury mass ratio]{hansen09,kaib15,raymond17b}. Perhaps planetesimal formation was simply more efficient at 1 au and 5 au than in between.  ALMA has indeed found a number of young circumstellar disks containing rings of dust~\citep{alma15,Andrews2018b}. Given that planetesimal formation is only triggered in regions of high dust density~\citep{carrera15,yang17}, it is plausible to imagine that planetesimals also form in rings.  The Low-mass Asteroid Belt model can plausibly match the terrestrial planets and asteroid belt and is on the same footing as the Early Instability and Grand Tack models.

The second mechanism of water delivery by external pollution depends on {\em outward} migration in the framework of the Grand Tack model~\citep[][; see also Fig.~\ref{fig:cartoon}]{Walsh2011,jacobson14b,raymond14c,brasser16}.  The growing Jupiter would have carved a gap in the gaseous disk and migrated inward in the type 2 regime.  Saturn would have grown later farther out, migrated inward and become trapped in mean motion resonance with Jupiter~\citep{morby07,pierens08}. \textcolor{black}{Two planets are in mean motion resonance when their orbital periods form the ratio of small integers; e.g., in the 3:2 resonance the inner planet orbits the star three times for every two orbits of the outer planet, at which time the planets re-align.} After this point the two planets would have shared a common gap in the disk and tilted the torque balance so as to migrate outward together~\citep{masset01,morby07,crida09}. This outward migration mechanism operates when two planets share a common gap with the innermost being more massive. Hydrodynamical simulations find that Jupiter and Saturn can migrate outward in the 3:2 or 2:1 resonances~\citep{masset01,morby07,zhang10,pierens11,pierens14}. Jupiter and Saturn would thus have migrated outward together until the disk dissipated.
 
Jupiter's migration would have dramatically sculpted small body populations~\citep{walsh11,walsh12}.  During its inward migration, Jupiter pushed most of the inner rocky material further inward by resonant shepherding~\citep[see][]{fogg05,raymond06c}, which acted to compress a broad disk of rocky material into a narrow annulus (similar to the one invoked for the Low-mass Asteroid Belt model). A fraction of rocky planetesimals were scattered onto wider orbits, some scattered out to the Oort cloud \citep{Meech2016}. At the turnaround point of its migration Jupiter was 1.5-2 au from the Sun~\citep{walsh11,brasser16}. Then, during Jupiter and Saturn's outward migration the giant planets first encountered the scattered rocky planetesimals and then pristine outer-disk planetesimals. As the giant planets migrated through these small bodies most were ejected, but a small fraction were scattered inward and left behind on stable orbits once the planets migrated past. This mechanism is less dependent on the planetesimal size than the one illustrated in Fig.~\ref{fig:injection} because the orbits of scattered planetesimals are stabilized by the giant planets migrating away rather than by gas drag. In the Grand Tack model, the surviving planetesimals are trapped in the asteroid belt with a similar orbital distribution to the observed one~\citep{deienno16}. 

In the Grand Tack model, water-rich planetesimals are scattered into the terrestrial planet-forming region from beyond Jupiter's orbit~\citep{walsh11,obrien14}.  The mass in water-delivering planetesimals can be calibrated to the mass in planetesimals trapped in the asteroid belt. Taking into account later depletion of the belt~\citep{minton10,nesvorny15}, the mass in polluting planetesimals is a few to ten percent of an Earth mass. The growing Earth accretes enough water to account for its current water budget~\citep{walsh11,obrien14}.

External pollution represents a viable scenario for water delivery. Inward scattering of planetesimals is an inevitable byproduct of giant planet formation~\citep[see Fig.~\ref{fig:injection} and ][]{raymond17}. The Grand Tack model matches a number of characteristics of the inner solar system. The main uncertainty lies in the outward migration mechanism~\citep[see][for a discussion]{raymond14b}, which requires that Jupiter remain substantially more massive than Saturn during the entire accretion phase~\citep{masset01}.

The astute reader may ask themself how the external pollution mechanism differs from the old comet-delivery model.  That model proposed that Earth grew locally and  later received its water via a bombardment of comets~\citep[e.g.,][]{delsemme92,Owen1995}. In contrast with the cometary model, the external pollution model invokes (1) self-consistent, global dynamical scenarios that explain the origin of Earth's water in the context of models that match the architecture of the inner solar system; and (2) does not invoke comets but rather polluting objects that originate from the same reservoir as C-type asteroids, which match Earth's D/H and $^{15}$/$^{14}$N ratios (comets do not match the Earth's $^{15}$N/$^{14}$N ratio; see Fig.~\ref{fig:N}).

\subsubsection{Inward Migration} 
\label{sec:migration}

Orbital migration is a ubiquitous process in planet formation.  Given that planets form in gaseous disks, gas-driven migration is simply inevitable once planets reach a critical mass~\citep{kley12,baruteau14}.

Migration is almost certainly a key process in the formation of so-called {\em super-Earths} and {\em sub-Neptunes}, which exist around 30-50\% of main sequence stars~\citep{howard12,mayor11,petigura13,mulders18}. For virtually any disk profile, these planets should have migration timescales that are far shorter than the disk lifetime~\citep[e.g.,][]{ogihara15}. Models that invoke the migration of growing planetary embryos can quantitatively match the observed super-Earth distributions~\citep[e.g.,][]{ida10,cossou14,izidoro17,izidoro19,ogihara18}.

There is good reason to think that planetary embryos massive enough to undergo long-range migration form past the snow line.  Planetesimals may form by the streaming instability most readily just past the snow line~\citep{armitage16,drazkowska17}.  Pebble accretion is also much more efficient past the snow line; by the time $10 \mearth$ ice-rich cores have formed at 5 au, rocky planetary embryos in the inner disk may only reach $\sim 0.1 \mearth$~\citep{morby15}. This fits nicely with our picture of solar system formation, which requires a population of $\sim$Mars-mass rocky embryos in the inner disk and a handful of giant planet cores in the outer disk.  

If embryos large enough to migrate do indeed form past the snow line, then many are likely to be ice-rich.  However, if embryos form at the snow line and migrate inward while continuing to accrete, they may only be $\sim 5-10\%$ water by mass, a far cry from the 50\% that is often assumed. \textcolor{black}{That assumption is questionable given that the most water-rich \textcolor{black}{meteorites (including both components, the chondrites and matrix)} have water-to-rock ratios of $\sim0.4$ despite appearing to be outer Solar System objects~\citep{alexander19a,alexander19b}.} In addition, while the inward migration of icy super-Earths perturbs the growth of terrestrial planets~\citep{izidoro14}, it may lead to the formation of very close-in planets that are entirely rocky~\citep{raymond18b}.  By all of these avenues, migration tends to produce planets whose feeding zones are disconnected from their final orbital radii~\citep[e.g.,][]{kuchner03}. 

Could inward migration explain the origin of Earth water?  Probably not. Multiple lines of (admittedly circumstantial) evidence suggest that the building blocks of the terrestrial planets were roughly Mars-sized~\citep[see][]{morby12}. Mars is below the mass required for long-range migration.  In addition, if Earth or its building blocks did migrate inward then it is hard to understand why they would have stopped where they did rather than migrating closer to the Sun.

However, migration may indeed play a role in delivering water to Earth-like planets in other systems.  This may be particularly important for planets orbiting low-mass stars (see Section~\ref{sec:summary}).  

\section{Discussion}
\label{sec:discuss}

As discussed in Section~\ref{sec:watermodel}, a number of different scenarios can in principle match the amount and isotopic composition of Earth's water.  However, some of them do not fit in a clear way within a self-consistent picture of the dynamical and chemical evolution of the solar system.  On the other hand, other mechanisms are essentially inevitable, as they are simple byproducts of planet formation~\citep[for instance, the growth of a giant planet invariably pollutes its inner regions with water-rich planetesimals:][]{raymond17}.

\subsection{Evaluation of Water Models}
\label{sec:summary}

We now evaluate critically the six scenarios from Section~\ref{sec:watermodel}.  We find that the external pollution mechanism is currently the most likely candidate.

While physically motivated, it is hard to see how the two scenarios for local water accretion could fit in a bigger picture of solar system formation.  The adsorption of water vapor onto grains struggles because (1) the gas in the inner solar system was likely mainly dry; (2) the D/H ratio of adsorbed water should in principle be nebular, not Earth-like; and (3) the existence of dry chondritic meteorites (e.g., enstatite chondrites) restricts the plausible parameter space for the mechanism to operate. However, it should be noted that the D/H ratio for the primordial mantle material is only an upper limit and it could be lower, \textcolor{black}{or the measurements did not measure the D/H of the original/indigenous water } \citep{Hallis2015}, so more measurements are needed.

Earth and its constituent planetary embryos may have accreted primordial hydrogen-rich atmospheres.  Oxidation during the magma ocean phase may have produced water~\citep{Ikoma2006}, after which extensive atmospheric loss could have increased the D/H ratio to Earth ocean-like values~\citep{Genda2008}. However, it seems a great coincidence for the surviving water on Earth to match carbonaceous chondrites in their D/H ratios. \textcolor{black}{It is also unclear whether this model could explain Earth's $^{15}$N/$^{14}$N ratio \textcolor{black}{given that the Nitrogen would have come from a different source}}.  

The Sun's snow line may have been interior to Earth's orbit during a significant fraction of the disk lifetime~\citep[e.g.,][]{oka11,martin12}.  However, it seems unlikely that inward-drifting pebbles provided Earth's water.  \textcolor{black}{The age distributions of the two, isotopically-distinct classes of meteorites -- carbonaceous and non-carbonaceous -- suggest that those reservoirs were kept separate as of $\sim$1 Myr after CAIs, and this segregation has been interpreted as being caused by the growing giant planets (perhaps Jupiter's core) blocking the inward drift of carbonaceous pebbles~\citep{budde16,kruijer17,desch18}.} Since the carbonaceous pebbles represent the source of water, it is hard to imagine how they could have delivered water to Earth without producing a population of meteorites intermediate in composition between the two known classes. \textcolor{black}{Yet the ordinary and R chondrites formed at $\sim 2$Myr after CAIs but show signs of having accreted water ice~\citep{alexander18}. The origin of that ice is hard to understand, and may be linked with pebble recycling in the inner disk or Jupiter's core acting as an imperfect pebble barrier~\citep[e.g.,][]{morby16}. }

In the classical model of solar system formation~\citep{wetherill92}, Earth's water is a byproduct of its broad feeding zone~\citep{morby00,raymond04,raymond07a}. However, the classical model cannot easily match the large Earth/Mars mass ratio~\citep[e.g.,][]{raymond09c,morishima10} and is therefore suspect.  The Early Instability model can reproduce Mars' mass and also delivers water to Earth due to its broad feeding zone~\citep{clement18,clement19}.  Yet the processes that shaped the initial water distribution remain unexplained by such models.

At present, the external pollution model provides the most complete explanation for the origin of Earth's water.  A population of planetesimals on high-eccentricity orbits crossing the terrestrial zone is naturally produced by the giant planets' growth \citep{raymond17} and migration~\citep{walsh11,obrien14}. This fits within the Grand Tack and Low-mass Asteroid Belt models for terrestrial planet formation (see Fig.~\ref{fig:3models}).  The same dynamical processes also implant objects into the outer asteroid belt. The objects that delivered water to Earth should then have had the same chemical signature as carbonaceous chondrites, which do indeed provide a good match to Earth's isotopic composition in terms of water and nitrogen~\citep{Lecuyer1998,Marty2012}.  The amount of water delivered depends on unconstrained parameters (e.g., the disk properties and planetary migration rates) but plausible values can match Earth.  Of course, these dynamical models remain a matter of debate~\citep{Raymond2018}.  Nonetheless, there are no obvious problems with this delivery mechanism.

Finally, inward-migrating planetary embryos can indeed deliver water to inner rocky planets~\citep[or themselves become inner ice-rich planets; e.g., ][]{terquem07,izidoro19,bitsch19b}. However, the building blocks of our solar system's terrestrial planets are likely to have been $\sim$Mars-mass~\citep[see][]{morby12}, below the mass for substantial orbital migration.

\subsection{Water Loss Processes}
\label{sec:losses}

Several mechanisms exist that may significantly dry out planetesimals and planets that were not accounted for in the models presented in Section~\ref{sec:watermodel}.

The short-lived radionuclide $^{26}$Al (half-life of $\sim$700,000 years) provided a huge amount of heat to the early solar system.  As a result, any planetesimals that formed within roughly 2 million years of CAIs would have been completely dehydrated~\citep{grimm93,monteux18}. In addition, the presence of $^{26}$Al can lead to a bifurcation between very water-rich planets with tens of percent of water by mass and relatively dry, rocky planets like Earth~\citep{lichtenberg19}. Because $^{26}$Al is produced in massive stars, its abundance in planet-forming disks may vary considerably~\citep{hester04,gounelle08,gaidos09,lichtenberg16}, leading to a diversity in the water contents of terrestrial exoplanets. 

Impacts may also strip planets of water.  Ice-rich mantles can be stripped in giant impacts~\citep{marcus10}. Given that the impact impedance of water is lower than that of rock, giant impacts preferentially remove water from the surfaces of water-rich planets~\citep{genda05}. Planetesimal impacts may erode planetary atmospheres~\citep{svetsov07,schlichting15}, causing water loss if a large fraction of planets' water is in the atmosphere.

Improving our understanding of water loss may be as important as improving our understanding of water delivery.

\subsection{Extrapolation to Exoplanets}
\label{sec:exoplanets}

The bulk of extra-solar planetary systems look very different than our own~\citep[see Section~\ref{sec:intro} and discussion in][]{Raymond2018}. Yet we think that the same fundamental processes govern the formation of all systems.  In this section we evaluate which of the mechanisms from Section~\ref{sec:watermodel} are likely to be dominant in exoplanet systems.

Our analysis leads to two conclusions.  First, we expect water to be delivered to virtually all rocky planets.  Any completely dry planets are likely to have {\it lost} their water.  Second, while pebble snow, broad feeding zones and external pollution are likely to play a role, we expect migration to be the dominant mechanism of water delivery.  Migration is also the mechanism that should produce planets with the highest water contents. This conclusion is based on overwhelming empirical and theoretical evidence that the population of super-Earths form predominantly during the gaseous disk lifetime, leading to the inescapable conclusion that planet-disk interactions -- and therefore orbital migration -- must have played a role in shaping this population~\citep[e.g., as in the model of][]{izidoro17,izidoro19}.  Indeed, current thinking invokes migration as one of a handful of essential ingredients in planet formation models~\citep[see][]{Raymond2018}.

Imagine a planet growing in the hotter regions of a planet-forming disk, in its star's habitable zone. Ironically, while a planet in this region can maintain liquid water on its surface (with the right atmosphere), the local building blocks are dry. Given the diversity of water delivery mechanisms, it is hard to imagine such a planet remaining dry.  Of course, if water is sourced locally -- from adsorption onto grains or from the oxidation of a primordial H-rich envelope -- then it will be hydrated anyway.  If the planet accretes from a smooth disk of solid material then its feeding zone will widen in time to encompass more distant, volatile-rich bodies.  Even if the planet is growing from a ring of material, the snow line moves inward as the gaseous disk cools, such that pebbles can ``snow'' down and deliver water.  Pebble snow can be shut down by the growth of a large outer planet, but that planet would produce a rain of water-rich planetesimals that pollute the inner rocky zone. A large outer planet might even migrate inward and deliver water in bulk (and perhaps become the ocean-covered seed of a habitable zone planet).   
Most stars are lower in mass than the Sun~\citep{chabrier03}, with lower luminosities and correspondingly closer-in habitable zones~\citep{kasting93}. Certain factors argue that habitable zone planets around low-mass stars should be drier than around Sun-like stars. In dynamical terms, the snow line is farther away from the habitable zone around low-mass stars~\citep{kennedy08,mulders15c}. This would presumably reduce the efficiency of pebble snow and also of external pollution~\citep[which is also reduced by the lower frequency of gas giants around low-mass stars;][]{johnson07}. Since their habitable zones are closer-in, accretion timescales are shorter around low-mass stars (compared with Sun-like stars) and impact speeds higher~\citep{lissauer07,raymond07b}. Other factors are ambiguous with regards to water delivery. While fast accretion timescales imply an increase in the efficiency of migration, the masses of planet-forming disks appear to be lower around low-mass stars~\citep{pascucci16}. Low-mass stars are observed to have more ``super-Earths'' and fewer ``mini-Neptunes'' than Sun-like stars~\citep{mulders15} but it remains unclear how this connects with planet formation and water delivery.

Migration is likely a dominant process in determining planetary water contents.  Many models for the origin of super-Earths invoke large-scale migration of large planetary embryos~\citep{terquem07,mcneil10,ida10,cossou14,ogihara15,izidoro17,izidoro19}.  Given that embryos are thought to form faster past the snow line~\citep[by pebble accretion;][]{lambrechts14b,morby15}, most of the super-Earths formed by migration should be ice-rich~\citep[but see][]{raymond18b}. Some large embryos may grow close to the inner edge of the disk by accumulating drifting pebbles~\citep{chatterjee14,chatterjee15}. Embryos that grow large past the snow line before migrating inward have water contents of ten or more percent~\citep{bitsch19b,izidoro19}, whereas those that grow close-in should be purely rocky.  Despite the preferential stripping of icy mantles during giant impacts~\citep{marcus10}, collisions between these two populations will likely maintain the bimodal distribution of very water-rich planets and very dry ones.  Planets with intermediate water contents would only form after one ice-rich embryo underwent a series of giant impacts with pure rock embryos. 

None of the other water delivery mechanisms is likely to produce planets with more than $\sim1$\% water by mass.  Pebble snow is susceptible to being deactivated by the growth of any large core on an exterior orbit~\citep{lambrechts14,bitsch18}. In addition, the pebble flux may drop substantially when the pebble production front reaches the outer edge of the disk, limiting the overall amount of water that can be delivered~\citep{ida19}. The sweet spot for pebble snow may thus be early, before the growth of large cores and while the pebble flux is high.  Very water-rich planets may grow from ice-rich pebbles~\citep{lambrechts12,morby15}, and these are the planets likely to migrate inward~\citep[e.g.,][]{ormel17,bitsch19b}. 

\begin{figure*}[ht]
\begin{center}
\includegraphics[width=12cm]{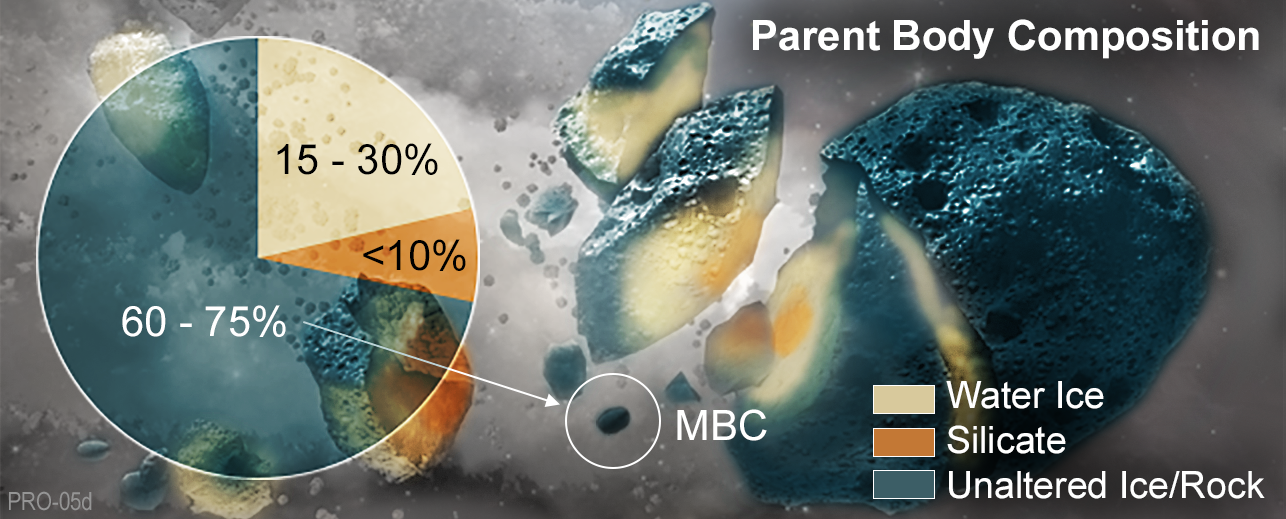}
\caption{Thermal models show that MBC parent bodies preserve unaltered materials throughout 60-75\% of the body for their history \citep{Castillo2010}. MBCs are from this unaltered fraction \citep{Marsset2016}.}
\label{fig:Family}
\end{center}
\end{figure*}

Water contents similar to Earth's ($\sim$0.1\% water by mass) are not detectable with present-day techniques. Our knowledge of the water contents of exoplanets is very limited.  Mass and radius measurements exist for dozens of known close-in low-mass planets~\citep[e.g.,][]{batalha13,marcy14}.  These data have revealed a dichotomy between small, solid {\em super-Earths} ($R \lesssim 1.5-2 \rearth$) and larger, gas-rich {\em sub-Neptunes}~\citep{rogers15,wolfgang16,chen17}. Some well-measured super-Earths have densities high enough to be scaled-up versions of Earth or even Mercury~\citep{howard13,santerne18,bonomo19}. However, constraining water contents from density measurements is fraught with uncertainty~\citep[e.g.,][]{adams08,selsis07b,dorn15}. The gap in the size distribution of super-Earths~\citep{fulton17} has been interpreted as an indication that most super-Earths are rocky~\citep{lopez17,owen17,jin18} but there is considerable debate~\citep[e.g.,][]{zeng19}. Given measurement uncertainties, the meaning of ``rocky'' vs ''ice-rich'' is ambiguous; some models can match the radius gap with planets with anything up to 20\% water by mass~\citep{gupta19}.  

\subsection{What New Measurements are Needed?}
\label{sec:newobs}

What measurements do we need to move forward? As discussed previously, the external pollution mechanism is currently the most likely candidate. However, this still leaves a wide range of possible distances from where the volatiles may have originated. The source location in the disk will depend upon where and when Jupiter's core formed. The key will be to match the signatures from volatiles in primitive objects that preserve the early solar system record and whose dynamical provenance is reasonably well understood. This region is the ``wet'' outer asteroid belt. Material scattered into the inner solar system will be both implanted into the asteroid belt, and impact the growing terrestrial planets. While the material that built the terrestrial planets no longer exists, the material scattered inward still exists today in the asteroid belt, and volatiles have been preserved in the outer belt.

Direct evidence for water in the outer asteroid belt stems from the {\it Herschel} detection of water vapor from Ceres \citep{Kuppers2014} coupled with water ice detections on Ceres' surface by the {\it Dawn} mission \citep{Combe2016}, and a possible H$_2$O-frost signature on the asteroids 90 Antiope \citep{Hargrove2015} and 24 Themis \citep{Campins2010,Rivkin2010}. Main Belt comets (MBCs) are the primitive representatives of accessible ice from the outer asteroid belt. Orbiting in the outer asteroid belt since the early stages of solar system formation \citep{Hsieh2006,Jewitt2012}, they exhibit comet-like tails, attributed to outgassing of volatiles from ices in their interiors. The ices, preserved by a layer of dust, carry a signature of the early solar system. 

All objects whose semimajor axes are less than Jupiter's, with an orbit dynamically decoupled from Jupiter \citep{Vaghi1973,Kresak1980}, and exhibiting mass loss with a cometary appearance, are termed ``Active Asteroids''. They are dynamically unrelated to comets \citep{Levison2006,Haghighipour2009}. There are 25 currently known. Several causes for their comet-like activity have been postulated, including electrostatic ejection \citep{Criswell1977}, mass shedding from collisional and radiation torques (YORP) \citep{Drahus2011,Jacobson2011} rotational instability and volatile sublimation \citep{Jewitt2012}.

Only a thermal process, e.g., sublimation, can explain the observed repeated perihelion activity. The twelve believed to be water ice sublimation-driven (seven with repeat activity \citep{Hsieh2018a}) are termed MBCs \citep{Snodgrass2017}, of value for their primordial ices. At the distance of the asteroid belt, the interior temperatures are warm enough that only water and its trapped volatiles remain over the age of the solar system \citep{Prialnik2009}. MBC sublimation is believed to be triggered by impacts of meter-sized objects that remove part of the surface layer, allowing heat to reach the ice \citep{Hsieh2009,Haghighipour2016}. Models predict activity will continue for 100’s of orbits, and will show significant fading before ceasing \citep{Prialnik2009}. Conversely, non-MBC active asteroids exhibit very different, short-lived dust structures \citep{Jewitt2012,Hainaut2012,kleyna2013}. All attempts to directly detect gas\footnote{H$_2$O and its more readily observable proxy, CN, a minor species that fluoresces strongly and is dragged out of the nucleus with water as it sublimates.} have been unsuccessful--unsurprising, given that the detection limit for a 10m telescope is one to two orders of magnitude above the amount of gas required to lift the observed dust. Indeed, Kuiper belt and Oort cloud comets at the same distance routinely show no spectral evidence for gas, although their dust comae are strong. 

MBCs are small (radii $<$ 2 km), with low albedos and with flat featureless spectra in the visible. Most MBCs are related to collisional asteroid families \citep{Hsieh2018b} since this is an excellent way to bring interior ices closer to the surface. For example, the Themis family represents the collisional remnants of an icy protoplanet a few 100 km in size (Fig.~\ref{fig:Family}). A weak or absent hydration signature suggests limited aqueous alteration in the Themis parent interior \citep{Marsset2016}, consistent with thermal models predicting a thick, unaltered outer layer of primitive ice and rock \citep{Castillo2010}. Diversity in the Themis family spectral properties is interpreted as a gradient in the parent body composition, supporting that scenario \citep{Fornasier2016}. Stripping of Themis’ outer layer formed a large fraction of the family members, including the MBCs, without re-processing their volatile content \citep[][Fig.~\ref{fig:Family}]{Durda2007,Rivkin2014}. 

MBCs are samples from the unexplored icy asteroids that are small enough and/or formed late enough to have escaped complete hydrothermal processing and still have primitive outer layers (Fig.~\ref{fig:Family}), unlike Ceres, an icy asteroid that has suffered intensive aqueous alteration \citep{Ammannito2016}. An alternate explanation for the origin of some MBCs is that they are primordial planetesimals that never accreted into larger bodies. In both cases, MBCs offer accessible pristine volatiles. Thermal models show that these ices can survive over the age of the solar system due to an insulating dust layer \citep{Prialnik2009,Schorghofer2008}. Thus, these represent the best source of material from which measurements can help distinguish between solar system formation models and the origin of inner solar system water. The key measurements to make will be to obtain the multiple volatile isotopic fingerprints: D/H in water, $^{17}$O/$^{16}$O, $^{18}$O/$^{16}$O, $^{15}$N/$^{14}$N, and $^{84}$Kr/$^{36}$Ar.

The next generation of sub-millimeter telescopes envisioned for the next decade will have 10$\times$ the sensitivity and resolution of the exquisite performance of ALMA \citep{Murphy2018}. These facilities will be able probe planet formation inside 10 au \citep{Andrews2018b}, unveiling the chemistry and structures in the habitable zones in nearby protoplanetary disks \citep{Ricci2018,McGuire2018}. Careful isotopic measurements of primitive volatiles in the outer asteroid belt gives us the ability to explore these past processes in our own solar system. 

\vskip .5in
\noindent \textbf{Acknowledgments.} \\
We thank referees Hidenori Genda and Conel Alexander for helpful reports that greatly improved this review. 
We acknowledge extensive discussions that shaped our thinking over the past many years with researchers in the field, including Edwin Bergin, Gary Huss, Alexander Krot, Alessandro Morbidelli, Jonathan Lunine, Bernard Marty, Andre Izidoro, John Chambers, Kevin Walsh, Seth Jacobson, and Conel Alexander.
K.J.M. acknowledges support from the NASA Astrobiology Institute under Cooperative Agreement No. NNA09DA77A issued through the Office of Space Science, NASA Grant 80-NSSC18K0853 and a grant from the National Science Foundation AST-1617015.
S.N.R. thanks NASA Astrobiology Institute's Virtual Planetary Laboratory Lead Team, funded under solicitation NNH12ZDA002C and cooperative agreement no. NNA13AA93A.
We would particularly like to thank James E. Polk (Jet Propulsion Laboratory) for helping to edit this chapter.

\bibliographystyle{sss-full.bst}
\bibliography{refs.bib,refsSR.bib}

\begin{thebibliography}{365}
\providecommand{\natexlab}[1]{#1}
\parskip=0pt \itemsep=0pt \small \baselineskip=11pt

\bibitem[{\emph{{Abe} et~al.}(2000)\emph{{Abe}, {Ohtani}, {Okuchi}, {Righter},
  and {Drake}}}]{abe00}
{Abe} Y., {Ohtani} E., {Okuchi} T., {Righter} K., and {Drake} M. (2000)
  \emph{{Water in the Early Earth}}, pp. 413--433.

\bibitem[{\emph{{Adams} et~al.}(2008)\emph{{Adams}, {Seager}, and
  {Elkins-Tanton}}}]{adams08}
{Adams} E.~R., {Seager} S., and {Elkins-Tanton} L. (2008) \emph{{Ocean Planet
  or Thick Atmosphere: On the Mass-Radius Relationship for Solid Exoplanets
  with Massive Atmospheres}}, \emph{\apj}, \emph{673}, 1160--1164.

\bibitem[{\emph{{Agnor} et~al.}(1999)\emph{{Agnor}, {Canup}, and
  {Levison}}}]{agnor99}
{Agnor} C.~B., {Canup} R.~M., and {Levison} H.~F. (1999) \emph{{On the
  Character and Consequences of Large Impacts in the Late Stage of Terrestrial
  Planet Formation}}, \emph{Icarus}, \emph{142}, 219--237.

\bibitem[{\emph{{A'Hearn} et~al.}(1995)\emph{{A'Hearn}, {Millis}, {Schleicher},
  {Osip}, and {Birch}}}]{AHearn1995}
{A'Hearn} M.~F., {Millis} R.~C., {Schleicher} D.~O., {Osip} D.~J., and {Birch}
  P.~V. (1995) \emph{{The ensemble properties of comets: Results from
  narrowband photometry of 85 comets, 1976-1992.}}, \emph{\icarus}, \emph{118},
  223--270.

\bibitem[{\emph{{Aikawa} et~al.}(2002)\emph{{Aikawa}, {van Zadelhoff}, {van
  Dishoeck}, and {Herbst}}}]{Aikawa2002}
{Aikawa} Y., {van Zadelhoff} G.~J., {van Dishoeck} E.~F., and {Herbst} E.
  (2002) \emph{{Warm molecular layers in protoplanetary disks}}, \emph{\aap},
  \emph{386}, 622--632.

\bibitem[{\emph{{Alexander}}(2019{\natexlab{a}})}]{alexander19a}
{Alexander} C. M.~O. (2019{\natexlab{a}}) \emph{{Quantitative models for the
  elemental and isotopic fractionations in chondrites: The carbonaceous
  chondrites}}, \emph{\gca}, \emph{254}, 277--309.

\bibitem[{\emph{{Alexander}}(2019{\natexlab{b}})}]{alexander19b}
{Alexander} C. M.~O. (2019{\natexlab{b}}) \emph{{Quantitative models for the
  elemental and isotopic fractionations in the chondrites: The non-carbonaceous
  chondrites}}, \emph{\gca}, \emph{254}, 246--276.

\bibitem[{\emph{{Alexander} et~al.}(2012{\natexlab{a}})\emph{{Alexander},
  {Bowden}, {Fogel}, {Howard}, {Herd}, and {Nittler}}}]{alexander12}
{Alexander} C.~M.~O.~., {Bowden} R., {Fogel} M.~L., {Howard} K.~T., {Herd}
  C.~D.~K., and {Nittler} L.~R. (2012{\natexlab{a}}) \emph{{The Provenances of
  Asteroids, and Their Contributions to the Volatile Inventories of the
  Terrestrial Planets}}, \emph{Science}, \emph{337}, 721.

\bibitem[{\emph{{Alexander} et~al.}(2012{\natexlab{b}})\emph{{Alexander},
  {Bowden}, {Fogel}, {Howard}, {Herd}, and {Nittler}}}]{Alexander2012}
{Alexander} C.~M.~O., {Bowden} R., {Fogel} M.~L., {Howard} K.~T., {Herd}
  C.~D.~K., and {Nittler} L.~R. (2012{\natexlab{b}}) \emph{{The Provenances of
  Asteroids, and Their Contributions to the Volatile Inventories of the
  Terrestrial Planets}}, \emph{Science}, \emph{337}, 721.

\bibitem[{\emph{{Alexander} et~al.}(2018{\natexlab{a}})\emph{{Alexander},
  {McKeegan}, and {Altwegg}}}]{Alexander2018}
{Alexander} C. M.~O., {McKeegan} K.~D., and {Altwegg} K. (2018{\natexlab{a}})
  \emph{{Water Reservoirs in Small Planetary Bodies: Meteorites, Asteroids, and
  Comets}}, \emph{\ssr}, \emph{214}, 36.

\bibitem[{\emph{{Alexander} et~al.}(2018{\natexlab{b}})\emph{{Alexander},
  {McKeegan}, and {Altwegg}}}]{alexander18}
{Alexander} C.~M.~O., {McKeegan} K.~D., and {Altwegg} K. (2018{\natexlab{b}})
  \emph{{Water Reservoirs in Small Planetary Bodies: Meteorites, Asteroids, and
  Comets}}, \emph{\ssr}, \emph{214}, 36.

\bibitem[{\emph{{Alexander} et~al.}(2017)\emph{{Alexander}, {Nittler},
  {Davidson}, and {Ciesla}}}]{Alexander2017}
{Alexander} C. M.~O., {Nittler} L.~R., {Davidson} J., and {Ciesla} F.~J. (2017)
  \emph{{Measuring the level of interstellar inheritance in the solar
  protoplanetary disk}}, \emph{Meteoritics and Planetary Science}, \emph{52},
  1797--1821.

\bibitem[{\emph{{ALMA Partnership} et~al.}(2015)\emph{{ALMA Partnership},
  {Brogan}, {P{\'e}rez}, {Hunter}, {Dent}, {Hales}, {Hills}, {Corder},
  {Fomalont}, {Vlahakis}, {Asaki}, {Barkats}, {Hirota}, {Hodge},
  {Impellizzeri}, {Kneissl}, {Liuzzo}, {Lucas}, {Marcelino}, {Matsushita},
  {Nakanishi}, {Phillips}, {Richards}, {Toledo}, {Aladro}, {Broguiere},
  {Cortes}, {Cortes}, {Espada}, {Galarza}, {Garcia-Appadoo}, {Guzman-Ramirez},
  {Humphreys}, {Jung}, {Kameno}, {Laing}, {Leon}, {Marconi}, {Mignano},
  {Nikolic}, {Nyman}, {Radiszcz}, {Remijan}, {Rod{\'o}n}, {Sawada},
  {Takahashi}, {Tilanus}, {Vila Vilaro}, {Watson}, {Wiklind}, {Akiyama},
  {Chapillon}, {de Gregorio-Monsalvo}, {Di Francesco}, {Gueth}, {Kawamura},
  {Lee}, {Nguyen Luong}, {Mangum}, {Pietu}, {Sanhueza}, {Saigo}, {Takakuwa},
  {Ubach}, {van Kempen}, {Wootten}, {Castro-Carrizo}, {Francke}, {Gallardo},
  {Garcia}, {Gonzalez}, {Hill}, {Kaminski}, {Kurono}, {Liu}, {Lopez},
  {Morales}, {Plarre}, {Schieven}, {Testi}, {Videla}, {Villard}, {Andreani},
  {Hibbard}, and {Tatematsu}}}]{alma15}
{ALMA Partnership}, {Brogan} C.~L., {P{\'e}rez} L.~M., {Hunter} T.~R., {Dent}
  W.~R.~F., {Hales} A.~S., {Hills} R.~E., {Corder} S., {Fomalont} E.~B.,
  {Vlahakis} C., {Asaki} Y., {Barkats} D., {Hirota} A., {Hodge} J.~A.,
  {Impellizzeri} C.~M.~V., {Kneissl} R., {Liuzzo} E., {Lucas} R., {Marcelino}
  N., {Matsushita} S., {Nakanishi} K., {Phillips} N., {Richards} A.~M.~S.,
  {Toledo} I., {Aladro} R., {Broguiere} D., {Cortes} J.~R., {Cortes} P.~C.,
  {Espada} D., {Galarza} F., {Garcia-Appadoo} D., {Guzman-Ramirez} L.,
  {Humphreys} E.~M., {Jung} T., {Kameno} S., {Laing} R.~A., {Leon} S.,
  {Marconi} G., {Mignano} A., {Nikolic} B., {Nyman} L.-A., {Radiszcz} M.,
  {Remijan} A., {Rod{\'o}n} J.~A., {Sawada} T., {Takahashi} S., {Tilanus}
  R.~P.~J., {Vila Vilaro} B., {Watson} L.~C., {Wiklind} T., {Akiyama} E.,
  {Chapillon} E., {de Gregorio-Monsalvo} I., {Di Francesco} J., {Gueth} F.,
  {Kawamura} A., {Lee} C.-F., {Nguyen Luong} Q., {Mangum} J., {Pietu} V.,
  {Sanhueza} P., {Saigo} K., {Takakuwa} S., {Ubach} C., {van Kempen} T.,
  {Wootten} A., {Castro-Carrizo} A., {Francke} H., {Gallardo} J., {Garcia} J.,
  {Gonzalez} S., {Hill} T., {Kaminski} T., {Kurono} Y., {Liu} H.-Y., {Lopez}
  C., {Morales} F., {Plarre} K., {Schieven} G., {Testi} L., {Videla} L.,
  {Villard} E., {Andreani} P., {Hibbard} J.~E., and {Tatematsu} K. (2015)
  \emph{{The 2014 ALMA Long Baseline Campaign: First Results from High Angular
  Resolution Observations toward the HL Tau Region}}, \emph{\apjl}, \emph{808},
  L3.

\bibitem[{\emph{{Altwegg} et~al.}(2015)\emph{{Altwegg}, {Balsiger}, {Bar-Nun},
  {Berthelier}, {Bieler}, {Bochsler}, {Briois}, {Calmonte}, {Combi}, and {De
  Keyser}}}]{Altwegg2015}
{Altwegg} K., {Balsiger} H., {Bar-Nun} A., {Berthelier} J.~J., {Bieler} A.,
  {Bochsler} P., {Briois} C., {Calmonte} U., {Combi} M., and {De Keyser} J.
  (2015) \emph{{67P/Churyumov-Gerasimenko, a Jupiter family comet with a high
  D/H ratio}}, \emph{Science}, \emph{347}, 1261952.

\bibitem[{\emph{{Amelin} et~al.}(2002)\emph{{Amelin}, {Krot}, {Hutcheon}, and
  {Ulyanov}}}]{Amelin2002}
{Amelin} Y., {Krot} A.~N., {Hutcheon} I.~D., and {Ulyanov} A.~A. (2002)
  \emph{{Lead Isotopic Ages of Chondrules and Calcium-Aluminum-Rich
  Inclusions}}, \emph{Science}, \emph{297}, 1678--1683.

\bibitem[{\emph{{Ammannito} et~al.}(2016)\emph{{Ammannito}, {DeSanctis},
  {Ciarniello}, {Frigeri}, {Carrozzo}, {Combe}, {Ehlmann}, {Marchi}, {McSween},
  and {Raponi}}}]{Ammannito2016}
{Ammannito} E., {DeSanctis} M.~C., {Ciarniello} M., {Frigeri} A., {Carrozzo}
  F.~G., {Combe} J.~P., {Ehlmann} B.~L., {Marchi} S., {McSween} H.~Y., and
  {Raponi} A. (2016) \emph{{Distribution of phyllosilicates on the surface of
  Ceres}}, \emph{Science}, \emph{353}, aaf4279.

\bibitem[{\emph{{Anders} and {Grevesse}}(1989)}]{Anders1989}
{Anders} E. and {Grevesse} N. (1989) \emph{{Abundances of the elements:
  Meteoritic and solar}}, \emph{\gca}, \emph{53}, 197--214.

\bibitem[{\emph{{Andrews} et~al.}(2018)\emph{{Andrews}, {Wilner},
  {Mac{\'\i}as}, {Carrasco-Gonz{\'a}lez}, and {Isella}}}]{Andrews2018b}
{Andrews} S.~M., {Wilner} D.~J., {Mac{\'\i}as} E., {Carrasco-Gonz{\'a}lez} C.,
  and {Isella} A. (2018) in \emph{Science with a Next Generation Very Large
  Array} (E.~{Murphy}, ed.), vol. 517 of \emph{Astronomical Society of the
  Pacific Conference Series}, p. 137.

\bibitem[{\emph{{Armitage} et~al.}(2016)\emph{{Armitage}, {Eisner}, and
  {Simon}}}]{armitage16}
{Armitage} P.~J., {Eisner} J.~A., and {Simon} J.~B. (2016) \emph{{Prompt
  Planetesimal Formation beyond the Snow Line}}, \emph{\apjl}, \emph{828}, L2.

\bibitem[{\emph{{Asaduzzaman} et~al.}(2014)\emph{{Asaduzzaman}, {Zega},
  {Laref}, {Runge}, {Deymier}, and {Muralidharan}}}]{asaduzzaman14}
{Asaduzzaman} A.~M., {Zega} T.~J., {Laref} S., {Runge} K., {Deymier} P.~A., and
  {Muralidharan} K. (2014) \emph{{A computational investigation of adsorption
  of organics on mineral surfaces: Implications for organics delivery in the
  early solar system}}, \emph{Earth and Planetary Science Letters}, \emph{408},
  355--361.

\bibitem[{\emph{{Badro} et~al.}(2014)\emph{{Badro}, {C{\^o}t{\'e}}, and
  {Brodholt}}}]{Badro2014}
{Badro} J., {C{\^o}t{\'e}} A.~S., and {Brodholt} J.~P. (2014) \emph{{A
  seismologically consistent compositional model of Earth's core}},
  \emph{Proceedings of the National Academy of Science}, \emph{111},
  7542--7545.

\bibitem[{\emph{{Bar-Nun} et~al.}(1985)\emph{{Bar-Nun}, {Herman}, {Laufer}, and
  {Rappaport}}}]{BarNun1985}
{Bar-Nun} A., {Herman} G., {Laufer} D., and {Rappaport} M.~L. (1985)
  \emph{{Trapping and release of gases by water ice and implications for icy
  bodies}}, \emph{\icarus}, \emph{63}, 317--332.

\bibitem[{\emph{{Bar-Nun} et~al.}(1988)\emph{{Bar-Nun}, {Kleinfeld}, and
  {Kochavi}}}]{BarNun1988}
{Bar-Nun} A., {Kleinfeld} I., and {Kochavi} E. (1988) \emph{{Trapping of gas
  mixtures by amorphous water ice}}, \emph{\prb}, \emph{38}, 7749--7754.

\bibitem[{\emph{{Baruteau} et~al.}(2014)\emph{{Baruteau}, {Crida},
  {Paardekooper}, {Masset}, {Guilet}, {Bitsch}, {Nelson}, {Kley}, and
  {Papaloizou}}}]{baruteau14}
{Baruteau} C., {Crida} A., {Paardekooper} S.-J., {Masset} F., {Guilet} J.,
  {Bitsch} B., {Nelson} R., {Kley} W., and {Papaloizou} J. (2014)
  \emph{{Planet-Disk Interactions and Early Evolution of Planetary Systems}},
  \emph{Protostars and Planets VI}, pp. 667--689.

\bibitem[{\emph{{Batalha} et~al.}(2013)\emph{{Batalha}, {Rowe}, {Bryson},
  {Barclay}, {Burke}, {Caldwell}, {Christiansen}, {Mullally}, {Thompson},
  {Brown}, {Dupree}, {Fabrycky}, {Ford}, {Fortney}, {Gilliland}, {Isaacson},
  {Latham}, {Marcy}, {Quinn}, {Ragozzine}, {Shporer}, {Borucki}, {Ciardi},
  {Gautier}, {Haas}, {Jenkins}, {Koch}, {Lissauer}, {Rapin}, {Basri}, {Boss},
  {Buchhave}, {Carter}, {Charbonneau}, {Christensen-Dalsgaard}, {Clarke},
  {Cochran}, {Demory}, {Desert}, {Devore}, {Doyle}, {Esquerdo}, {Everett},
  {Fressin}, {Geary}, {Girouard}, {Gould}, {Hall}, {Holman}, {Howard},
  {Howell}, {Ibrahim}, {Kinemuchi}, {Kjeldsen}, {Klaus}, {Li}, {Lucas},
  {Meibom}, {Morris}, {Pr{\v s}a}, {Quintana}, {Sanderfer}, {Sasselov},
  {Seader}, {Smith}, {Steffen}, {Still}, {Stumpe}, {Tarter}, {Tenenbaum},
  {Torres}, {Twicken}, {Uddin}, {Van Cleve}, {Walkowicz}, and
  {Welsh}}}]{batalha13}
{Batalha} N.~M., {Rowe} J.~F., {Bryson} S.~T., {Barclay} T., {Burke} C.~J.,
  {Caldwell} D.~A., {Christiansen} J.~L., {Mullally} F., {Thompson} S.~E.,
  {Brown} T.~M., {Dupree} A.~K., {Fabrycky} D.~C., {Ford} E.~B., {Fortney}
  J.~J., {Gilliland} R.~L., {Isaacson} H., {Latham} D.~W., {Marcy} G.~W.,
  {Quinn} S.~N., {Ragozzine} D., {Shporer} A., {Borucki} W.~J., {Ciardi} D.~R.,
  {Gautier} T.~N., III, {Haas} M.~R., {Jenkins} J.~M., {Koch} D.~G., {Lissauer}
  J.~J., {Rapin} W., {Basri} G.~S., {Boss} A.~P., {Buchhave} L.~A., {Carter}
  J.~A., {Charbonneau} D., {Christensen-Dalsgaard} J., {Clarke} B.~D.,
  {Cochran} W.~D., {Demory} B.-O., {Desert} J.-M., {Devore} E., {Doyle} L.~R.,
  {Esquerdo} G.~A., {Everett} M., {Fressin} F., {Geary} J.~C., {Girouard}
  F.~R., {Gould} A., {Hall} J.~R., {Holman} M.~J., {Howard} A.~W., {Howell}
  S.~B., {Ibrahim} K.~A., {Kinemuchi} K., {Kjeldsen} H., {Klaus} T.~C., {Li}
  J., {Lucas} P.~W., {Meibom} S., {Morris} R.~L., {Pr{\v s}a} A., {Quintana}
  E., {Sanderfer} D.~T., {Sasselov} D., {Seader} S.~E., {Smith} J.~C.,
  {Steffen} J.~H., {Still} M., {Stumpe} M.~C., {Tarter} J.~C., {Tenenbaum} P.,
  {Torres} G., {Twicken} J.~D., {Uddin} K., {Van Cleve} J., {Walkowicz} L., and
  {Welsh} W.~F. (2013) \emph{{Planetary Candidates Observed by Kepler. III.
  Analysis of the First 16 Months of Data}}, \emph{\apjs}, \emph{204}, 24.

\bibitem[{\emph{{Bate}}(2018)}]{bate18}
{Bate} M.~R. (2018) \emph{{On the diversity and statistical properties of
  protostellar discs}}, \emph{\mnras}, \emph{475}, 5618--5658.

\bibitem[{\emph{{Benz} et~al.}(1989)\emph{{Benz}, {Cameron}, and
  {Melosh}}}]{benz89}
{Benz} W., {Cameron} A.~G.~W., and {Melosh} H.~J. (1989) \emph{{The origin of
  the moon and the single impact hypothesis. III}}, \emph{\icarus}, \emph{81},
  113--131.

\bibitem[{\emph{{Benz} et~al.}(1986)\emph{{Benz}, {Slattery}, and
  {Cameron}}}]{benz86}
{Benz} W., {Slattery} W.~L., and {Cameron} A.~G.~W. (1986) \emph{{The origin of
  the moon and the single-impact hypothesis. I}}, \emph{\icarus}, \emph{66},
  515--535.

\bibitem[{\emph{{Bergin} and {Tafalla}}(2007b)}]{Bergin2007b}
{Bergin} E.~A. and {Tafalla} M. (2007b) \emph{{Cold Dark Clouds: The Initial
  Conditions for Star Formation}}, \emph{\araa}, \emph{45}, 339--396.

\bibitem[{\emph{{Birnstiel} et~al.}(2016)\emph{{Birnstiel}, {Fang}, and
  {Johansen}}}]{birnstiel16}
{Birnstiel} T., {Fang} M., and {Johansen} A. (2016) \emph{{Dust Evolution and
  the Formation of Planetesimals}}, \emph{\ssr}, \emph{205}, 41--75.

\bibitem[{\emph{{Birnstiel} et~al.}(2012)\emph{{Birnstiel}, {Klahr}, and
  {Ercolano}}}]{birnstiel12}
{Birnstiel} T., {Klahr} H., and {Ercolano} B. (2012) \emph{{A simple model for
  the evolution of the dust population in protoplanetary disks}}, \emph{\aap},
  \emph{539}, A148.

\bibitem[{\emph{{Bitsch} et~al.}(2015)\emph{{Bitsch}, {Johansen}, {Lambrechts},
  and {Morbidelli}}}]{bitsch15}
{Bitsch} B., {Johansen} A., {Lambrechts} M., and {Morbidelli} A. (2015)
  \emph{{The structure of protoplanetary discs around evolving young stars}},
  \emph{\aap}, \emph{575}, A28.

\bibitem[{\emph{{Bitsch} et~al.}(2018)\emph{{Bitsch}, {Morbidelli}, {Johansen},
  {Lega}, {Lambrechts}, and {Crida}}}]{bitsch18}
{Bitsch} B., {Morbidelli} A., {Johansen} A., {Lega} E., {Lambrechts} M., and
  {Crida} A. (2018) \emph{{Pebble-isolation mass: Scaling law and implications
  for the formation of super-Earths and gas giants}}, \emph{\aap}, \emph{612},
  A30.

\bibitem[{\emph{{Bitsch} et~al.}(2019)\emph{{Bitsch}, {Raymond}, and
  {Izidoro}}}]{bitsch19b}
{Bitsch} B., {Raymond} S.~N., and {Izidoro} A. (2019) \emph{{Rocky super-Earths
  or waterworlds: the interplay of planet migration, pebble accretion, and disc
  evolution}}, \emph{\aap}, \emph{624}, A109.

\bibitem[{\emph{{Bockel{\'e}e-Morvan} et~al.}(2015)\emph{{Bockel{\'e}e-Morvan},
  {Calmonte}, {Charnley}, {Duprat}, {Engrand}, {Gicquel}, {H{\"a}ssig},
  {Jehin}, {Kawakita}, {Marty}, {Milam}, {Morse}, {Rousselot}, {Sheridan}, and
  {Wirstr{\"o}m}}}]{Bockelee2015}
{Bockel{\'e}e-Morvan} D., {Calmonte} U., {Charnley} S., {Duprat} J., {Engrand}
  C., {Gicquel} A., {H{\"a}ssig} M., {Jehin} E., {Kawakita} H., {Marty} B.,
  {Milam} S., {Morse} A., {Rousselot} P., {Sheridan} S., and {Wirstr{\"o}m} E.
  (2015) \emph{{Cometary Isotopic Measurements}}, \emph{\ssr}, \emph{197},
  47--83.

\bibitem[{\emph{{Boe} et~al.}(2019)\emph{{Boe}, {Jedicke}, {Meech}, {Wiegert},
  {Weryk}, {Chambers}, {Denneau}, {Kaiser}, {Kudritzki}, {Magnier},
  {Wainscoat}, and {Waters}}}]{Boe2019}
{Boe} B., {Jedicke} R., {Meech} K.~J., {Wiegert} P., {Weryk} R.~J., {Chambers}
  K.~C., {Denneau} L., {Kaiser} N., {Kudritzki} R.~P., {Magnier} E.~A.,
  {Wainscoat} R.~J., and {Waters} C. (2019) \emph{{The orbit and size-frequency
  distribution of long period comets observed by Pan-STARRS1}}, \emph{\icarus},
  \emph{333}, 252--272.

\bibitem[{\emph{{Boehnke} and {Harrison}}(2016)}]{boehnke16}
{Boehnke} P. and {Harrison} T.~M. (2016) \emph{{Illusory Late Heavy
  Bombardments}}, \emph{Proceedings of the National Academy of Science},
  \emph{113}, 10802--10806.

\bibitem[{\emph{{Boley}}(2009)}]{boley09}
{Boley} A.~C. (2009) \emph{{The Two Modes of Gas Giant Planet Formation}},
  \emph{\apjl}, \emph{695}, L53--L57.

\bibitem[{\emph{{Bonomo} et~al.}(2019)\emph{{Bonomo}, {Zeng}, {Damasso},
  {Leinhardt}, {Justesen}, {Lopez}, {Lund}, {Malavolta}, {Silva Aguirre}, and
  {Buchhave}}}]{bonomo19}
{Bonomo} A.~S., {Zeng} L., {Damasso} M., {Leinhardt} Z.~M., {Justesen} A.~B.,
  {Lopez} E., {Lund} M.~N., {Malavolta} L., {Silva Aguirre} V., and {Buchhave}
  L.~A. (2019) \emph{{A giant impact as the likely origin of different twins in
  the Kepler-107 exoplanet system}}, \emph{Nature Astronomy}, \emph{3},
  416--423.

\bibitem[{\emph{{Boss}}(1997)}]{boss97}
{Boss} A.~P. (1997) \emph{{Giant planet formation by gravitational
  instability.}}, \emph{Science}, \emph{276}, 1836--1839.

\bibitem[{\emph{{Bouvier} and {Wadhwa}}(2010)}]{Bouvier2010}
{Bouvier} A. and {Wadhwa} M. (2010) \emph{{The age of the Solar System
  redefined by the oldest Pb-Pb age of a meteoritic inclusion}}, \emph{Nature
  Geoscience}, \emph{3}, 637--641.

\bibitem[{\emph{{Brasser} et~al.}(2016)\emph{{Brasser}, {Matsumura}, {Ida},
  {Mojzsis}, and {Werner}}}]{brasser16}
{Brasser} R., {Matsumura} S., {Ida} S., {Mojzsis} S.~J., and {Werner} S.~C.
  (2016) \emph{{Analysis of Terrestrial Planet Formation by the Grand Tack
  Model: System Architecture and Tack Location}}, \emph{\apj}, \emph{821}, 75.

\bibitem[{\emph{{Brasser} et~al.}(2013)\emph{{Brasser}, {Walsh}, and
  {Nesvorn{\'y}}}}]{brasser13}
{Brasser} R., {Walsh} K.~J., and {Nesvorn{\'y}} D. (2013) \emph{{Constraining
  the primordial orbits of the terrestrial planets}}, \emph{\mnras}.

\bibitem[{\emph{{Budde} et~al.}(2016)\emph{{Budde}, {Burkhardt}, {Brennecka},
  {Fischer-G{\"o}dde}, {Kruijer}, and {Kleine}}}]{budde16}
{Budde} G., {Burkhardt} C., {Brennecka} G.~A., {Fischer-G{\"o}dde} M.,
  {Kruijer} T.~S., and {Kleine} T. (2016) \emph{{Molybdenum isotopic evidence
  for the origin of chondrules and a distinct genetic heritage of carbonaceous
  and non-carbonaceous meteorites}}, \emph{Earth and Planetary Science
  Letters}, \emph{454}, 293--303.

\bibitem[{\emph{{Butler} et~al.}(2006)\emph{{Butler}, {Wright}, {Marcy},
  {Fischer}, {Vogt}, {Tinney}, {Jones}, {Carter}, {Johnson}, {McCarthy}, and
  {Penny}}}]{butler06}
{Butler} R.~P., {Wright} J.~T., {Marcy} G.~W., {Fischer} D.~A., {Vogt} S.~S.,
  {Tinney} C.~G., {Jones} H.~R.~A., {Carter} B.~D., {Johnson} J.~A., {McCarthy}
  C., and {Penny} A.~J. (2006) \emph{{Catalog of Nearby Exoplanets}},
  \emph{\apj}, \emph{646}, 505--522.

\bibitem[{\emph{{Cabrera} et~al.}(2014)\emph{{Cabrera}, {Csizmadia}, {Lehmann},
  {Dvorak}, {Gandolfi}, {Rauer}, {Erikson}, {Dreyer}, {Eigm{\"u}ller}, and
  {Hatzes}}}]{cabrera14}
{Cabrera} J., {Csizmadia} S., {Lehmann} H., {Dvorak} R., {Gandolfi} D., {Rauer}
  H., {Erikson} A., {Dreyer} C., {Eigm{\"u}ller} P., and {Hatzes} A. (2014)
  \emph{{The Planetary System to KIC 11442793: A Compact Analogue to the Solar
  System}}, \emph{\apj}, \emph{781}, 18.

\bibitem[{\emph{{Campins} et~al.}(2010)\emph{{Campins}, {Hargrove},
  {Pinilla-Alonso}, {Howell}, {Kelley}, {Licandro}, {Moth{\'e}-Diniz},
  {Fern{\'a}ndez}, and {Ziffer}}}]{Campins2010}
{Campins} H., {Hargrove} K., {Pinilla-Alonso} N., {Howell} E.~S., {Kelley}
  M.~S., {Licandro} J., {Moth{\'e}-Diniz} T., {Fern{\'a}ndez} Y., and {Ziffer}
  J. (2010) \emph{{Water ice and organics on the surface of the asteroid 24
  Themis}}, \emph{\nat}, \emph{464}, 1320--1321.

\bibitem[{\emph{{Canup} and {Asphaug}}(2001)}]{canup01}
{Canup} R.~M. and {Asphaug} E. (2001) \emph{{Origin of the Moon in a giant
  impact near the end of the Earth's formation}}, \emph{\nat}, \emph{412},
  708--712.

\bibitem[{\emph{{Carrera} et~al.}(2017)\emph{{Carrera}, {Gorti}, {Johansen},
  and {Davies}}}]{carrera17}
{Carrera} D., {Gorti} U., {Johansen} A., and {Davies} M.~B. (2017)
  \emph{{Planetesimal Formation by the Streaming Instability in a
  Photoevaporating Disk}}, \emph{\apj}, \emph{839}, 16.

\bibitem[{\emph{{Carrera} et~al.}(2015)\emph{{Carrera}, {Johansen}, and
  {Davies}}}]{carrera15}
{Carrera} D., {Johansen} A., and {Davies} M.~B. (2015) \emph{{How to form
  planetesimals from mm-sized chondrules and chondrule aggregates}},
  \emph{\aap}, \emph{579}, A43.

\bibitem[{\emph{{Castillo-Rogez} and {Schmidt}}(2010)}]{Castillo2010}
{Castillo-Rogez} J.~C. and {Schmidt} B.~E. (2010) \emph{{Geophysical evolution
  of the Themis family parent body}}, \emph{\grl}, \emph{37}, L10202.

\bibitem[{\emph{{Cazaux} et~al.}(2011)\emph{{Cazaux}, {Caselli}, and
  {Spaans}}}]{Cazaux2011}
{Cazaux} S., {Caselli} P., and {Spaans} M. (2011) \emph{{Interstellar Ices as
  Witnesses of Star Formation: Selective Deuteration of Water and Organic
  Molecules Unveiled}}, \emph{\apjl}, \emph{741}, L34.

\bibitem[{\emph{{Chabrier}}(2003)}]{chabrier03}
{Chabrier} G. (2003) \emph{{Galactic Stellar and Substellar Initial Mass
  Function}}, \emph{\pasp}, \emph{115}, 763--795.

\bibitem[{\emph{{Chambers}}(2001)}]{chambers01}
{Chambers} J.~E. (2001) \emph{{Making More Terrestrial Planets}},
  \emph{Icarus}, \emph{152}, 205--224.

\bibitem[{\emph{{Chatterjee} et~al.}(2008)\emph{{Chatterjee}, {Ford},
  {Matsumura}, and {Rasio}}}]{chatterjee08}
{Chatterjee} S., {Ford} E.~B., {Matsumura} S., and {Rasio} F.~A. (2008)
  \emph{{Dynamical Outcomes of Planet-Planet Scattering}}, \emph{\apj},
  \emph{686}, 580--602.

\bibitem[{\emph{{Chatterjee} and {Tan}}(2014)}]{chatterjee14}
{Chatterjee} S. and {Tan} J.~C. (2014) \emph{{Inside-out Planet Formation}},
  \emph{\apj}, \emph{780}, 53.

\bibitem[{\emph{{Chatterjee} and {Tan}}(2015)}]{chatterjee15}
{Chatterjee} S. and {Tan} J.~C. (2015) \emph{{Vulcan Planets: Inside-out
  Formation of the Innermost Super-Earths}}, \emph{\apjl}, \emph{798}, L32.

\bibitem[{\emph{{Chen} and {Kipping}}(2017)}]{chen17}
{Chen} J. and {Kipping} D. (2017) \emph{{Probabilistic Forecasting of the
  Masses and Radii of Other Worlds}}, \emph{\apj}, \emph{834}, 17.

\bibitem[{\emph{{Cieza} et~al.}(2016)\emph{{Cieza}, {Casassus}, {Tobin}, {Bos},
  {Williams}, {Perez}, {Zhu}, {Caceres}, {Canovas}, {Dunham}, {Hales},
  {Prieto}, {Principe}, {Schreiber}, {Ruiz-Rodriguez}, and {Zurlo}}}]{cieza16}
{Cieza} L.~A., {Casassus} S., {Tobin} J., {Bos} S.~P., {Williams} J.~P.,
  {Perez} S., {Zhu} Z., {Caceres} C., {Canovas} H., {Dunham} M.~M., {Hales} A.,
  {Prieto} J.~L., {Principe} D.~A., {Schreiber} M.~R., {Ruiz-Rodriguez} D., and
  {Zurlo} A. (2016) \emph{{Imaging the water snow-line during a protostellar
  outburst}}, \emph{\nat}, \emph{535}, 258--261.

\bibitem[{\emph{{Clark} et~al.}(2019)\emph{{Clark}, {Brown}, {Cruikshank}, and
  {Swayze}}}]{Clark2019}
{Clark} R.~N., {Brown} R.~H., {Cruikshank} D.~P., and {Swayze} G.~A. (2019)
  \emph{{Isotopic ratios of Saturn's rings and satellites: Implications for the
  origin of water and Phoebe}}, \emph{\icarus}, \emph{321}, 791--802.

\bibitem[{\emph{{Clayton} et~al.}(1973)\emph{{Clayton}, {Grossman}, and
  {Mayeda}}}]{Clayton1973}
{Clayton} R.~N., {Grossman} L., and {Mayeda} T.~K. (1973) \emph{{A Component of
  Primitive Nuclear Composition in Carbonaceous Meteorites}}, \emph{Science},
  \emph{182}, 485--488.

\bibitem[{\emph{{Cleeves} et~al.}(2014a)\emph{{Cleeves}, {Bergin}, and
  {Adams}}}]{Cleeves2014a}
{Cleeves} L.~I., {Bergin} E.~A., and {Adams} F.~C. (2014a) \emph{{Exclusion of
  Cosmic Rays in Protoplanetary Disks. II. Chemical Gradients and Observational
  Signatures}}, \emph{\apj}, \emph{794}, 123.

\bibitem[{\emph{{Cleeves} et~al.}(2014b)\emph{{Cleeves}, {Bergin}, {Alexand
  er}, {Du}, {Graninger}, {{\"O}berg}, and {Harries}}}]{Cleeves2014b}
{Cleeves} L.~I., {Bergin} E.~A., {Alexand er} C. M.~O.~D., {Du} F., {Graninger}
  D., {{\"O}berg} K.~I., and {Harries} T.~J. (2014b) \emph{{The ancient
  heritage of water ice in the solar system}}, \emph{Science}, \emph{345},
  1590--1593.

\bibitem[{\emph{{Cleeves} et~al.}(2016)\emph{{Cleeves}, {Bergin}, {O'D. Alexand
  er}, {Du}, {Graninger}, {{\"O}berg}, and {Harries}}}]{cleeves16}
{Cleeves} L.~I., {Bergin} E.~A., {O'D. Alexand er} C.~M., {Du} F., {Graninger}
  D., {{\"O}berg} K.~I., and {Harries} T.~J. (2016) \emph{{Exploring the
  Origins of Deuterium Enrichments in Solar Nebular Organics}}, \emph{\apj},
  \emph{819}, 13.

\bibitem[{\emph{{Cleeves} et~al.}(2015)\emph{{Cleeves}, {Bergin}, {Qi},
  {Adams}, and {{\"O}berg}}}]{Cleeves2015}
{Cleeves} L.~I., {Bergin} E.~A., {Qi} C., {Adams} F.~C., and {{\"O}berg} K.~I.
  (2015) \emph{{Constraining the X-Ray and Cosmic-Ray Ionization Chemistry of
  the TW Hya Protoplanetary Disk: Evidence for a Sub-interstellar Cosmic-Ray
  Rate}}, \emph{\apj}, \emph{799}, 204.

\bibitem[{\emph{{Clement} et~al.}(2019{\natexlab{a}})\emph{{Clement}, {Kaib},
  {Raymond}, {Chambers}, and {Walsh}}}]{clement19}
{Clement} M.~S., {Kaib} N.~A., {Raymond} S.~N., {Chambers} J.~E., and {Walsh}
  K.~J. (2019{\natexlab{a}}) \emph{{The early instability scenario: Terrestrial
  planet formation during the giant planet instability, and the effect of
  collisional fragmentation}}, \emph{\icarus}, \emph{321}, 778--790.

\bibitem[{\emph{{Clement} et~al.}(2018)\emph{{Clement}, {Kaib}, {Raymond}, and
  {Walsh}}}]{clement18}
{Clement} M.~S., {Kaib} N.~A., {Raymond} S.~N., and {Walsh} K.~J. (2018)
  \emph{{Mars' Growth Stunted by an Early Giant Planet Instability}},
  \emph{ArXiv e-prints}.

\bibitem[{\emph{{Clement} et~al.}(2019{\natexlab{b}})\emph{{Clement},
  {Raymond}, and {Kaib}}}]{clement19b}
{Clement} M.~S., {Raymond} S.~N., and {Kaib} N.~A. (2019{\natexlab{b}})
  \emph{{Excitation and Depletion of the Asteroid Belt in the Early Instability
  Scenario}}, \emph{\aj}, \emph{157}, 38.

\bibitem[{\emph{{Clesi} et~al.}(2018)\emph{{Clesi}, {Bouhifd},
  {Bolfan-Casanova}, {Manthilake}, {Schiavi}, {Raepsaet}, {Bureau}, {Khodja},
  and {Andrault}}}]{clesi18}
{Clesi} V., {Bouhifd} M.~A., {Bolfan-Casanova} N., {Manthilake} G., {Schiavi}
  F., {Raepsaet} C., {Bureau} H., {Khodja} H., and {Andrault} D. (2018)
  \emph{{Low hydrogen contents in the cores of terrestrial planets}},
  \emph{Science Advances}, \emph{4}, e1701876.

\bibitem[{\emph{{Combe} et~al.}(2016)\emph{{Combe}, {McCord}, {Tosi},
  {Ammannito}, {Carrozzo}, {De Sanctis}, {Raponi}, {Byrne}, {Landis}, and
  {Hughson}}}]{Combe2016}
{Combe} J.-P., {McCord} T.~B., {Tosi} F., {Ammannito} E., {Carrozzo} F.~G., {De
  Sanctis} M.~C., {Raponi} A., {Byrne} S., {Landis} M.~E., and {Hughson} K.
  H.~G. (2016) \emph{{Detection of local H$_{2}$O exposed at the surface of
  Ceres}}, \emph{Science}, \emph{353}, aaf3010.

\bibitem[{\emph{{Cossou} et~al.}(2014)\emph{{Cossou}, {Raymond}, {Hersant}, and
  {Pierens}}}]{cossou14}
{Cossou} C., {Raymond} S.~N., {Hersant} F., and {Pierens} A. (2014) \emph{{Hot
  super-Earths and giant planet cores from different migration histories}},
  \emph{\aap}, \emph{569}, A56.

\bibitem[{\emph{{Coutens} et~al.}(2014)\emph{{Coutens}, {Vastel}, {Hincelin},
  {Herbst}, {Lis}, {Chavarr{\'\i}a}, {G{\'e}rin}, {van der Tak}, {Persson},
  {Goldsmith}, and {Caux}}}]{Coutens2014}
{Coutens} A., {Vastel} C., {Hincelin} U., {Herbst} E., {Lis} D.~C.,
  {Chavarr{\'\i}a} L., {G{\'e}rin} M., {van der Tak} F.~F.~S., {Persson} C.~M.,
  {Goldsmith} P.~F., and {Caux} E. (2014) \emph{{Water deuterium fractionation
  in the high-mass star-forming region G34.26+0.15 based on Herschel/HIFI
  data}}, \emph{\mnras}, \emph{445}, 1299--1313.

\bibitem[{\emph{{Crida} et~al.}(2009)\emph{{Crida}, {Masset}, and
  {Morbidelli}}}]{crida09}
{Crida} A., {Masset} F., and {Morbidelli} A. (2009) \emph{{Long Range Outward
  Migration of Giant Planets, with Application to Fomalhaut b}}, \emph{\apjl},
  \emph{705}, L148--L152.

\bibitem[{\emph{{Crida} et~al.}(2006)\emph{{Crida}, {Morbidelli}, and
  {Masset}}}]{crida06}
{Crida} A., {Morbidelli} A., and {Masset} F. (2006) \emph{{On the width and
  shape of gaps in protoplanetary disks}}, \emph{Icarus}, \emph{181}, 587--604.

\bibitem[{\emph{{Criswell} and {De}}(1977)}]{Criswell1977}
{Criswell} D.~R. and {De} B.~R. (1977) \emph{{Intense localized photoelectric
  charging in the lunar sunset terminator region, 2. Supercharging at the
  progression of sunset}}, \emph{\jgr}, \emph{82}, 1005.

\bibitem[{\emph{{Cumming} et~al.}(2008)\emph{{Cumming}, {Butler}, {Marcy},
  {Vogt}, {Wright}, and {Fischer}}}]{cumming08}
{Cumming} A., {Butler} R.~P., {Marcy} G.~W., {Vogt} S.~S., {Wright} J.~T., and
  {Fischer} D.~A. (2008) \emph{{The Keck Planet Search: Detectability and the
  Minimum Mass and Orbital Period Distribution of Extrasolar Planets}},
  \emph{\pasp}, \emph{120}, 531--554.

\bibitem[{\emph{{Dauphas}}(2003)}]{dauphas03}
{Dauphas} N. (2003) \emph{{The dual origin of the terrestrial atmosphere}},
  \emph{\icarus}, \emph{165}, 326--339.

\bibitem[{\emph{{Dauphas}}(2017)}]{dauphas17}
{Dauphas} N. (2017) \emph{{The isotopic nature of the Earth's accreting
  material through time}}, \emph{\nat}, \emph{541}, 521--524.

\bibitem[{\emph{{Dauphas} and {Pourmand}}(2011)}]{dauphas11}
{Dauphas} N. and {Pourmand} A. (2011) \emph{{Hf-W-Th evidence for rapid growth
  of Mars and its status as a planetary embryo}}, \emph{\nat}, \emph{473},
  489--492.

\bibitem[{\emph{{Day} et~al.}(2007)\emph{{Day}, {Pearson}, and
  {Taylor}}}]{day07}
{Day} J.~M.~D., {Pearson} D.~G., and {Taylor} L.~A. (2007) \emph{{Highly
  Siderophile Element Constraints on Accretion and Differentiation of the
  Earth-Moon System}}, \emph{Science}, \emph{315}, 217--.

\bibitem[{\emph{{Deienno} et~al.}(2016)\emph{{Deienno}, {Gomes}, {Walsh},
  {Morbidelli}, and {Nesvorn{\'y}}}}]{deienno16}
{Deienno} R., {Gomes} R.~S., {Walsh} K.~J., {Morbidelli} A., and {Nesvorn{\'y}}
  D. (2016) \emph{{Is the Grand Tack model compatible with the orbital
  distribution of main belt asteroids?}}, \emph{\icarus}, \emph{272}, 114--124.

\bibitem[{\emph{{Delsemme}}(1992)}]{delsemme92}
{Delsemme} A.~H. (1992) \emph{{Cometary origin of carbon, nitrogen, and water
  on the earth}}, \emph{Origins of Life and Evolution of the Biosphere},
  \emph{21}, 279--298.

\bibitem[{\emph{{DeMeo} and {Carry}}(2013)}]{demeo13}
{DeMeo} F.~E. and {Carry} B. (2013) \emph{{The taxonomic distribution of
  asteroids from multi-filter all-sky photometric surveys}}, \emph{Icarus},
  \emph{226}, 723--741.

\bibitem[{\emph{{Desch} et~al.}(2018{\natexlab{a}})\emph{{Desch}, {Kalyaan},
  and {Alexander}}}]{desch18}
{Desch} S.~J., {Kalyaan} A., and {Alexander} C.~M. (2018{\natexlab{a}})
  \emph{{The Effect of Jupiter's Formation on the Distribution of Refractory
  Elements and Inclusions in Meteorites}}, \emph{\apjs}, \emph{238}, 11.

\bibitem[{\emph{{Desch} et~al.}(2018{\natexlab{b}})\emph{{Desch}, {Kalyaan},
  and {O'D. Alexander}}}]{Desch2018}
{Desch} S.~J., {Kalyaan} A., and {O'D. Alexander} C.~M. (2018{\natexlab{b}})
  \emph{{The Effect of Jupiter's Formation on the Distribution of Refractory
  Elements and Inclusions in Meteorites}}, \emph{\apjs}, \emph{238}, 11.

\bibitem[{\emph{{Deutsch} et~al.}(2019)\emph{{Deutsch}, {Head}, and
  {Neumann}}}]{Deutsch2019}
{Deutsch} A.~N., {Head} J.~W., and {Neumann} G.~A. (2019) \emph{{Age
  constraints of Mercury's polar deposits suggest recent delivery of ice}},
  \emph{Earth and Planetary Science Letters}, \emph{520}, 26--33.

\bibitem[{\emph{{Dodson-Robinson} and {Bodenheimer}}(2010)}]{dodson-robinson10}
{Dodson-Robinson} S.~E. and {Bodenheimer} P. (2010) \emph{{The formation of
  Uranus and Neptune in solid-rich feeding zones: Connecting chemistry and
  dynamics}}, \emph{Icarus}, \emph{207}, 491--498.

\bibitem[{\emph{{Donahue} et~al.}(1982)\emph{{Donahue}, {Hoffman}, {Hodges},
  and {Watson}}}]{Donahue1982}
{Donahue} T.~M., {Hoffman} J.~H., {Hodges} R.~R., and {Watson} A.~J. (1982)
  \emph{{Venus Was Wet: A Measurement of the Ratio of Deuterium to Hydrogen}},
  \emph{Science}, \emph{216}, 630--633.

\bibitem[{\emph{{Dorn} et~al.}(2015)\emph{{Dorn}, {Khan}, {Heng}, {Connolly},
  {Alibert}, {Benz}, and {Tackley}}}]{dorn15}
{Dorn} C., {Khan} A., {Heng} K., {Connolly} J. A.~D., {Alibert} Y., {Benz} W.,
  and {Tackley} P. (2015) \emph{{Can we constrain the interior structure of
  rocky exoplanets from mass and radius measurements?}}, \emph{\aap},
  \emph{577}, A83.

\bibitem[{\emph{{Drahus} et~al.}(2011)\emph{{Drahus}, {Jewitt},
  {Guilbert-Lepoutre}, {Waniak}, {Hoge}, {Lis}, {Yoshida}, {Peng}, and
  {Sievers}}}]{Drahus2011}
{Drahus} M., {Jewitt} D., {Guilbert-Lepoutre} A., {Waniak} W., {Hoge} J., {Lis}
  D.~C., {Yoshida} H., {Peng} R., and {Sievers} A. (2011) \emph{{Rotation State
  of Comet 103P/Hartley 2 from Radio Spectroscopy at 1 mm}}, \emph{\apjl},
  \emph{734}, L4.

\bibitem[{\emph{{Drake}}(2004)}]{Drake2004}
{Drake} M.~J. (2004) \emph{{Origin of Water in the Terrestrial Planets}},
  \emph{Meteoritics and Planetary Science Supplement}, \emph{39}, 5031.

\bibitem[{\emph{{Dr{\c a}{\.z}kowska} and {Alibert}}(2017)}]{drazkowska17}
{Dr{\c a}{\.z}kowska} J. and {Alibert} Y. (2017) \emph{{Planetesimal formation
  starts at the snow line}}, \emph{\aap}, \emph{608}, A92.

\bibitem[{\emph{{Du} and {Bergin}}(2014)}]{Du2014}
{Du} F. and {Bergin} E.~A. (2014) \emph{{Water Vapor Distribution in
  Protoplanetary Disks}}, \emph{\apj}, \emph{792}, 2.

\bibitem[{\emph{{Durda} et~al.}(2007)\emph{{Durda}, {Bottke}, {Nesvorn{\'y}},
  {Enke}, {Merline}, {Asphaug}, and {Richardson}}}]{Durda2007}
{Durda} D.~D., {Bottke} W.~F., {Nesvorn{\'y}} D., {Enke} B.~L., {Merline}
  W.~J., {Asphaug} E., and {Richardson} D.~C. (2007) \emph{{Size-frequency
  distributions of fragments from SPH/ N-body simulations of asteroid impacts:
  Comparison with observed asteroid families}}, \emph{\icarus}, \emph{186},
  498--516.

\bibitem[{\emph{{Eberhardt} et~al.}(1987)\emph{{Eberhardt}, {Dolder},
  {Schulte}, {Krankowsky}, {Lammerzahl}, {Hoffman}, {Hodges}, {Berthelier}, and
  {Illiano}}}]{Eberhardt1987}
{Eberhardt} P., {Dolder} U., {Schulte} W., {Krankowsky} D., {Lammerzahl} P.,
  {Hoffman} J.~H., {Hodges} R.~R., {Berthelier} J.~J., and {Illiano} J.~M.
  (1987) \emph{{The D/h Ratio in Water from Comet p/ Halley}}, \emph{\aap},
  \emph{187}, 435.

\bibitem[{\emph{{Eberhardt} et~al.}(1995)\emph{{Eberhardt}, {Reber},
  {Krankowsky}, and {Hodges}}}]{Eberhardt1995}
{Eberhardt} P., {Reber} M., {Krankowsky} D., and {Hodges} R.~R. (1995)
  \emph{{The D/H and \^18\^O/\^16\^O ratios in water from comet P/Halley.}},
  \emph{\aap}, \emph{302}, 301.

\bibitem[{\emph{{Elkins-Tanton}}(2008)}]{Elkins-Tanton2008}
{Elkins-Tanton} L.~T. (2008) \emph{{Linked magma ocean solidification and
  atmospheric growth for Earth and Mars}}, \emph{Earth and Planetary Science
  Letters}, \emph{271}, 181--191.

\bibitem[{\emph{{Elkins-Tanton}}(2012)}]{Elkins2012}
{Elkins-Tanton} L.~T. (2012) \emph{{Magma Oceans in the Inner Solar System}},
  \emph{Annual Review of Earth and Planetary Sciences}, \emph{40}, 113--139.

\bibitem[{\emph{{Ercolano} and {Pascucci}}(2017)}]{Ercolano2017}
{Ercolano} B. and {Pascucci} I. (2017) \emph{{The dispersal of planet-forming
  discs: theory confronts observations}}, \emph{Royal Society Open Science},
  \emph{4}, 170114.

\bibitem[{\emph{{Fernandes} et~al.}(2019)\emph{{Fernandes}, {Mulders},
  {Pascucci}, {Mordasini}, and {Emsenhuber}}}]{fernandes19}
{Fernandes} R.~B., {Mulders} G.~D., {Pascucci} I., {Mordasini} C., and
  {Emsenhuber} A. (2019) \emph{{Hints for a Turnover at the Snow Line in the
  Giant Planet Occurrence Rate}}, \emph{\apj}, \emph{874}, 81.

\bibitem[{\emph{{Feuchtgruber} et~al.}(2013)\emph{{Feuchtgruber}, {Lellouch},
  {Orton}, {de Graauw}, {Vandenbussche}, {Swinyard}, {Moreno}, {Jarchow},
  {Billebaud}, {Cavali{\'e}}, {Sidher}, and {Hartogh}}}]{Feuchtgruber2013}
{Feuchtgruber} H., {Lellouch} E., {Orton} G., {de Graauw} T., {Vandenbussche}
  B., {Swinyard} B., {Moreno} R., {Jarchow} C., {Billebaud} F., {Cavali{\'e}}
  T., {Sidher} S., and {Hartogh} P. (2013) \emph{{The D/H ratio in the
  atmospheres of Uranus and Neptune from Herschel-PACS observations}},
  \emph{\aap}, \emph{551}, A126.

\bibitem[{\emph{{Fischer} et~al.}(2014)\emph{{Fischer}, {Howard}, {Laughlin},
  {Macintosh}, {Mahadevan}, {Sahlmann}, and {Yee}}}]{fischer14}
{Fischer} D.~A., {Howard} A.~W., {Laughlin} G.~P., {Macintosh} B., {Mahadevan}
  S., {Sahlmann} J., and {Yee} J.~C. (2014) \emph{{Exoplanet Detection
  Techniques}}, \emph{Protostars and Planets VI}, pp. 715--737.

\bibitem[{\emph{{Fischer} and {Nimmo}}(2018)}]{Fischer2018}
{Fischer} R.~A. and {Nimmo} F. (2018) \emph{{Effects of core formation on the
  Hf-W isotopic composition of the Earth and dating of the Moon-forming
  impact}}, \emph{Earth and Planetary Science Letters}, \emph{499}, 257--265.

\bibitem[{\emph{{Fogg} and {Nelson}}(2005)}]{fogg05}
{Fogg} M.~J. and {Nelson} R.~P. (2005) \emph{{Oligarchic and giant impact
  growth of terrestrial planets in the presence of gas giant planet
  migration}}, \emph{\aap}, \emph{441}, 791--806.

\bibitem[{\emph{{Ford} and {Rasio}}(2008)}]{ford08}
{Ford} E.~B. and {Rasio} F.~A. (2008) \emph{{Origins of Eccentric Extrasolar
  Planets: Testing the Planet-Planet Scattering Model}}, \emph{\apj},
  \emph{686}, 621--636.

\bibitem[{\emph{{Fornasier} et~al.}(2016)\emph{{Fornasier}, {Lantz}, {Perna},
  {Campins}, {Barucci}, and {Nesvorny}}}]{Fornasier2016}
{Fornasier} S., {Lantz} C., {Perna} D., {Campins} H., {Barucci} M.~A., and
  {Nesvorny} D. (2016) \emph{{Spectral variability on primitive asteroids of
  the Themis and Beagle families: Space weathering effects or parent body
  heterogeneity?}}, \emph{\icarus}, \emph{269}, 1--14.

\bibitem[{\emph{{Fulton} et~al.}(2017)\emph{{Fulton}, {Petigura}, {Howard},
  {Isaacson}, {Marcy}, {Cargile}, {Hebb}, {Weiss}, {Johnson}, {Morton},
  {Sinukoff}, {Crossfield}, and {Hirsch}}}]{fulton17}
{Fulton} B.~J., {Petigura} E.~A., {Howard} A.~W., {Isaacson} H., {Marcy} G.~W.,
  {Cargile} P.~A., {Hebb} L., {Weiss} L.~M., {Johnson} J.~A., {Morton} T.~D.,
  {Sinukoff} E., {Crossfield} I.~J.~M., and {Hirsch} L.~A. (2017) \emph{{The
  California-Kepler Survey. III. A Gap in the Radius Distribution of Small
  Planets}}, \emph{\aj}, \emph{154}, 109.

\bibitem[{\emph{{Gaidos} et~al.}(2009)\emph{{Gaidos}, {Krot}, {Williams}, and
  {Raymond}}}]{gaidos09}
{Gaidos} E., {Krot} A.~N., {Williams} J.~P., and {Raymond} S.~N. (2009)
  \emph{{$^{26}$Al and the Formation of the Solar System from a Molecular Cloud
  Contaminated by Wolf-Rayet Winds}}, \emph{\apj}, \emph{696}, 1854--1863.

\bibitem[{\emph{{Garaud} and {Lin}}(2007)}]{Garaud2007}
{Garaud} P. and {Lin} D.~N.~C. (2007) \emph{{The Effect of Internal Dissipation
  and Surface Irradiation on the Structure of Disks and the Location of the
  Snow Line around Sun-like Stars}}, \emph{\apj}, \emph{654}, 606--624.

\bibitem[{\emph{{Geiss} and {Gloeckler}}(1998)}]{Geiss1998}
{Geiss} J. and {Gloeckler} G. (1998) \emph{{Abundances of Deuterium and
  Helium-3 in the Protosolar Cloud}}, \emph{\ssr}, \emph{84}, 239--250.

\bibitem[{\emph{{Genda} and {Abe}}(2005)}]{genda05}
{Genda} H. and {Abe} Y. (2005) \emph{{Enhanced atmospheric loss on protoplanets
  at the giant impact phase in the presence of oceans}}, \emph{\nat},
  \emph{433}, 842--844.

\bibitem[{\emph{{Genda} and {Ikoma}}(2008)}]{Genda2008}
{Genda} H. and {Ikoma} M. (2008) \emph{{Origin of the ocean on the Earth: Early
  evolution of water D/H in a hydrogen-rich atmosphere}}, \emph{\icarus},
  \emph{194}, 42--52.

\bibitem[{\emph{{Gillon} et~al.}(2017)\emph{{Gillon}, {Triaud}, {Demory},
  {Jehin}, {Agol}, {Deck}, {Lederer}, {de Wit}, {Burdanov}, {Ingalls},
  {Bolmont}, {Leconte}, {Raymond}, {Selsis}, {Turbet}, {Barkaoui}, {Burgasser},
  {Burleigh}, {Carey}, {Chaushev}, {Copperwheat}, {Delrez}, {Fernandes},
  {Holdsworth}, {Kotze}, {Van Grootel}, {Almleaky}, {Benkhaldoun}, {Magain},
  and {Queloz}}}]{gillon17}
{Gillon} M., {Triaud} A.~H.~M.~J., {Demory} B.-O., {Jehin} E., {Agol} E.,
  {Deck} K.~M., {Lederer} S.~M., {de Wit} J., {Burdanov} A., {Ingalls} J.~G.,
  {Bolmont} E., {Leconte} J., {Raymond} S.~N., {Selsis} F., {Turbet} M.,
  {Barkaoui} K., {Burgasser} A., {Burleigh} M.~R., {Carey} S.~J., {Chaushev}
  A., {Copperwheat} C.~M., {Delrez} L., {Fernandes} C.~S., {Holdsworth} D.~L.,
  {Kotze} E.~J., {Van Grootel} V., {Almleaky} Y., {Benkhaldoun} Z., {Magain}
  P., and {Queloz} D. (2017) \emph{{Seven temperate terrestrial planets around
  the nearby ultracool dwarf star TRAPPIST-1}}, \emph{\nat}, \emph{542},
  456--460.

\bibitem[{\emph{{Gladman} et~al.}(2001)\emph{{Gladman}, {Kavelaars}, {Petit},
  {Morbidelli}, {Holman}, and {Loredo}}}]{gladman01}
{Gladman} B., {Kavelaars} J.~J., {Petit} J., {Morbidelli} A., {Holman} M.~J.,
  and {Loredo} T. (2001) \emph{{The Structure of the Kuiper Belt: Size
  Distribution and Radial Extent}}, \emph{\aj}, \emph{122}, 1051--1066.

\bibitem[{\emph{{Gomes} et~al.}(2005)\emph{{Gomes}, {Levison}, {Tsiganis}, and
  {Morbidelli}}}]{gomes05}
{Gomes} R., {Levison} H.~F., {Tsiganis} K., and {Morbidelli} A. (2005)
  \emph{{Origin of the cataclysmic Late Heavy Bombardment period of the
  terrestrial planets}}, \emph{\nat}, \emph{435}, 466--469.

\bibitem[{\emph{{Gould} et~al.}(2010)\emph{{Gould}, {Dong}, {Gaudi}, {Udalski},
  {Bond}, {Greenhill}, {Street}, {Dominik}, {Sumi}, {Szyma{\'n}ski}, {Han},
  {Allen}, {Bolt}, {Bos}, {Christie}, {DePoy}, {Drummond}, {Eastman},
  {Gal-Yam}, {Higgins}, {Janczak}, {Kaspi}, {Koz{\l}owski}, {Lee}, {Mallia},
  {Maury}, {Maoz}, {McCormick}, {Monard}, {Moorhouse}, {Morgan}, {Natusch},
  {Ofek}, {Park}, {Pogge}, {Polishook}, {Santallo}, {Shporer}, {Spector},
  {Thornley}, {Yee}, {{$\mu$}FUN Collaboration}, {Kubiak}, {Pietrzy{\'n}ski},
  {Soszy{\'n}ski}, {Szewczyk}, {Wyrzykowski}, {Ulaczyk}, {Poleski}, {OGLE
  Collaboration}, {Abe}, {Bennett}, {Botzler}, {Douchin}, {Freeman}, {Fukui},
  {Furusawa}, {Hearnshaw}, {Hosaka}, {Itow}, {Kamiya}, {Kilmartin}, {Korpela},
  {Lin}, {Ling}, {Makita}, {Masuda}, {Matsubara}, {Miyake}, {Muraki}, {Nagaya},
  {Nishimoto}, {Ohnishi}, {Okumura}, {Perrott}, {Philpott}, {Rattenbury},
  {Saito}, {Sako}, {Sullivan}, {Sweatman}, {Tristram}, {von Seggern}, {Yock},
  {MOA Collaboration}, {Albrow}, {Batista}, {Beaulieu}, {Brillant}, {Caldwell},
  {Calitz}, {Cassan}, {Cole}, {Cook}, {Coutures}, {Dieters}, {Dominis Prester},
  {Donatowicz}, {Fouqu{\'e}}, {Hill}, {Hoffman}, {Jablonski}, {Kane}, {Kains},
  {Kubas}, {Marquette}, {Martin}, {Martioli}, {Meintjes}, {Menzies},
  {Pedretti}, {Pollard}, {Sahu}, {Vinter}, {Wambsganss}, {Watson}, {Williams},
  {Zub}, {PLANET Collaboration}, {Allan}, {Bode}, {Bramich}, {Burgdorf},
  {Clay}, {Fraser}, {Hawkins}, {Horne}, {Kerins}, {Lister}, {Mottram},
  {Saunders}, {Snodgrass}, {Steele}, {Tsapras}, {RoboNet Collaboration},
  {J{\o}rgensen}, {Anguita}, {Bozza}, {Calchi Novati}, {Harps{\o}e}, {Hinse},
  {Hundertmark}, {Kj{\ae}rgaard}, {Liebig}, {Mancini}, {Masi}, {Mathiasen},
  {Rahvar}, {Ricci}, {Scarpetta}, {Southworth}, {Surdej}, {Th{\"o}ne}, and
  {MiNDSTEp Consortium}}}]{gould10}
{Gould} A., {Dong} S., {Gaudi} B.~S., {Udalski} A., {Bond} I.~A., {Greenhill}
  J., {Street} R.~A., {Dominik} M., {Sumi} T., {Szyma{\'n}ski} M.~K., {Han} C.,
  {Allen} W., {Bolt} G., {Bos} M., {Christie} G.~W., {DePoy} D.~L., {Drummond}
  J., {Eastman} J.~D., {Gal-Yam} A., {Higgins} D., {Janczak} J., {Kaspi} S.,
  {Koz{\l}owski} S., {Lee} C., {Mallia} F., {Maury} A., {Maoz} D., {McCormick}
  J., {Monard} L.~A.~G., {Moorhouse} D., {Morgan} N., {Natusch} T., {Ofek}
  E.~O., {Park} B., {Pogge} R.~W., {Polishook} D., {Santallo} R., {Shporer} A.,
  {Spector} O., {Thornley} G., {Yee} J.~C., {{$\mu$}FUN Collaboration},
  {Kubiak} M., {Pietrzy{\'n}ski} G., {Soszy{\'n}ski} I., {Szewczyk} O.,
  {Wyrzykowski} {\L}., {Ulaczyk} K., {Poleski} R., {OGLE Collaboration}, {Abe}
  F., {Bennett} D.~P., {Botzler} C.~S., {Douchin} D., {Freeman} M., {Fukui} A.,
  {Furusawa} K., {Hearnshaw} J.~B., {Hosaka} S., {Itow} Y., {Kamiya} K.,
  {Kilmartin} P.~M., {Korpela} A., {Lin} W., {Ling} C.~H., {Makita} S.,
  {Masuda} K., {Matsubara} Y., {Miyake} N., {Muraki} Y., {Nagaya} M.,
  {Nishimoto} K., {Ohnishi} K., {Okumura} T., {Perrott} Y.~C., {Philpott} L.,
  {Rattenbury} N., {Saito} T., {Sako} T., {Sullivan} D.~J., {Sweatman} W.~L.,
  {Tristram} P.~J., {von Seggern} E., {Yock} P.~C.~M., {MOA Collaboration},
  {Albrow} M., {Batista} V., {Beaulieu} J.~P., {Brillant} S., {Caldwell} J.,
  {Calitz} J.~J., {Cassan} A., {Cole} A., {Cook} K., {Coutures} C., {Dieters}
  S., {Dominis Prester} D., {Donatowicz} J., {Fouqu{\'e}} P., {Hill} K.,
  {Hoffman} M., {Jablonski} F., {Kane} S.~R., {Kains} N., {Kubas} D.,
  {Marquette} J., {Martin} R., {Martioli} E., {Meintjes} P., {Menzies} J.,
  {Pedretti} E., {Pollard} K., {Sahu} K.~C., {Vinter} C., {Wambsganss} J.,
  {Watson} R., {Williams} A., {Zub} M., {PLANET Collaboration}, {Allan} A.,
  {Bode} M.~F., {Bramich} D.~M., {Burgdorf} M.~J., {Clay} N., {Fraser} S.,
  {Hawkins} E., {Horne} K., {Kerins} E., {Lister} T.~A., {Mottram} C.,
  {Saunders} E.~S., {Snodgrass} C., {Steele} I.~A., {Tsapras} Y., {RoboNet
  Collaboration}, {J{\o}rgensen} U.~G., {Anguita} T., {Bozza} V., {Calchi
  Novati} S., {Harps{\o}e} K., {Hinse} T.~C., {Hundertmark} M., {Kj{\ae}rgaard}
  P., {Liebig} C., {Mancini} L., {Masi} G., {Mathiasen} M., {Rahvar} S.,
  {Ricci} D., {Scarpetta} G., {Southworth} J., {Surdej} J., {Th{\"o}ne} C.~C.,
  and {MiNDSTEp Consortium} (2010) \emph{{Frequency of Solar-like Systems and
  of Ice and Gas Giants Beyond the Snow Line from High-magnification
  Microlensing Events in 2005-2008}}, \emph{\apj}, \emph{720}, 1073--1089.

\bibitem[{\emph{{Gounelle} and {Meibom}}(2008)}]{gounelle08}
{Gounelle} M. and {Meibom} A. (2008) \emph{{The Origin of Short-lived
  Radionuclides and the Astrophysical Environment of Solar System Formation}},
  \emph{\apj}, \emph{680}, 781--792.

\bibitem[{\emph{{Gradie} and {Tedesco}}(1982)}]{gradie82}
{Gradie} J. and {Tedesco} E. (1982) \emph{{Compositional structure of the
  asteroid belt}}, \emph{Science}, \emph{216}, 1405--1407.

\bibitem[{\emph{{Grimm} and {McSween}}(1993)}]{grimm93}
{Grimm} R.~E. and {McSween} H.~Y. (1993) \emph{{Heliocentric zoning of the
  asteroid belt by aluminum-26 heating}}, \emph{Science}, \emph{259}, 653--655.

\bibitem[{\emph{{Grossman}}(1972)}]{Grossman1972}
{Grossman} L. (1972) \emph{{Condensation in the primitive solar nebula}},
  \emph{\gca}, \emph{36}, 597--619.

\bibitem[{\emph{{Gupta} and {Schlichting}}(2019)}]{gupta19}
{Gupta} A. and {Schlichting} H.~E. (2019) \emph{{Sculpting the valley in the
  radius distribution of small exoplanets as a by-product of planet formation:
  the core-powered mass-loss mechanism}}, \emph{\mnras}, \emph{487}, 24--33.

\bibitem[{\emph{{Haghighipour}}(2009)}]{Haghighipour2009}
{Haghighipour} N. (2009) \emph{{Dynamical constraints on the origin of Main
  Belt comets}}, \emph{Meteoritics and Planetary Science}, \emph{44},
  1863--1869.

\bibitem[{\emph{{Haghighipour} and {Boss}}(2003)}]{haghighipour03}
{Haghighipour} N. and {Boss} A.~P. (2003) \emph{{On Pressure Gradients and
  Rapid Migration of Solids in a Nonuniform Solar Nebula}}, \emph{\apj},
  \emph{583}, 996--1003.

\bibitem[{\emph{{Haghighipour} et~al.}(2016)\emph{{Haghighipour}, {Maindl},
  {Sch{\"a}fer}, {Speith}, and {Dvorak}}}]{Haghighipour2016}
{Haghighipour} N., {Maindl} T.~I., {Sch{\"a}fer} C., {Speith} R., and {Dvorak}
  R. (2016) \emph{{Triggering Sublimation-driven Activity of Main Belt
  Comets}}, \emph{\apj}, \emph{830}, 22.

\bibitem[{\emph{{Hainaut} et~al.}(2012)\emph{{Hainaut}, {Kleyna}, {Sarid},
  {Hermalyn}, {Zenn}, {Meech}, {Schultz}, {Hsieh}, {Trancho}, and
  {Pittichov{\'a}}}}]{Hainaut2012}
{Hainaut} O.~R., {Kleyna} J., {Sarid} G., {Hermalyn} B., {Zenn} A., {Meech}
  K.~J., {Schultz} P.~H., {Hsieh} H., {Trancho} G., and {Pittichov{\'a}} J.
  (2012) \emph{{P/2010 A2 LINEAR. I. An impact in the asteroid main belt}},
  \emph{\aap}, \emph{537}, A69.

\bibitem[{\emph{{Haisch} et~al.}(2001)\emph{{Haisch}, {Lada}, and
  {Lada}}}]{haisch01}
{Haisch} K.~E., Jr., {Lada} E.~A., and {Lada} C.~J. (2001) \emph{{Disk
  Frequencies and Lifetimes in Young Clusters}}, \emph{\apjl}, \emph{553},
  L153--L156.

\bibitem[{\emph{{Halliday}}(2013)}]{halliday13}
{Halliday} A.~N. (2013) \emph{{The origins of volatiles in the terrestrial
  planets}}, \emph{\gca}, \emph{105}, 146--171.

\bibitem[{\emph{{Hallis}}(2017)}]{Hallis2017}
{Hallis} L.~J. (2017) \emph{{D/H ratios of the inner Solar System}},
  \emph{Philosophical Transactions of the Royal Society of London Series A},
  \emph{375}, 20150390.

\bibitem[{\emph{{Hallis} et~al.}(2015)\emph{{Hallis}, {Huss}, {Nagashima},
  {Taylor}, {Halld{\'o}rsson}, {Hilton}, {Mottl}, and {Meech}}}]{Hallis2015}
{Hallis} L.~J., {Huss} G.~R., {Nagashima} K., {Taylor} G.~J., {Halld{\'o}rsson}
  S.~A., {Hilton} D.~R., {Mottl} M.~J., and {Meech} K.~J. (2015)
  \emph{{Evidence for primordial water in Earth{\textquoteright}s deep
  mantle}}, \emph{Science}, \emph{350}, 795--797.

\bibitem[{\emph{{Hamano} et~al.}(2013)\emph{{Hamano}, {Abe}, and
  {Genda}}}]{Hamano2013}
{Hamano} K., {Abe} Y., and {Genda} H. (2013) \emph{{Emergence of two types of
  terrestrial planet on solidification of magma ocean}}, \emph{\nat},
  \emph{497}, 607--610.

\bibitem[{\emph{{Hansen}}(2009)}]{hansen09}
{Hansen} B.~M.~S. (2009) \emph{{Formation of the Terrestrial Planets from a
  Narrow Annulus}}, \emph{\apj}, \emph{703}, 1131--1140.

\bibitem[{\emph{{Hargrove} et~al.}(2015)\emph{{Hargrove}, {Emery}, {Campins},
  and {Kelley}}}]{Hargrove2015}
{Hargrove} K.~D., {Emery} J.~P., {Campins} H., and {Kelley} M. S.~P. (2015)
  \emph{{Asteroid (90) Antiope: Another icy member of the Themis family?}},
  \emph{\icarus}, \emph{254}, 150--156.

\bibitem[{\emph{{Harper} and {Jacobsen}}(1996)}]{harper96}
{Harper} C.~L. and {Jacobsen} S.~B. (1996) \emph{{Evidence for $^{182}$Hf in
  the early Solar System and constraints on the timescale of terrestrial
  accretion and core formation}}, \emph{\gca}, \emph{60}, 1131--1153.

\bibitem[{\emph{{Hartmann}}(2019)}]{hartmann19}
{Hartmann} W.~K. (2019) in \emph{Lunar and Planetary Science Conference}, Lunar
  and Planetary Science Conference, p. 1064.

\bibitem[{\emph{{Hartogh} et~al.}(2011)\emph{{Hartogh}, {Lis},
  {Bockel{\'e}e-Morvan}, {de Val-Borro}, {Biver}, {K{\"u}ppers},
  {Emprechtinger}, {Bergin}, {Crovisier}, and {Rengel}}}]{Hartogh2011}
{Hartogh} P., {Lis} D.~C., {Bockel{\'e}e-Morvan} D., {de Val-Borro} M., {Biver}
  N., {K{\"u}ppers} M., {Emprechtinger} M., {Bergin} E.~A., {Crovisier} J., and
  {Rengel} M. (2011) \emph{{Ocean-like water in the Jupiter-family comet
  103P/Hartley 2}}, \emph{\nat}, \emph{478}, 218--220.

\bibitem[{\emph{{Heays} et~al.}(2014)\emph{{Heays}, {Visser}, {Gredel},
  {Ubachs}, {Lewis}, {Gibson}, and {van Dishoeck}}}]{Heays2014}
{Heays} A.~N., {Visser} R., {Gredel} R., {Ubachs} W., {Lewis} B.~R., {Gibson}
  S.~T., and {van Dishoeck} E.~F. (2014) \emph{{Isotope selective
  photodissociation of N$_{2}$ by the interstellar radiation field and cosmic
  rays}}, \emph{\aap}, \emph{562}, A61.

\bibitem[{\emph{{Herbst}}(2003)}]{Herbst2003}
{Herbst} E. (2003) \emph{{Isotopic Fractionation by Ion-Molecule Reactions}},
  \emph{\ssr}, \emph{106}, 293--304.

\bibitem[{\emph{{Herbst} and {van Dishoeck}}(2009)}]{herb2009}
{Herbst} E. and {van Dishoeck} E.~F. (2009) \emph{{Complex Organic Interstellar
  Molecules}}, \emph{\araa}, \emph{47}, 427--480.

\bibitem[{\emph{{Hester} et~al.}(2004)\emph{{Hester}, {Desch}, {Healy}, and
  {Leshin}}}]{hester04}
{Hester} J.~J., {Desch} S.~J., {Healy} K.~R., and {Leshin} L.~A. (2004)
  \emph{{The Cradle of the Solar System}}, \emph{Science}, \emph{304}.

\bibitem[{\emph{{Hillenbrand} et~al.}(2008)\emph{{Hillenbrand}, {Carpenter},
  {Kim}, {Meyer}, {Backman}, {Moro-Mart{\'{\i}}n}, {Hollenbach}, {Hines},
  {Pascucci}, and {Bouwman}}}]{hillenbrand08}
{Hillenbrand} L.~A., {Carpenter} J.~M., {Kim} J.~S., {Meyer} M.~R., {Backman}
  D.~E., {Moro-Mart{\'{\i}}n} A., {Hollenbach} D.~J., {Hines} D.~C., {Pascucci}
  I., and {Bouwman} J. (2008) \emph{{The Complete Census of 70 {$\mu$}m-bright
  Debris Disks within ``the Formation and Evolution of Planetary Systems''
  Spitzer Legacy Survey of Sun-like Stars}}, \emph{\apj}, \emph{677}, 630--656.

\bibitem[{\emph{{Hirschmann}}(2006)}]{hirschmann06}
{Hirschmann} M.~M. (2006) \emph{{Water, Melting, and the Deep Earth H2O
  Cycle}}, \emph{Annual Review of Earth and Planetary Sciences}, \emph{34},
  629--653.

\bibitem[{\emph{{Howard} et~al.}(2012)\emph{{Howard}, {Marcy}, {Bryson},
  {Jenkins}, {Rowe}, {Batalha}, {Borucki}, {Koch}, {Dunham}, {Gautier}, {Van
  Cleve}, {Cochran}, {Latham}, {Lissauer}, {Torres}, {Brown}, {Gilliland},
  {Buchhave}, {Caldwell}, {Christensen-Dalsgaard}, {Ciardi}, {Fressin}, {Haas},
  {Howell}, {Kjeldsen}, {Seager}, {Rogers}, {Sasselov}, {Steffen}, {Basri},
  {Charbonneau}, {Christiansen}, {Clarke}, {Dupree}, {Fabrycky}, {Fischer},
  {Ford}, {Fortney}, {Tarter}, {Girouard}, {Holman}, {Johnson}, {Klaus},
  {Machalek}, {Moorhead}, {Morehead}, {Ragozzine}, {Tenenbaum}, {Twicken},
  {Quinn}, {Isaacson}, {Shporer}, {Lucas}, {Walkowicz}, {Welsh}, {Boss},
  {Devore}, {Gould}, {Smith}, {Morris}, {Prsa}, {Morton}, {Still}, {Thompson},
  {Mullally}, {Endl}, and {MacQueen}}}]{howard12}
{Howard} A.~W., {Marcy} G.~W., {Bryson} S.~T., {Jenkins} J.~M., {Rowe} J.~F.,
  {Batalha} N.~M., {Borucki} W.~J., {Koch} D.~G., {Dunham} E.~W., {Gautier}
  T.~N., III, {Van Cleve} J., {Cochran} W.~D., {Latham} D.~W., {Lissauer}
  J.~J., {Torres} G., {Brown} T.~M., {Gilliland} R.~L., {Buchhave} L.~A.,
  {Caldwell} D.~A., {Christensen-Dalsgaard} J., {Ciardi} D., {Fressin} F.,
  {Haas} M.~R., {Howell} S.~B., {Kjeldsen} H., {Seager} S., {Rogers} L.,
  {Sasselov} D.~D., {Steffen} J.~H., {Basri} G.~S., {Charbonneau} D.,
  {Christiansen} J., {Clarke} B., {Dupree} A., {Fabrycky} D.~C., {Fischer}
  D.~A., {Ford} E.~B., {Fortney} J.~J., {Tarter} J., {Girouard} F.~R., {Holman}
  M.~J., {Johnson} J.~A., {Klaus} T.~C., {Machalek} P., {Moorhead} A.~V.,
  {Morehead} R.~C., {Ragozzine} D., {Tenenbaum} P., {Twicken} J.~D., {Quinn}
  S.~N., {Isaacson} H., {Shporer} A., {Lucas} P.~W., {Walkowicz} L.~M., {Welsh}
  W.~F., {Boss} A., {Devore} E., {Gould} A., {Smith} J.~C., {Morris} R.~L.,
  {Prsa} A., {Morton} T.~D., {Still} M., {Thompson} S.~E., {Mullally} F.,
  {Endl} M., and {MacQueen} P.~J. (2012) \emph{{Planet Occurrence within 0.25
  AU of Solar-type Stars from Kepler}}, \emph{\apjs}, \emph{201}, 15.

\bibitem[{\emph{{Howard} et~al.}(2013)\emph{{Howard}, {Sanchis-Ojeda}, {Marcy},
  {Johnson}, {Winn}, {Isaacson}, {Fischer}, {Fulton}, {Sinukoff}, and
  {Fortney}}}]{howard13}
{Howard} A.~W., {Sanchis-Ojeda} R., {Marcy} G.~W., {Johnson} J.~A., {Winn}
  J.~N., {Isaacson} H., {Fischer} D.~A., {Fulton} B.~J., {Sinukoff} E., and
  {Fortney} J.~J. (2013) \emph{{A rocky composition for an Earth-sized
  exoplanet}}, \emph{\nat}, \emph{503}, 381--384.

\bibitem[{\emph{{Hsieh} et~al.}(2018a)\emph{{Hsieh}, {Ishiguro}, {Kim},
  {Knight}, {Lin}, {Micheli}, {Moskovitz}, {Sheppard}, {Thirouin}, and
  {Trujillo}}}]{Hsieh2018a}
{Hsieh} H.~H., {Ishiguro} M., {Kim} Y., {Knight} M.~M., {Lin} Z.-Y., {Micheli}
  M., {Moskovitz} N.~A., {Sheppard} S.~S., {Thirouin} A., and {Trujillo} C.~A.
  (2018a) \emph{{The 2016 Reactivations of the Main-belt Comets 238P/Read and
  288P/(300163) 2006 VW$_{139}$}}, \emph{\aj}, \emph{156}, 223.

\bibitem[{\emph{{Hsieh} and {Jewitt}}(2006)}]{Hsieh2006}
{Hsieh} H.~H. and {Jewitt} D. (2006) \emph{{A Population of Comets in the Main
  Asteroid Belt}}, \emph{Science}, \emph{312}, 561--563.

\bibitem[{\emph{{Hsieh} et~al.}(2009)\emph{{Hsieh}, {Jewitt}, and
  {Ishiguro}}}]{Hsieh2009}
{Hsieh} H.~H., {Jewitt} D., and {Ishiguro} M. (2009) \emph{{Physical Properties
  of Main-Belt Comet P/2005 U1 (Read)}}, \emph{\aj}, \emph{137}, 157--168.

\bibitem[{\emph{{Hsieh} et~al.}(2018b)\emph{{Hsieh}, {Novakovi{\'c}}, {Kim},
  and {Brasser}}}]{Hsieh2018b}
{Hsieh} H.~H., {Novakovi{\'c}} B., {Kim} Y., and {Brasser} R. (2018b)
  \emph{{Asteroid Family Associations of Active Asteroids}}, \emph{\aj},
  \emph{155}, 96.

\bibitem[{\emph{{Ida} et~al.}(2016)\emph{{Ida}, {Guillot}, and
  {Morbidelli}}}]{ida16}
{Ida} S., {Guillot} T., and {Morbidelli} A. (2016) \emph{{The radial dependence
  of pebble accretion rates: A source of diversity in planetary systems. I.
  Analytical formulation}}, \emph{\aap}, \emph{591}, A72.

\bibitem[{\emph{{Ida} and {Lin}}(2010)}]{ida10}
{Ida} S. and {Lin} D.~N.~C. (2010) \emph{{Toward a Deterministic Model of
  Planetary Formation. VI. Dynamical Interaction and Coagulation of Multiple
  Rocky Embryos and Super-Earth Systems around Solar-type Stars}}, \emph{\apj},
  \emph{719}, 810--830.

\bibitem[{\emph{{Ida} et~al.}(2013)\emph{{Ida}, {Lin}, and {Nagasawa}}}]{ida13}
{Ida} S., {Lin} D.~N.~C., and {Nagasawa} M. (2013) \emph{{Toward a
  Deterministic Model of Planetary Formation. VII. Eccentricity Distribution of
  Gas Giants}}, \emph{\apj}, \emph{775}, 42.

\bibitem[{\emph{{Ida} et~al.}(2019)\emph{{Ida}, {Yamamura}, and
  {Okuzumi}}}]{ida19}
{Ida} S., {Yamamura} T., and {Okuzumi} S. (2019) \emph{{Water delivery by
  pebble accretion to rocky planets in habitable zones in evolving disks}},
  \emph{\aap}, \emph{624}, A28.

\bibitem[{\emph{{Ikoma} and {Genda}}(2006)}]{Ikoma2006}
{Ikoma} M. and {Genda} H. (2006) \emph{{Constraints on the Mass of a Habitable
  Planet with Water of Nebular Origin}}, \emph{\apj}, \emph{648}, 696--706.

\bibitem[{\emph{{Izidoro} et~al.}(2019)\emph{{Izidoro}, {Bitsch}, {Raymond},
  {Johansen}, {Morbidelli}, {Lambrechts}, and {Jacobson}}}]{izidoro19}
{Izidoro} A., {Bitsch} B., {Raymond} S.~N., {Johansen} A., {Morbidelli} A.,
  {Lambrechts} M., and {Jacobson} S.~A. (2019) \emph{{Formation of planetary
  systems by pebble accretion and migration: Hot super-Earth systems from
  breaking compact resonant chains}}, \emph{arXiv e-prints}, arXiv:1902.08772.

\bibitem[{\emph{{Izidoro} et~al.}(2014)\emph{{Izidoro}, {Morbidelli}, and
  {Raymond}}}]{izidoro14}
{Izidoro} A., {Morbidelli} A., and {Raymond} S.~N. (2014) \emph{{Terrestrial
  Planet Formation in the Presence of Migrating Super-Earths}}, \emph{\apj},
  \emph{794}, 11.

\bibitem[{\emph{{Izidoro} et~al.}(2015{\natexlab{a}})\emph{{Izidoro},
  {Morbidelli}, {Raymond}, {Hersant}, and {Pierens}}}]{izidoro15b}
{Izidoro} A., {Morbidelli} A., {Raymond} S.~N., {Hersant} F., and {Pierens} A.
  (2015{\natexlab{a}}) \emph{{Accretion of Uranus and Neptune from
  inward-migrating planetary embryos blocked by Jupiter and Saturn}},
  \emph{\aap}, \emph{582}, A99.

\bibitem[{\emph{{Izidoro} et~al.}(2017)\emph{{Izidoro}, {Ogihara}, {Raymond},
  {Morbidelli}, {Pierens}, {Bitsch}, {Cossou}, and {Hersant}}}]{izidoro17}
{Izidoro} A., {Ogihara} M., {Raymond} S.~N., {Morbidelli} A., {Pierens} A.,
  {Bitsch} B., {Cossou} C., and {Hersant} F. (2017) \emph{{Breaking the chains:
  hot super-Earth systems from migration and disruption of compact resonant
  chains}}, \emph{\mnras}, \emph{470}, 1750--1770.

\bibitem[{\emph{{Izidoro} et~al.}(2015{\natexlab{b}})\emph{{Izidoro},
  {Raymond}, {Morbidelli}, and {Winter}}}]{izidoro15c}
{Izidoro} A., {Raymond} S.~N., {Morbidelli} A., and {Winter} O.~C.
  (2015{\natexlab{b}}) \emph{{Terrestrial planet formation constrained by Mars
  and the structure of the asteroid belt}}, \emph{\mnras}, \emph{453},
  3619--3634.

\bibitem[{\emph{{Jackson} et~al.}(2010)\emph{{Jackson}, {Carlson}, {Kurz},
  {Kempton}, {Francis}, and {Blusztajn}}}]{Jackson2010}
{Jackson} M.~G., {Carlson} R.~W., {Kurz} M.~D., {Kempton} P.~D., {Francis} D.,
  and {Blusztajn} J. (2010) \emph{{Evidence for the survival of the oldest
  terrestrial mantle reservoir}}, \emph{\nat}, \emph{466}, 853--856.

\bibitem[{\emph{{Jacobson} and {Morbidelli}}(2014)}]{jacobson14b}
{Jacobson} S.~A. and {Morbidelli} A. (2014) \emph{{Lunar and terrestrial planet
  formation in the Grand Tack scenario}}, \emph{Philosophical Transactions of
  the Royal Society of London Series A}, \emph{372}, 0174.

\bibitem[{\emph{{Jacobson} et~al.}(2014)\emph{{Jacobson}, {Morbidelli},
  {Raymond}, {O'Brien}, {Walsh}, and {Rubie}}}]{jacobson14}
{Jacobson} S.~A., {Morbidelli} A., {Raymond} S.~N., {O'Brien} D.~P., {Walsh}
  K.~J., and {Rubie} D.~C. (2014) \emph{{Highly siderophile elements in Earth's
  mantle as a clock for the Moon-forming impact}}, \emph{\nat}, \emph{508},
  84--87.

\bibitem[{\emph{{Jacobson} and {Scheeres}}(2011)}]{Jacobson2011}
{Jacobson} S.~A. and {Scheeres} D.~J. (2011) \emph{{Dynamics of rotationally
  fissioned asteroids: Source of observed small asteroid systems}},
  \emph{\icarus}, \emph{214}, 161--178.

\bibitem[{\emph{{Jacquet} and {Robert}}(2013)}]{Jacquet2013}
{Jacquet} E. and {Robert} F. (2013) \emph{{Water transport in protoplanetary
  disks and the hydrogen isotopic composition of chondrites}}, \emph{\icarus},
  \emph{223}, 722--732.

\bibitem[{\emph{{Jakosky} et~al.}(2017)\emph{{Jakosky}, {Slipski}, {Benna},
  {Mahaffy}, {Elrod}, {Yelle}, {Stone}, and {Alsaeed}}}]{Jakosky2017}
{Jakosky} B.~M., {Slipski} M., {Benna} M., {Mahaffy} P., {Elrod} M., {Yelle}
  R., {Stone} S., and {Alsaeed} N. (2017) \emph{{Mars{\textquoteright}
  atmospheric history derived from upper-atmosphere measurements of
  $^{38}$Ar/$^{36}$Ar}}, \emph{Science}, \emph{355}, 1408--1410.

\bibitem[{\emph{{Jewitt}}(2012)}]{Jewitt2012}
{Jewitt} D. (2012) \emph{{The Active Asteroids}}, \emph{\aj}, \emph{143}, 66.

\bibitem[{\emph{{Jin} and {Mordasini}}(2018)}]{jin18}
{Jin} S. and {Mordasini} C. (2018) \emph{{Compositional Imprints in
  Density-Distance-Time: A Rocky Composition for Close-in Low-mass Exoplanets
  from the Location of the Valley of Evaporation}}, \emph{\apj}, \emph{853},
  163.

\bibitem[{\emph{{Jing} et~al.}(2011)\emph{{Jing}, {He}, {Brucato}, {De Sio},
  {Tozzetti}, and {Vidali}}}]{Jing2011}
{Jing} D., {He} J., {Brucato} J., {De Sio} A., {Tozzetti} L., and {Vidali} G.
  (2011) \emph{{On Water Formation in the Interstellar Medium: Laboratory Study
  of the O+D Reaction on Surfaces}}, \emph{\apjl}, \emph{741}, L9.

\bibitem[{\emph{{Johansen} et~al.}(2014)\emph{{Johansen}, {Blum}, {Tanaka},
  {Ormel}, {Bizzarro}, and {Rickman}}}]{johansen14}
{Johansen} A., {Blum} J., {Tanaka} H., {Ormel} C., {Bizzarro} M., and {Rickman}
  H. (2014) \emph{{The Multifaceted Planetesimal Formation Process}},
  \emph{Protostars and Planets VI}, pp. 547--570.

\bibitem[{\emph{{Johansen} and {Lambrechts}}(2017)}]{johansen17}
{Johansen} A. and {Lambrechts} M. (2017) \emph{{Forming Planets via Pebble
  Accretion}}, \emph{Annual Review of Earth and Planetary Sciences}, \emph{45},
  359--387.

\bibitem[{\emph{{Johansen} et~al.}(2009)\emph{{Johansen}, {Youdin}, and
  {Klahr}}}]{johansen09}
{Johansen} A., {Youdin} A., and {Klahr} H. (2009) \emph{{Zonal Flows and
  Long-lived Axisymmetric Pressure Bumps in Magnetorotational Turbulence}},
  \emph{\apj}, \emph{697}, 1269--1289.

\bibitem[{\emph{{Johnson} et~al.}(2007)\emph{{Johnson}, {Butler}, {Marcy},
  {Fischer}, {Vogt}, {Wright}, and {Peek}}}]{johnson07}
{Johnson} J.~A., {Butler} R.~P., {Marcy} G.~W., {Fischer} D.~A., {Vogt} S.~S.,
  {Wright} J.~T., and {Peek} K.~M.~G. (2007) \emph{{A New Planet around an M
  Dwarf: Revealing a Correlation between Exoplanets and Stellar Mass}},
  \emph{\apj}, \emph{670}, 833--840.

\bibitem[{\emph{{Juri{\'c}} and {Tremaine}}(2008)}]{juric08}
{Juri{\'c}} M. and {Tremaine} S. (2008) \emph{{Dynamical Origin of Extrasolar
  Planet Eccentricity Distribution}}, \emph{\apj}, \emph{686}, 603--620.

\bibitem[{\emph{{Kaib} and {Cowan}}(2015)}]{kaib15}
{Kaib} N.~A. and {Cowan} N.~B. (2015) \emph{{The feeding zones of terrestrial
  planets and insights into Moon formation}}, \emph{\icarus}, \emph{252},
  161--174.

\bibitem[{\emph{{Kalyaan} and {Desch}}(2019)}]{kalyaan19}
{Kalyaan} A. and {Desch} S.~J. (2019) \emph{{Effect of Different Angular
  Momentum Transport Mechanisms on the Distribution of Water in Protoplanetary
  Disks}}, \emph{\apj}, \emph{875}, 43.

\bibitem[{\emph{{Kasting} and {Pollack}}(1983)}]{Kasting1983}
{Kasting} J.~F. and {Pollack} J.~B. (1983) \emph{{Loss of water from Venus. I.
  Hydrodynamic escape of hydrogen}}, \emph{\icarus}, \emph{53}, 479--508.

\bibitem[{\emph{{Kasting} et~al.}(1993)\emph{{Kasting}, {Whitmire}, and
  {Reynolds}}}]{kasting93}
{Kasting} J.~F., {Whitmire} D.~P., and {Reynolds} R.~T. (1993) \emph{{Habitable
  Zones around Main Sequence Stars}}, \emph{Icarus}, \emph{101}, 108--128.

\bibitem[{\emph{{Kennedy} and {Kenyon}}(2008)}]{kennedy08}
{Kennedy} G.~M. and {Kenyon} S.~J. (2008) \emph{{Planet Formation around Stars
  of Various Masses: The Snow Line and the Frequency of Giant Planets}},
  \emph{\apj}, \emph{673}, 502--512.

\bibitem[{\emph{{Kimura} et~al.}(1974)\emph{{Kimura}, {Lewis}, and
  {Anders}}}]{kimura74}
{Kimura} K., {Lewis} R.~S., and {Anders} E. (1974) \emph{{Distribution of gold
  and rhenium between nickel-iron and silicate melts: implications for the
  abundance of siderophile elements on the Earth and Moon}}, \emph{\gca},
  \emph{38}, 683--701.

\bibitem[{\emph{{King} et~al.}(2010)\emph{{King}, {Stimpfl}, {Deymier},
  {Drake}, {Catlow}, {Putnis}, and {de Leeuw}}}]{King2010}
{King} H.~E., {Stimpfl} M., {Deymier} P., {Drake} M.~J., {Catlow} C.~R.~A.,
  {Putnis} A., and {de Leeuw} N.~H. (2010) \emph{{Computer simulations of water
  interactions with low-coordinated forsterite surface sites: Implications for
  the origin of water in the inner solar system}}, \emph{Earth and Planetary
  Science Letters}, \emph{300}, 11--18.

\bibitem[{\emph{{Kleine} et~al.}(2009)\emph{{Kleine}, {Touboul}, {Bourdon},
  {Nimmo}, {Mezger}, {Palme}, {Jacobsen}, {Yin}, and {Halliday}}}]{kleine09}
{Kleine} T., {Touboul} M., {Bourdon} B., {Nimmo} F., {Mezger} K., {Palme} H.,
  {Jacobsen} S.~B., {Yin} Q.-Z., and {Halliday} A.~N. (2009) \emph{{Hf-W
  chronology of the accretion and early evolution of asteroids and terrestrial
  planets}}, \emph{\gca}, \emph{73}, 5150--5188.

\bibitem[{\emph{{Kley} and {Crida}}(2008)}]{kley08}
{Kley} W. and {Crida} A. (2008) \emph{{Migration of protoplanets in radiative
  discs}}, \emph{\aap}, \emph{487}, L9--L12.

\bibitem[{\emph{{Kley} and {Nelson}}(2012)}]{kley12}
{Kley} W. and {Nelson} R.~P. (2012) \emph{{Planet-Disk Interaction and Orbital
  Evolution}}, \emph{\araa}, \emph{50}, 211--249.

\bibitem[{\emph{{Kleyna} et~al.}(2013)\emph{{Kleyna}, {Hainaut}, and
  {Meech}}}]{kleyna2013}
{Kleyna} J., {Hainaut} O.~R., and {Meech} K.~J. (2013) \emph{{P/2010 A2 LINEAR.
  II. Dynamical dust modelling}}, \emph{\aap}, \emph{549}, A13.

\bibitem[{\emph{{Kokubo} and {Ida}}(2000)}]{kokubo00}
{Kokubo} E. and {Ida} S. (2000) \emph{{Formation of Protoplanets from
  Planetesimals in the Solar Nebula}}, \emph{Icarus}, \emph{143}, 15--27.

\bibitem[{\emph{{Kresak}}(1980)}]{Kresak1980}
{Kresak} L. (1980) \emph{{Dynamics, interrelations and evolution of the systems
  of asteroids and comets}}, \emph{Moon and Planets}, \emph{22}, 83--98.

\bibitem[{\emph{{Krot} and {Bizzarro}}(2009)}]{Krot2009}
{Krot} A.~N. and {Bizzarro} M. (2009) \emph{{Chronology of meteorites and the
  early solar system}}, \emph{\gca}, \emph{73}, 4919--4921.

\bibitem[{\emph{{Krot} et~al.}(2010)\emph{{Krot}, {Nagashima}, {Ciesla},
  {Meyer}, {Hutcheon}, {Davis}, {Huss}, and {Scott}}}]{Krot2010}
{Krot} A.~N., {Nagashima} K., {Ciesla} F.~J., {Meyer} B.~S., {Hutcheon} I.~D.,
  {Davis} A.~M., {Huss} G.~R., and {Scott} E. R.~D. (2010) \emph{{Oxygen
  Isotopic Composition of the Sun and Mean Oxygen Isotopic Composition of the
  Protosolar Silicate Dust: Evidence from Refractory Inclusions}}, \emph{\apj},
  \emph{713}, 1159--1166.

\bibitem[{\emph{{Kruijer} et~al.}(2017)\emph{{Kruijer}, {Burkhardt}, {Budde},
  and {Kleine}}}]{kruijer17}
{Kruijer} T.~S., {Burkhardt} C., {Budde} C., and {Kleine} T. (2017) \emph{{Age
  of Jupiter inferred from the distinct genetics and formation times of
  meteorites}}, \emph{PNAS}.

\bibitem[{\emph{{Kruijer} et~al.}(2014)\emph{{Kruijer}, {Touboul},
  {Fischer-G{\"o}dde}, {Bermingham}, {Walker}, and {Kleine}}}]{kruijer14}
{Kruijer} T.~S., {Touboul} M., {Fischer-G{\"o}dde} M., {Bermingham} K.~R.,
  {Walker} R.~J., and {Kleine} T. (2014) \emph{{Protracted core formation and
  rapid accretion of protoplanets}}, \emph{Science}, \emph{344}, 1150--1154.

\bibitem[{\emph{{Kuchner}}(2003)}]{kuchner03}
{Kuchner} M.~J. (2003) \emph{{Volatile-rich Earth-Mass Planets in the Habitable
  Zone}}, \emph{\apjl}, \emph{596}, L105--L108.

\bibitem[{\emph{{K{\"u}ppers} et~al.}(2014)\emph{{K{\"u}ppers}, {O'Rourke},
  {Bockel{\'e}e-Morvan}, {Zakharov}, {Lee}, {von Allmen}, {Carry}, {Teyssier},
  {Marston}, and {M{\"u}ller}}}]{Kuppers2014}
{K{\"u}ppers} M., {O'Rourke} L., {Bockel{\'e}e-Morvan} D., {Zakharov} V., {Lee}
  S., {von Allmen} P., {Carry} B., {Teyssier} D., {Marston} A., and
  {M{\"u}ller} T. (2014) in \emph{Asteroids, Comets, Meteors 2014}
  (K.~{Muinonen}, A.~{Penttil{\"a}}, M.~{Granvik}, A.~{Virkki}, G.~{Fedorets},
  O.~{Wilkman}, and T.~{Kohout}, eds.), p. 298.

\bibitem[{\emph{{Kurosawa}}(2015)}]{Kurosawa2015}
{Kurosawa} K. (2015) \emph{{Impact-driven planetary desiccation: The origin of
  the dry Venus}}, \emph{Earth and Planetary Science Letters}, \emph{429},
  181--190.

\bibitem[{\emph{{Lambrechts} and {Johansen}}(2012)}]{lambrechts12}
{Lambrechts} M. and {Johansen} A. (2012) \emph{{Rapid growth of gas-giant cores
  by pebble accretion}}, \emph{\aap}, \emph{544}, A32.

\bibitem[{\emph{{Lambrechts} and {Johansen}}(2014)}]{lambrechts14}
{Lambrechts} M. and {Johansen} A. (2014) \emph{{Forming the cores of giant
  planets from the radial pebble flux in protoplanetary discs}}, \emph{\aap},
  \emph{572}, A107.

\bibitem[{\emph{{Lambrechts} et~al.}(2014)\emph{{Lambrechts}, {Johansen}, and
  {Morbidelli}}}]{lambrechts14b}
{Lambrechts} M., {Johansen} A., and {Morbidelli} A. (2014) \emph{{Separating
  gas-giant and ice-giant planets by halting pebble accretion}}, \emph{\aap},
  \emph{572}, A35.

\bibitem[{\emph{{Laufer} et~al.}(2017)\emph{{Laufer}, {Bar-Nun}, and {Ninio
  Greenberg}}}]{Laufer2017}
{Laufer} D., {Bar-Nun} A., and {Ninio Greenberg} A. (2017) \emph{{Trapping
  mechanism of O$_{2}$ in water ice as first measured by Rosetta spacecraft}},
  \emph{\mnras}, \emph{469}, S818--S823.

\bibitem[{\emph{{Lecar} et~al.}(2006)\emph{{Lecar}, {Podolak}, {Sasselov}, and
  {Chiang}}}]{lecar06}
{Lecar} M., {Podolak} M., {Sasselov} D., and {Chiang} E. (2006) \emph{{On the
  Location of the Snow Line in a Protoplanetary Disk}}, \emph{\apj},
  \emph{640}, 1115--1118.

\bibitem[{\emph{{L{\'e}cuyer} et~al.}(1998)\emph{{L{\'e}cuyer}, {Gillet}, and
  {Robert}}}]{Lecuyer1998}
{L{\'e}cuyer} C., {Gillet} P., and {Robert} F. (1998) \emph{{The hydrogen
  isotope composition of seawater and the global water cycle}}, \emph{Chemical
  Geology}, \emph{145}, 249--261.

\bibitem[{\emph{{Lellouch} et~al.}(2001)\emph{{Lellouch}, {B{\'e}zard},
  {Fouchet}, {Feuchtgruber}, {Encrenaz}, and {de Graauw}}}]{Lellouch2001}
{Lellouch} E., {B{\'e}zard} B., {Fouchet} T., {Feuchtgruber} H., {Encrenaz} T.,
  and {de Graauw} T. (2001) \emph{{The deuterium abundance in Jupiter and
  Saturn from ISO-SWS observations}}, \emph{\aap}, \emph{370}, 610--622.

\bibitem[{\emph{{Levison} and {Duncan}}(1997)}]{levison97}
{Levison} H.~F. and {Duncan} M.~J. (1997) \emph{{From the Kuiper Belt to
  Jupiter-Family Comets: The Spatial Distribution of Ecliptic Comets}},
  \emph{Icarus}, \emph{127}, 13--32.

\bibitem[{\emph{{Levison} et~al.}(2011)\emph{{Levison}, {Morbidelli},
  {Tsiganis}, {Nesvorn{\'y}}, and {Gomes}}}]{levison11}
{Levison} H.~F., {Morbidelli} A., {Tsiganis} K., {Nesvorn{\'y}} D., and {Gomes}
  R. (2011) \emph{{Late Orbital Instabilities in the Outer Planets Induced by
  Interaction with a Self-gravitating Planetesimal Disk}}, \emph{\aj},
  \emph{142}, 152.

\bibitem[{\emph{{Levison} et~al.}(2008)\emph{{Levison}, {Morbidelli},
  {Vanlaerhoven}, {Gomes}, and {Tsiganis}}}]{levison08}
{Levison} H.~F., {Morbidelli} A., {Vanlaerhoven} C., {Gomes} R., and {Tsiganis}
  K. (2008) \emph{{Origin of the structure of the Kuiper belt during a
  dynamical instability in the orbits of Uranus and Neptune}}, \emph{Icarus},
  \emph{196}, 258--273.

\bibitem[{\emph{{Levison} et~al.}(2006)\emph{{Levison}, {Terrell}, {Wiegert},
  {Dones}, and {Duncan}}}]{Levison2006}
{Levison} H.~F., {Terrell} D., {Wiegert} P.~A., {Dones} L., and {Duncan} M.~J.
  (2006) \emph{{On the origin of the unusual orbit of Comet 2P/Encke}},
  \emph{\icarus}, \emph{182}, 161--168.

\bibitem[{\emph{{Lichtenberg} et~al.}(2019)\emph{{Lichtenberg}, {Golabek},
  {Burn}, {Meyer}, {Alibert}, {Gerya}, and {Mordasini}}}]{lichtenberg19}
{Lichtenberg} T., {Golabek} G.~J., {Burn} R., {Meyer} M.~R., {Alibert} Y.,
  {Gerya} T.~V., and {Mordasini} C. (2019) \emph{{A water budget dichotomy of
  rocky protoplanets from $^{26}$Al-heating}}, \emph{Nature Astronomy},
  \emph{3}, 307--313.

\bibitem[{\emph{{Lichtenberg} et~al.}(2016)\emph{{Lichtenberg}, {Parker}, and
  {Meyer}}}]{lichtenberg16}
{Lichtenberg} T., {Parker} R.~J., and {Meyer} M.~R. (2016) \emph{{Isotopic
  enrichment of forming planetary systems from supernova pollution}},
  \emph{\mnras}, \emph{462}, 3979--3992.

\bibitem[{\emph{{Lin} and {Papaloizou}}(1986)}]{lin86}
{Lin} D.~N.~C. and {Papaloizou} J. (1986) \emph{{On the tidal interaction
  between protoplanets and the protoplanetary disk. III - Orbital migration of
  protoplanets}}, \emph{\apj}, \emph{309}, 846--857.

\bibitem[{\emph{{Linsky}}(2007)}]{Linsky2007}
{Linsky} J.~L. (2007) \emph{{D/H and Nearby Interstellar Cloud Structures}},
  \emph{\ssr}, \emph{130}, 367--375.

\bibitem[{\emph{{Lis} et~al.}(2019)\emph{{Lis}, {Bockel{\'e}e-Morvan},
  {G{\"u}sten}, {Biver}, {Stutzki}, {Delorme}, {Dur{\'a}n}, {Wiesemeyer}, and
  {Okada}}}]{Lis2019}
{Lis} D.~C., {Bockel{\'e}e-Morvan} D., {G{\"u}sten} R., {Biver} N., {Stutzki}
  J., {Delorme} Y., {Dur{\'a}n} C., {Wiesemeyer} H., and {Okada} Y. (2019)
  \emph{{Terrestrial deuterium-to-hydrogen ratio in water in hyperactive
  comets}}, \emph{\aap}, \emph{625}, L5.

\bibitem[{\emph{{Lissauer}}(1987)}]{lissauer87}
{Lissauer} J.~J. (1987) \emph{{Timescales for planetary accretion and the
  structure of the protoplanetary disk}}, \emph{Icarus}, \emph{69}, 249--265.

\bibitem[{\emph{{Lissauer} et~al.}(2009)\emph{{Lissauer}, {Hubickyj},
  {D'Angelo}, and {Bodenheimer}}}]{lissauer09}
{Lissauer} J.~J., {Hubickyj} O., {D'Angelo} G., and {Bodenheimer} P. (2009)
  \emph{{Models of Jupiter's growth incorporating thermal and hydrodynamic
  constraints}}, \emph{Icarus}, \emph{199}, 338--350.

\bibitem[{\emph{{Lissauer} and {Stevenson}}(2007)}]{lissauer07}
{Lissauer} J.~J. and {Stevenson} D.~J. (2007) \emph{{Formation of Giant
  Planets}}, \emph{Protostars and Planets V}, pp. 591--606.

\bibitem[{\emph{{Lopez}}(2017)}]{lopez17}
{Lopez} E.~D. (2017) \emph{{Born dry in the photoevaporation desert: Kepler's
  ultra-short-period planets formed water-poor}}, \emph{\mnras}, \emph{472},
  245--253.

\bibitem[{\emph{{Lugaro} et~al.}(2012)\emph{{Lugaro}, {Liffman}, {Ireland}, and
  {Maddison}}}]{Lugaro2012}
{Lugaro} M., {Liffman} K., {Ireland} T.~R., and {Maddison} S.~T. (2012)
  \emph{{Can Galactic Chemical Evolution Explain the Oxygen Isotopic Variations
  in the Solar System?}}, \emph{\apj}, \emph{759}, 51.

\bibitem[{\emph{{Lyons}}(2012)}]{Lyons2012}
{Lyons} J.~R. (2012) in \emph{Lunar and Planetary Science Conference}, p. 2858.

\bibitem[{\emph{{Lyons} and {Young}}(2005)}]{Lyons2005}
{Lyons} J.~R. and {Young} E.~D. (2005) \emph{{CO self-shielding as the origin
  of oxygen isotope anomalies in the early solar nebula}}, \emph{\nat},
  \emph{435}, 317--320.

\bibitem[{\emph{{Lyra} et~al.}(2010)\emph{{Lyra}, {Paardekooper}, and {Mac
  Low}}}]{lyra10}
{Lyra} W., {Paardekooper} S.-J., and {Mac Low} M.-M. (2010) \emph{{Orbital
  Migration of Low-mass Planets in Evolutionary Radiative Models: Avoiding
  Catastrophic Infall}}, \emph{\apjl}, \emph{715}, L68--L73.

\bibitem[{\emph{{Marcus} et~al.}(2010)\emph{{Marcus}, {Sasselov}, {Stewart},
  and {Hernquist}}}]{marcus10}
{Marcus} R.~A., {Sasselov} D., {Stewart} S.~T., and {Hernquist} L. (2010)
  \emph{{Water/Icy Super-Earths: Giant Impacts and Maximum Water Content}},
  \emph{\apjl}, \emph{719}, L45--L49.

\bibitem[{\emph{{Marcy} et~al.}(2014)\emph{{Marcy}, {Isaacson}, {Howard},
  {Rowe}, {Jenkins}, {Bryson}, {Latham}, {Howell}, {Gautier}, {Batalha},
  {Rogers}, {Ciardi}, {Fischer}, {Gilliland}, {Kjeldsen},
  {Christensen-Dalsgaard}, {Huber}, {Chaplin}, {Basu}, {Buchhave}, {Quinn},
  {Borucki}, {Koch}, {Hunter}, {Caldwell}, {Van Cleve}, {Kolbl}, {Weiss},
  {Petigura}, {Seager}, {Morton}, {Johnson}, {Ballard}, {Burke}, {Cochran},
  {Endl}, {MacQueen}, {Everett}, {Lissauer}, {Ford}, {Torres}, {Fressin},
  {Brown}, {Steffen}, {Charbonneau}, {Basri}, {Sasselov}, {Winn},
  {Sanchis-Ojeda}, {Christiansen}, {Adams}, {Henze}, {Dupree}, {Fabrycky},
  {Fortney}, {Tarter}, {Holman}, {Tenenbaum}, {Shporer}, {Lucas}, {Welsh},
  {Orosz}, {Bedding}, {Campante}, {Davies}, {Elsworth}, {Handberg}, {Hekker},
  {Karoff}, {Kawaler}, {Lund}, {Lundkvist}, {Metcalfe}, {Miglio}, {Silva
  Aguirre}, {Stello}, {White}, {Boss}, {Devore}, {Gould}, {Prsa}, {Agol},
  {Barclay}, {Coughlin}, {Brugamyer}, {Mullally}, {Quintana}, {Still},
  {Thompson}, {Morrison}, {Twicken}, {D{\'e}sert}, {Carter}, {Crepp},
  {H{\'e}brard}, {Santerne}, {Moutou}, {Sobeck}, {Hudgins}, {Haas},
  {Robertson}, {Lillo-Box}, and {Barrado}}}]{marcy14}
{Marcy} G.~W., {Isaacson} H., {Howard} A.~W., {Rowe} J.~F., {Jenkins} J.~M.,
  {Bryson} S.~T., {Latham} D.~W., {Howell} S.~B., {Gautier} T.~N., III,
  {Batalha} N.~M., {Rogers} L., {Ciardi} D., {Fischer} D.~A., {Gilliland}
  R.~L., {Kjeldsen} H., {Christensen-Dalsgaard} J., {Huber} D., {Chaplin}
  W.~J., {Basu} S., {Buchhave} L.~A., {Quinn} S.~N., {Borucki} W.~J., {Koch}
  D.~G., {Hunter} R., {Caldwell} D.~A., {Van Cleve} J., {Kolbl} R., {Weiss}
  L.~M., {Petigura} E., {Seager} S., {Morton} T., {Johnson} J.~A., {Ballard}
  S., {Burke} C., {Cochran} W.~D., {Endl} M., {MacQueen} P., {Everett} M.~E.,
  {Lissauer} J.~J., {Ford} E.~B., {Torres} G., {Fressin} F., {Brown} T.~M.,
  {Steffen} J.~H., {Charbonneau} D., {Basri} G.~S., {Sasselov} D.~D., {Winn}
  J., {Sanchis-Ojeda} R., {Christiansen} J., {Adams} E., {Henze} C., {Dupree}
  A., {Fabrycky} D.~C., {Fortney} J.~J., {Tarter} J., {Holman} M.~J.,
  {Tenenbaum} P., {Shporer} A., {Lucas} P.~W., {Welsh} W.~F., {Orosz} J.~A.,
  {Bedding} T.~R., {Campante} T.~L., {Davies} G.~R., {Elsworth} Y., {Handberg}
  R., {Hekker} S., {Karoff} C., {Kawaler} S.~D., {Lund} M.~N., {Lundkvist} M.,
  {Metcalfe} T.~S., {Miglio} A., {Silva Aguirre} V., {Stello} D., {White}
  T.~R., {Boss} A., {Devore} E., {Gould} A., {Prsa} A., {Agol} E., {Barclay}
  T., {Coughlin} J., {Brugamyer} E., {Mullally} F., {Quintana} E.~V., {Still}
  M., {Thompson} S.~E., {Morrison} D., {Twicken} J.~D., {D{\'e}sert} J.-M.,
  {Carter} J., {Crepp} J.~R., {H{\'e}brard} G., {Santerne} A., {Moutou} C.,
  {Sobeck} C., {Hudgins} D., {Haas} M.~R., {Robertson} P., {Lillo-Box} J., and
  {Barrado} D. (2014) \emph{{Masses, Radii, and Orbits of Small Kepler Planets:
  The Transition from Gaseous to Rocky Planets}}, \emph{\apjs}, \emph{210}, 20.

\bibitem[{\emph{{Marsset} et~al.}(2016)\emph{{Marsset}, {Vernazza}, {Birlan},
  {DeMeo}, {Binzel}, {Dumas}, {Milli}, and {Popescu}}}]{Marsset2016}
{Marsset} M., {Vernazza} P., {Birlan} M., {DeMeo} F., {Binzel} R.~P., {Dumas}
  C., {Milli} J., and {Popescu} M. (2016) \emph{{Compositional characterisation
  of the Themis family}}, \emph{\aap}, \emph{586}, A15.

\bibitem[{\emph{{Martin} and {Livio}}(2012)}]{martin12}
{Martin} R.~G. and {Livio} M. (2012) \emph{{On the evolution of the snow line
  in protoplanetary discs}}, \emph{\mnras}, \emph{425}, L6--L9.

\bibitem[{\emph{{Martin} and {Livio}}(2015)}]{martin15}
{Martin} R.~G. and {Livio} M. (2015) \emph{{The Solar System as an Exoplanetary
  System}}, \emph{\apj}, \emph{810}, 105.

\bibitem[{\emph{{Marty}}(2012)}]{Marty2012}
{Marty} B. (2012) \emph{{The origins and concentrations of water, carbon,
  nitrogen and noble gases on Earth}}, \emph{Earth and Planetary Science
  Letters}, \emph{313}, 56--66.

\bibitem[{\emph{{Marty} et~al.}(2017)\emph{{Marty}, {Altwegg}, {Balsiger},
  {Bar-Nun}, {Bekaert}, {Berthelier}, {Bieler}, {Briois}, {Calmonte}, and
  {Combi}}}]{Marty2017}
{Marty} B., {Altwegg} K., {Balsiger} H., {Bar-Nun} A., {Bekaert} D.~V.,
  {Berthelier} J.~J., {Bieler} A., {Briois} C., {Calmonte} U., and {Combi} M.
  (2017) \emph{{Xenon isotopes in 67P/Churyumov-Gerasimenko show that comets
  contributed to Earth's atmosphere}}, \emph{Science}, \emph{356}, 1069--1072.

\bibitem[{\emph{{Marty} et~al.}(2016)\emph{{Marty}, {Avice}, {Sano}, {Altwegg},
  {Balsiger}, {H{\"a}ssig}, {Morbidelli}, {Mousis}, and {Rubin}}}]{Marty2016}
{Marty} B., {Avice} G., {Sano} Y., {Altwegg} K., {Balsiger} H., {H{\"a}ssig}
  M., {Morbidelli} A., {Mousis} O., and {Rubin} M. (2016) \emph{{Origins of
  volatile elements (H, C, N, noble gases) on Earth and Mars in light of recent
  results from the ROSETTA cometary mission}}, \emph{Earth and Planetary
  Science Letters}, \emph{441}, 91--102.

\bibitem[{\emph{{Marty} and {Yokochi}}(2006)}]{marty06}
{Marty} B. and {Yokochi} R. (2006) \emph{{Water in the Early Earth}},
  \emph{Rev. Mineral Geophys.}, \emph{62}, 421--450.

\bibitem[{\emph{{Marty} et~al.}(2010)\emph{{Marty}, {Zimmermann}, {Burnard},
  {Wieler}, {Heber}, {Burnett}, {Wiens}, and {Bochsler}}}]{Marty2010}
{Marty} B., {Zimmermann} L., {Burnard} P.~G., {Wieler} R., {Heber} V.~S.,
  {Burnett} D.~L., {Wiens} R.~C., and {Bochsler} P. (2010) \emph{{Nitrogen
  isotopes in the recent solar wind from the analysis of Genesis targets:
  Evidence for large scale isotope heterogeneity in the early solar system}},
  \emph{\gca}, \emph{74}, 340--355.

\bibitem[{\emph{{Masset} and {Snellgrove}}(2001)}]{masset01}
{Masset} F. and {Snellgrove} M. (2001) \emph{{Reversing type II migration:
  resonance trapping of a lighter giant protoplanet}}, \emph{\mnras},
  \emph{320}, L55--L59.

\bibitem[{\emph{{Mayer} et~al.}(2002)\emph{{Mayer}, {Quinn}, {Wadsley}, and
  {Stadel}}}]{mayer02}
{Mayer} L., {Quinn} T., {Wadsley} J., and {Stadel} J. (2002) \emph{{Formation
  of Giant Planets by Fragmentation of Protoplanetary Disks}}, \emph{Science},
  \emph{298}, 1756--1759.

\bibitem[{\emph{{Mayor} et~al.}(2011)\emph{{Mayor}, {Marmier}, {Lovis}, {Udry},
  {S{\'e}gransan}, {Pepe}, {Benz}, {Bertaux}, {Bouchy}, {Dumusque}, {Lo Curto},
  {Mordasini}, {Queloz}, and {Santos}}}]{mayor11}
{Mayor} M., {Marmier} M., {Lovis} C., {Udry} S., {S{\'e}gransan} D., {Pepe} F.,
  {Benz} W., {Bertaux} J.~., {Bouchy} F., {Dumusque} X., {Lo Curto} G.,
  {Mordasini} C., {Queloz} D., and {Santos} N.~C. (2011) \emph{{The HARPS
  search for southern extra-solar planets XXXIV. Occurrence, mass distribution
  and orbital properties of super-Earths and Neptune-mass planets}},
  \emph{arXiv:1109.2497}.

\bibitem[{\emph{{McGuire} et~al.}(2018)\emph{{McGuire}, {Bergin}, {Blake},
  {Burkhardt}, {Cleeves}, {Loomis}, {Remijan}, {Shingledecker}, and
  {Willis}}}]{McGuire2018}
{McGuire} B.~A., {Bergin} E., {Blake} G.~A., {Burkhardt} A.~M., {Cleeves}
  L.~I., {Loomis} R.~A., {Remijan} A.~J., {Shingledecker} C.~N., and {Willis}
  E.~R. (2018) in \emph{Science with a Next Generation Very Large Array}
  (E.~{Murphy}, ed.), vol. 517 of \emph{Astronomical Society of the Pacific
  Conference Series}, p. 217.

\bibitem[{\emph{{McKeegan} et~al.}(2011)\emph{{McKeegan}, {Kallio}, {Heber},
  {Jarzebinski}, {Mao}, {Coath}, {Kunihiro}, {Wiens}, {Nordholt}, and
  {Moses}}}]{McKeegan2011}
{McKeegan} K.~D., {Kallio} A.~P.~A., {Heber} V.~S., {Jarzebinski} G., {Mao}
  P.~H., {Coath} C.~D., {Kunihiro} T., {Wiens} R.~C., {Nordholt} J.~E., and
  {Moses} R.~W. (2011) \emph{{The Oxygen Isotopic Composition of the Sun
  Inferred from Captured Solar Wind}}, \emph{Science}, \emph{332}, 1528.

\bibitem[{\emph{{McNeil} and {Nelson}}(2010)}]{mcneil10}
{McNeil} D.~S. and {Nelson} R.~P. (2010) \emph{{On the formation of hot
  Neptunes and super-Earths}}, \emph{\mnras}, \emph{401}, 1691--1708.

\bibitem[{\emph{{Meech} and {Svoren}}(2004)}]{Meech2004}
{Meech} K.~J. and {Svoren} J. (2004) \emph{{Using cometary activity to trace
  the physical and chemical evolution of cometary nuclei}}, p. 317.

\bibitem[{\emph{{Meech} et~al.}(2016)\emph{{Meech}, {Yang}, {Kleyna},
  {Hainaut}, {Berdyugina}, {Keane}, {Micheli}, {Morbidelli}, and
  {Wainscoat}}}]{Meech2016}
{Meech} K.~J., {Yang} B., {Kleyna} J., {Hainaut} O.~R., {Berdyugina} S.,
  {Keane} J.~V., {Micheli} M., {Morbidelli} A.~r., and {Wainscoat} R.~J. (2016)
  \emph{{Inner solar system material discovered in the Oort cloud}},
  \emph{Science Advances}, \emph{2}, e1600038.

\bibitem[{\emph{{Meibom} et~al.}(2007)\emph{{Meibom}, {Krot}, {Robert},
  {Mostefaoui}, {Russell}, {Petaev}, and {Gounelle}}}]{Meibom2007}
{Meibom} A., {Krot} A.~N., {Robert} F., {Mostefaoui} S., {Russell} S.~S.,
  {Petaev} M.~I., and {Gounelle} M. (2007) \emph{{Nitrogen and Carbon Isotopic
  Composition of the Sun Inferred from a High-Temperature Solar Nebular
  Condensate}}, \emph{\apjl}, \emph{656}, L33--L36.

\bibitem[{\emph{{Millar} et~al.}(1989)\emph{{Millar}, {Bennett}, and
  {Herbst}}}]{Millar1989}
{Millar} T.~J., {Bennett} A., and {Herbst} E. (1989) \emph{{Deuterium
  Fractionation in Dense Interstellar Clouds}}, \emph{\apj}, \emph{340}, 906.

\bibitem[{\emph{{Minton} and {Malhotra}}(2010)}]{minton10}
{Minton} D.~A. and {Malhotra} R. (2010) \emph{{Dynamical erosion of the
  asteroid belt and implications for large impacts in the inner Solar System}},
  \emph{Icarus}, \emph{207}, 744--757.

\bibitem[{\emph{{Mojzsis} et~al.}(2001)\emph{{Mojzsis}, {Harrison}, and
  {Pidgeon}}}]{Mojzsis2001}
{Mojzsis} S.~J., {Harrison} T.~M., and {Pidgeon} R.~T. (2001)
  \emph{{Oxygen-isotope evidence from ancient zircons for liquid water at the
  Earth's surface 4,300Myr ago}}, \emph{\nat}, \emph{409}, 178--181.

\bibitem[{\emph{{Monteux} et~al.}(2016)\emph{{Monteux}, {Andrault}, and
  {Samuel}}}]{monteux16}
{Monteux} J., {Andrault} D., and {Samuel} H. (2016) \emph{{On the cooling of a
  deep terrestrial magma ocean}}, \emph{Earth and Planetary Science Letters},
  \emph{448}, 140--149.

\bibitem[{\emph{{Monteux} et~al.}(2018)\emph{{Monteux}, {Golabek}, {Rubie},
  {Tobie}, and {Young}}}]{monteux18}
{Monteux} J., {Golabek} G.~J., {Rubie} D.~C., {Tobie} G., and {Young} E.~D.
  (2018) \emph{{Water and the Interior Structure of Terrestrial Planets and Icy
  Bodies}}, \emph{\ssr}, \emph{214}, 39.

\bibitem[{\emph{{Morbidelli} et~al.}(2016)\emph{{Morbidelli}, {Bitsch},
  {Crida}, {Gounelle}, {Guillot}, {Jacobson}, {Johansen}, {Lambrechts}, and
  {Lega}}}]{morby16}
{Morbidelli} A., {Bitsch} B., {Crida} A., {Gounelle} M., {Guillot} T.,
  {Jacobson} S., {Johansen} A., {Lambrechts} M., and {Lega} E. (2016)
  \emph{{Fossilized condensation lines in the Solar System protoplanetary
  disk}}, \emph{Icarus}, \emph{267}, 368--376.

\bibitem[{\emph{{Morbidelli} et~al.}(2000)\emph{{Morbidelli}, {Chambers},
  {Lunine}, {Petit}, {Robert}, {Valsecchi}, and {Cyr}}}]{morby00}
{Morbidelli} A., {Chambers} J., {Lunine} J.~I., {Petit} J.~M., {Robert} F.,
  {Valsecchi} G.~B., and {Cyr} K.~E. (2000) \emph{{Source regions and time
  scales for the delivery of water to Earth}}, \emph{Meteoritics and Planetary
  Science}, \emph{35}, 1309--1320.

\bibitem[{\emph{{Morbidelli} et~al.}(2015)\emph{{Morbidelli}, {Lambrechts},
  {Jacobson}, and {Bitsch}}}]{morby15}
{Morbidelli} A., {Lambrechts} M., {Jacobson} S., and {Bitsch} B. (2015)
  \emph{{The great dichotomy of the Solar System: Small terrestrial embryos and
  massive giant planet cores}}, \emph{Icarus}, \emph{258}, 418--429.

\bibitem[{\emph{{Morbidelli} and {Nesvorny}}(2012)}]{morby12}
{Morbidelli} A. and {Nesvorny} D. (2012) \emph{{Dynamics of pebbles in the
  vicinity of a growing planetary embryo: hydro-dynamical simulations}},
  \emph{\aap}, \emph{546}, A18.

\bibitem[{\emph{{Morbidelli} et~al.}(2018)\emph{{Morbidelli}, {Nesvorny},
  {Laurenz}, {Marchi}, {Rubie}, {Elkins-Tanton}, {Wieczorek}, and
  {Jacobson}}}]{morby18}
{Morbidelli} A., {Nesvorny} D., {Laurenz} V., {Marchi} S., {Rubie} D.~C.,
  {Elkins-Tanton} L., {Wieczorek} M., and {Jacobson} S. (2018) \emph{{The
  timeline of the lunar bombardment: Revisited}}, \emph{\icarus}, \emph{305},
  262--276.

\bibitem[{\emph{{Morbidelli} and {Raymond}}(2016)}]{morbyraymond16}
{Morbidelli} A. and {Raymond} S.~N. (2016) \emph{{Challenges in planet
  formation}}, \emph{Journal of Geophysical Research (Planets)}, \emph{121},
  1962--1980.

\bibitem[{\emph{{Morbidelli} et~al.}(2007)\emph{{Morbidelli}, {Tsiganis},
  {Crida}, {Levison}, and {Gomes}}}]{morby07}
{Morbidelli} A., {Tsiganis} K., {Crida} A., {Levison} H.~F., and {Gomes} R.
  (2007) \emph{{Dynamics of the Giant Planets of the Solar System in the
  Gaseous Protoplanetary Disk and Their Relationship to the Current Orbital
  Architecture}}, \emph{\aj}, \emph{134}, 1790--1798.

\bibitem[{\emph{{Morbidelli} and {Wood}}(2015)}]{morbywood15}
{Morbidelli} A. and {Wood} B.~J. (2015) \emph{{Late Accretion and the Late
  Veneer}}, \emph{Washington DC American Geophysical Union Geophysical
  Monograph Series}, \emph{212}, 71--82.

\bibitem[{\emph{{Morishima} et~al.}(2010)\emph{{Morishima}, {Stadel}, and
  {Moore}}}]{morishima10}
{Morishima} R., {Stadel} J., and {Moore} B. (2010) \emph{{From planetesimals to
  terrestrial planets: N-body simulations including the effects of nebular gas
  and giant planets}}, \emph{Icarus}, \emph{207}, 517--535.

\bibitem[{\emph{{Mottl} et~al.}(2007)\emph{{Mottl}, {Glazer}, {Kaiser}, and
  {Meech}}}]{Mottl2007}
{Mottl} M., {Glazer} B., {Kaiser} R., and {Meech} K. (2007) \emph{{Water and
  astrobiology}}, \emph{Chemie der Erde / Geochemistry}, \emph{67}, 253--282.

\bibitem[{\emph{{Mousis} et~al.}(2018)\emph{{Mousis}, {Ronnet}, {Lunine},
  {Luspay-Kuti}, {Mandt}, {Danger}, {Pauzat}, {Ellinger}, {Wurz}, and
  {Vernazza}}}]{Mousis2018}
{Mousis} O., {Ronnet} T., {Lunine} J.~I., {Luspay-Kuti} A., {Mandt} K.~E.,
  {Danger} G., {Pauzat} F., {Ellinger} Y., {Wurz} P., and {Vernazza} P. (2018)
  \emph{{Noble Gas Abundance Ratios Indicate the Agglomeration of
  67P/Churyumov-Gerasimenko from Warmed-up Ice}}, \emph{\apjl}, \emph{865},
  L11.

\bibitem[{\emph{{Mulders} et~al.}(2015{\natexlab{a}})\emph{{Mulders}, {Ciesla},
  {Min}, and {Pascucci}}}]{mulders15c}
{Mulders} G.~D., {Ciesla} F.~J., {Min} M., and {Pascucci} I.
  (2015{\natexlab{a}}) \emph{{The Snow Line in Viscous Disks around Low-mass
  Stars: Implications for Water Delivery to Terrestrial Planets in the
  Habitable Zone}}, \emph{\apj}, \emph{807}, 9.

\bibitem[{\emph{{Mulders} et~al.}(2015{\natexlab{b}})\emph{{Mulders},
  {Pascucci}, and {Apai}}}]{mulders15}
{Mulders} G.~D., {Pascucci} I., and {Apai} D. (2015{\natexlab{b}}) \emph{{A
  Stellar-mass-dependent Drop in Planet Occurrence Rates}}, \emph{\apj},
  \emph{798}, 112.

\bibitem[{\emph{{Mulders} et~al.}(2018)\emph{{Mulders}, {Pascucci}, {Apai}, and
  {Ciesla}}}]{mulders18}
{Mulders} G.~D., {Pascucci} I., {Apai} D., and {Ciesla} F.~J. (2018) \emph{{The
  Exoplanet Population Observation Simulator. I. The Inner Edges of Planetary
  Systems}}, \emph{\aj}, \emph{156}, 24.

\bibitem[{\emph{{Mumma} and {Charnley}}(2011)}]{Mumma2011}
{Mumma} M.~J. and {Charnley} S.~B. (2011) \emph{{The Chemical Composition of
  Comets{\textemdash}Emerging Taxonomies and Natal Heritage}}, \emph{\araa},
  \emph{49}, 471--524.

\bibitem[{\emph{{Muralidharan} et~al.}(2008{\natexlab{a}})\emph{{Muralidharan},
  {Deymier}, {Stimpfl}, {de Leeuw}, and {Drake}}}]{Muralidharan2008}
{Muralidharan} K., {Deymier} P., {Stimpfl} M., {de Leeuw} N.~H., and {Drake}
  M.~J. (2008{\natexlab{a}}) \emph{{Origin of water in the inner Solar System:
  A kinetic Monte Carlo study of water adsorption on forsterite}},
  \emph{\icarus}, \emph{198}, 400--407.

\bibitem[{\emph{{Muralidharan} et~al.}(2008{\natexlab{b}})\emph{{Muralidharan},
  {Deymier}, {Stimpfl}, {de Leeuw}, and {Drake}}}]{muralidharan08}
{Muralidharan} K., {Deymier} P., {Stimpfl} M., {de Leeuw} N.~H., and {Drake}
  M.~J. (2008{\natexlab{b}}) \emph{{Origin of water in the inner Solar System:
  A kinetic Monte Carlo study of water adsorption on forsterite}},
  \emph{\icarus}, \emph{198}, 400--407.

\bibitem[{\emph{{Murphy} et~al.}(2018)\emph{{Murphy}, {Bolatto}, {Chatterjee},
  {Casey}, {Chomiuk}, {Dale}, {de Pater}, {Dickinson}, {Francesco}, and
  {Hallinan}}}]{Murphy2018}
{Murphy} E.~J., {Bolatto} A., {Chatterjee} S., {Casey} C.~M., {Chomiuk} L.,
  {Dale} D., {de Pater} I., {Dickinson} M., {Francesco} J.~D., and {Hallinan}
  G. (2018) in \emph{Science with a Next Generation Very Large Array}
  (E.~{Murphy}, ed.), vol. 517 of \emph{Astronomical Society of the Pacific
  Conference Series}, p.~3.

\bibitem[{\emph{{Nanne} et~al.}(2019)\emph{{Nanne}, {Nimmo}, {Cuzzi}, and
  {Kleine}}}]{nanne19}
{Nanne} J. A.~M., {Nimmo} F., {Cuzzi} J.~N., and {Kleine} T. (2019)
  \emph{{Origin of the non-carbonaceous-carbonaceous meteorite dichotomy}},
  \emph{Earth and Planetary Science Letters}, \emph{511}, 44--54.

\bibitem[{\emph{{Nesvorn{\'y}}}(2015)}]{nesvorny15}
{Nesvorn{\'y}} D. (2015) \emph{{Evidence for Slow Migration of Neptune from the
  Inclination Distribution of Kuiper Belt Objects}}, \emph{\aj}, \emph{150},
  73.

\bibitem[{\emph{{Nesvorn{\'y}} et~al.}(2018{\natexlab{a}})\emph{{Nesvorn{\'y}},
  {Vokrouhlick{\'y}}, {Bottke}, and {Levison}}}]{nesvorny18}
{Nesvorn{\'y}} D., {Vokrouhlick{\'y}} D., {Bottke} W.~F., and {Levison} H.~F.
  (2018{\natexlab{a}}) \emph{{Evidence for very early migration of the Solar
  System planets from the Patroclus-Menoetius binary Jupiter Trojan}},
  \emph{Nature Astronomy}, \emph{2}, 878--882.

\bibitem[{\emph{{Nesvorn{\'y}} et~al.}(2018{\natexlab{b}})\emph{{Nesvorn{\'y}},
  {Vokrouhlick{\'y}}, {Bottke}, and {Levison}}}]{nesvorny18b}
{Nesvorn{\'y}} D., {Vokrouhlick{\'y}} D., {Bottke} W.~F., and {Levison} H.~F.
  (2018{\natexlab{b}}) \emph{{Evidence for very early migration of the Solar
  System planets from the Patroclus-Menoetius binary Jupiter Trojan}},
  \emph{Nature Astronomy}, \emph{2}, 878--882.

\bibitem[{\emph{{Nimmo} and {Kleine}}(2007)}]{nimmo07}
{Nimmo} F. and {Kleine} T. (2007) \emph{{How rapidly did Mars accrete?
  Uncertainties in the Hf W timing of core formation}}, \emph{Icarus},
  \emph{191}, 497--504.

\bibitem[{\emph{{Nomura} et~al.}(2014)\emph{{Nomura}, {Hirose}, {Uesegi},
  {Ohishi}, {Tsuchiyama}, {Miyake}, and {Ueno}}}]{nomura14}
{Nomura} R., {Hirose} K., {Uesegi} K., {Ohishi} Y., {Tsuchiyama} A., {Miyake}
  A., and {Ueno} Y. (2014) \emph{{Low Core-Mantle Boundary Temperature Inferred
  from the Solidus of Pyrolite}}, \emph{Science}, \emph{343}, 522--525.

\bibitem[{\emph{{O'Brien} et~al.}(2018)\emph{{O'Brien}, {Izidoro}, {Jacobson},
  {Raymond}, and {Rubie}}}]{obrien18}
{O'Brien} D.~P., {Izidoro} A., {Jacobson} S.~A., {Raymond} S.~N., and {Rubie}
  D.~C. (2018) \emph{{The Delivery of Water During Terrestrial Planet
  Formation}}, \emph{\ssr}, \emph{214}, 47.

\bibitem[{\emph{{O'Brien} et~al.}(2006)\emph{{O'Brien}, {Morbidelli}, and
  {Levison}}}]{obrien06}
{O'Brien} D.~P., {Morbidelli} A., and {Levison} H.~F. (2006) \emph{{Terrestrial
  planet formation with strong dynamical friction}}, \emph{Icarus}, \emph{184},
  39--58.

\bibitem[{\emph{{O'Brien} et~al.}(2014)\emph{{O'Brien}, {Walsh}, {Morbidelli},
  {Raymond}, and {Mandell}}}]{obrien14}
{O'Brien} D.~P., {Walsh} K.~J., {Morbidelli} A., {Raymond} S.~N., and {Mandell}
  A.~M. (2014) \emph{{Water delivery and giant impacts in the Grand Tack
  scenario}}, \emph{Icarus}, \emph{239}, 74--84.

\bibitem[{\emph{{Ogihara} et~al.}(2018)\emph{{Ogihara}, {Kokubo}, {Suzuki}, and
  {Morbidelli}}}]{ogihara18}
{Ogihara} M., {Kokubo} E., {Suzuki} T.~K., and {Morbidelli} A. (2018)
  \emph{{Formation of close-in super-Earths in evolving protoplanetary disks
  due to disk winds}}, \emph{\aap}, \emph{615}, A63.

\bibitem[{\emph{{Ogihara} et~al.}(2015)\emph{{Ogihara}, {Morbidelli}, and
  {Guillot}}}]{ogihara15}
{Ogihara} M., {Morbidelli} A., and {Guillot} T. (2015) \emph{{A reassessment of
  the in situ formation of close-in super-Earths}}, \emph{\aap}, \emph{578},
  A36.

\bibitem[{\emph{{Oka} et~al.}(2011)\emph{{Oka}, {Nakamoto}, and {Ida}}}]{oka11}
{Oka} A., {Nakamoto} T., and {Ida} S. (2011) \emph{{Evolution of Snow Line in
  Optically Thick Protoplanetary Disks: Effects of Water Ice Opacity and Dust
  Grain Size}}, \emph{\apj}, \emph{738}, 141.

\bibitem[{\emph{{Ormel} and {Klahr}}(2010)}]{ormel10}
{Ormel} C.~W. and {Klahr} H.~H. (2010) \emph{{The effect of gas drag on the
  growth of protoplanets. Analytical expressions for the accretion of small
  bodies in laminar disks}}, \emph{\aap}, \emph{520}, A43.

\bibitem[{\emph{{Ormel} et~al.}(2017)\emph{{Ormel}, {Liu}, and
  {Schoonenberg}}}]{ormel17}
{Ormel} C.~W., {Liu} B., and {Schoonenberg} D. (2017) \emph{{Formation of
  TRAPPIST-1 and other compact systems}}, \emph{\aap}, \emph{604}, A1.

\bibitem[{\emph{{Owen} and {Wu}}(2017)}]{owen17}
{Owen} J.~E. and {Wu} Y. (2017) \emph{{The Evaporation Valley in the Kepler
  Planets}}, \emph{\apj}, \emph{847}, 29.

\bibitem[{\emph{{Owen} and {Bar-Nun}}(1995)}]{Owen1995}
{Owen} T. and {Bar-Nun} A. (1995) \emph{{Comets, Impacts, and Atmospheres}},
  \emph{\icarus}, \emph{116}, 215--226.

\bibitem[{\emph{{Owen} et~al.}(2001)\emph{{Owen}, {Mahaffy}, {Niemann},
  {Atreya}, and {Wong}}}]{Owen2001}
{Owen} T., {Mahaffy} P.~R., {Niemann} H.~B., {Atreya} S., and {Wong} M. (2001)
  \emph{{Protosolar Nitrogen}}, \emph{\apjl}, \emph{553}, L77--L79.

\bibitem[{\emph{{Owen} and {Bar-Nun}}(2000)}]{Owen2000}
{Owen} T.~C. and {Bar-Nun} A. (2000) \emph{{Volatile Contributions from Icy
  Planetesimals}}, pp. 459--471.

\bibitem[{\emph{{Paardekooper} et~al.}(2011)\emph{{Paardekooper}, {Baruteau},
  and {Kley}}}]{paardekooper11}
{Paardekooper} S.-J., {Baruteau} C., and {Kley} W. (2011) \emph{{A torque
  formula for non-isothermal Type I planetary migration - II. Effects of
  diffusion}}, \emph{\mnras}, \emph{410}, 293--303.

\bibitem[{\emph{{Pascucci} et~al.}(2016)\emph{{Pascucci}, {Testi}, {Herczeg},
  {Long}, {Manara}, {Hendler}, {Mulders}, {Krijt}, {Ciesla}, {Henning},
  {Mohanty}, {Drabek-Maunder}, {Apai}, {Sz{\H u}cs}, {Sacco}, and
  {Olofsson}}}]{pascucci16}
{Pascucci} I., {Testi} L., {Herczeg} G.~J., {Long} F., {Manara} C.~F.,
  {Hendler} N., {Mulders} G.~D., {Krijt} S., {Ciesla} F., {Henning} T.,
  {Mohanty} S., {Drabek-Maunder} E., {Apai} D., {Sz{\H u}cs} L., {Sacco} G.,
  and {Olofsson} J. (2016) \emph{{A Steeper than Linear Disk Mass-Stellar Mass
  Scaling Relation}}, \emph{\apj}, \emph{831}, 125.

\bibitem[{\emph{{Petigura} et~al.}(2013)\emph{{Petigura}, {Howard}, and
  {Marcy}}}]{petigura13}
{Petigura} E.~A., {Howard} A.~W., and {Marcy} G.~W. (2013) \emph{{Prevalence of
  Earth-size planets orbiting Sun-like stars}}, \emph{Proceedings of the
  National Academy of Science}, \emph{110}, 19273--19278.

\bibitem[{\emph{{Pierel} et~al.}(2017)\emph{{Pierel}, {Nixon}, {Lellouch},
  {Fletcher}, {Bjoraker}, {Achterberg}, {B{\'e}zard}, {Hesman}, {Irwin}, and
  {Flasar}}}]{Pierel2017}
{Pierel} J.~D.~R., {Nixon} C.~A., {Lellouch} E., {Fletcher} L.~N., {Bjoraker}
  G.~L., {Achterberg} R.~K., {B{\'e}zard} B., {Hesman} B.~E., {Irwin} P.~G.~J.,
  and {Flasar} F.~M. (2017) \emph{{D/H Ratios on Saturn and Jupiter from
  Cassini CIRS}}, \emph{\aj}, \emph{154}, 178.

\bibitem[{\emph{{Pierens} and {Nelson}}(2008)}]{pierens08}
{Pierens} A. and {Nelson} R.~P. (2008) \emph{{Constraints on resonant-trapping
  for two planets embedded in a protoplanetary disc}}, \emph{\aap}, \emph{482},
  333--340.

\bibitem[{\emph{{Pierens} and {Raymond}}(2011)}]{pierens11}
{Pierens} A. and {Raymond} S.~N. (2011) \emph{{Two phase, inward-then-outward
  migration of Jupiter and Saturn in the gaseous solar nebula}}, \emph{\aap},
  \emph{533}, A131.

\bibitem[{\emph{{Pierens} et~al.}(2014)\emph{{Pierens}, {Raymond}, {Nesvorny},
  and {Morbidelli}}}]{pierens14}
{Pierens} A., {Raymond} S.~N., {Nesvorny} D., and {Morbidelli} A. (2014)
  \emph{{Outward Migration of Jupiter and Saturn in 3:2 or 2:1 Resonance in
  Radiative Disks: Implications for the Grand Tack and Nice models}},
  \emph{\apjl}, \emph{795}, L11.

\bibitem[{\emph{{Pirani} et~al.}(2019)\emph{{Pirani}, {Johansen}, {Bitsch},
  {Mustill}, and {Turrini}}}]{pirani19}
{Pirani} S., {Johansen} A., {Bitsch} B., {Mustill} A.~J., and {Turrini} D.
  (2019) \emph{{Consequences of planetary migration on the minor bodies of the
  early solar system}}, \emph{\aap}, \emph{623}, A169.

\bibitem[{\emph{{Podolak} et~al.}(2002)\emph{{Podolak}, {Mekler}, and
  {Prialnik}}}]{Podolak2002}
{Podolak} M., {Mekler} Y., and {Prialnik} D. (2002) \emph{{Is the D/H Ratio in
  the Comet Coma Equal to the D/H Ratio in the Comet Nucleus?}},
  \emph{\icarus}, \emph{160}, 208--211.

\bibitem[{\emph{{Pollack} et~al.}(1996)\emph{{Pollack}, {Hubickyj},
  {Bodenheimer}, {Lissauer}, {Podolak}, and {Greenzweig}}}]{pollack96}
{Pollack} J.~B., {Hubickyj} O., {Bodenheimer} P., {Lissauer} J.~J., {Podolak}
  M., and {Greenzweig} Y. (1996) \emph{{Formation of the Giant Planets by
  Concurrent Accretion of Solids and Gas}}, \emph{Icarus}, \emph{124}, 62--85.

\bibitem[{\emph{{Pontoppidan} et~al.}(2014)\emph{{Pontoppidan}, {Salyk},
  {Bergin}, {Brittain}, {Marty}, {Mousis}, and {{\"O}berg}}}]{Pontoppidan2014}
{Pontoppidan} K.~M., {Salyk} C., {Bergin} E.~A., {Brittain} S., {Marty} B.,
  {Mousis} O., and {{\"O}berg} K.~I. (2014) in \emph{Protostars and Planets VI}
  (H.~{Beuther}, R.~S. {Klessen}, C.~P. {Dullemond}, and T.~{Henning}, eds.),
  p. 363.

\bibitem[{\emph{{Prialnik} and {Rosenberg}}(2009)}]{Prialnik2009}
{Prialnik} D. and {Rosenberg} E.~D. (2009) \emph{{Can ice survive in main-belt
  comets? Long-term evolution models of comet 133P/Elst-Pizarro}},
  \emph{\mnras}, \emph{399}, L79--L83.

\bibitem[{\emph{{Qi} et~al.}(2013)\emph{{Qi}, {{\"O}berg}, {Wilner},
  {D'Alessio}, {Bergin}, {Andrews}, {Blake}, {Hogerheijde}, and {van
  Dishoeck}}}]{Qi2013}
{Qi} C., {{\"O}berg} K.~I., {Wilner} D.~J., {D'Alessio} P., {Bergin} E.,
  {Andrews} S.~M., {Blake} G.~A., {Hogerheijde} M.~R., and {van Dishoeck} E.~F.
  (2013) \emph{{Imaging of the CO Snow Line in a Solar Nebula Analog}},
  \emph{Science}, \emph{341}, 630--632.

\bibitem[{\emph{{Raymond} et~al.}(2010)\emph{{Raymond}, {Armitage}, and
  {Gorelick}}}]{raymond10}
{Raymond} S.~N., {Armitage} P.~J., and {Gorelick} N. (2010)
  \emph{{Planet-Planet Scattering in Planetesimal Disks. II. Predictions for
  Outer Extrasolar Planetary Systems}}, \emph{\apj}, \emph{711}, 772--795.

\bibitem[{\emph{{Raymond} et~al.}(2011)\emph{{Raymond}, {Armitage},
  {Moro-Mart{\'{\i}}n}, {Booth}, {Wyatt}, {Armstrong}, {Mandell}, {Selsis}, and
  {West}}}]{raymond11}
{Raymond} S.~N., {Armitage} P.~J., {Moro-Mart{\'{\i}}n} A., {Booth} M., {Wyatt}
  M.~C., {Armstrong} J.~C., {Mandell} A.~M., {Selsis} F., and {West} A.~A.
  (2011) \emph{{Debris disks as signposts of terrestrial planet formation}},
  \emph{\aap}, \emph{530}, A62.

\bibitem[{\emph{{Raymond} et~al.}(2012)\emph{{Raymond}, {Armitage},
  {Moro-Mart{\'{\i}}n}, {Booth}, {Wyatt}, {Armstrong}, {Mandell}, {Selsis}, and
  {West}}}]{raymond12}
{Raymond} S.~N., {Armitage} P.~J., {Moro-Mart{\'{\i}}n} A., {Booth} M., {Wyatt}
  M.~C., {Armstrong} J.~C., {Mandell} A.~M., {Selsis} F., and {West} A.~A.
  (2012) \emph{{Debris disks as signposts of terrestrial planet formation. II.
  Dependence of exoplanet architectures on giant planet and disk properties}},
  \emph{\aap}, \emph{541}, A11.

\bibitem[{\emph{{Raymond} et~al.}(2018{\natexlab{a}})\emph{{Raymond}, {Boulet},
  {Izidoro}, {Esteves}, and {Bitsch}}}]{raymond18b}
{Raymond} S.~N., {Boulet} T., {Izidoro} A., {Esteves} L., and {Bitsch} B.
  (2018{\natexlab{a}}) \emph{{Migration-driven diversity of super-Earth
  compositions}}, \emph{\mnras}, \emph{479}, L81--L85.

\bibitem[{\emph{{Raymond} and {Cossou}}(2014)}]{raymond14b}
{Raymond} S.~N. and {Cossou} C. (2014) \emph{{No universal minimum-mass
  extrasolar nebula: evidence against in situ accretion of systems of hot
  super-Earths}}, \emph{\mnras}, \emph{440}, L11--L15.

\bibitem[{\emph{{Raymond} and {Izidoro}}(2017{\natexlab{a}})}]{raymond17}
{Raymond} S.~N. and {Izidoro} A. (2017{\natexlab{a}}) \emph{{Origin of water in
  the inner Solar System: Planetesimals scattered inward during Jupiter and
  Saturn's rapid gas accretion}}, \emph{Icarus}, \emph{297}, 134--148.

\bibitem[{\emph{{Raymond} and {Izidoro}}(2017{\natexlab{b}})}]{raymond17b}
{Raymond} S.~N. and {Izidoro} A. (2017{\natexlab{b}}) \emph{{The empty
  primordial asteroid belt}}, \emph{Science Advances}, \emph{3}, e1701138.

\bibitem[{\emph{{Raymond} et~al.}(2018{\natexlab{b}})\emph{{Raymond},
  {Izidoro}, and {Morbidelli}}}]{Raymond2018}
{Raymond} S.~N., {Izidoro} A., and {Morbidelli} A. (2018{\natexlab{b}})
  \emph{{Solar System Formation in the Context of Extra-Solar Planets}},
  \emph{arXiv e-prints}, arXiv:1812.01033.

\bibitem[{\emph{{Raymond} et~al.}(2014)\emph{{Raymond}, {Kokubo}, {Morbidelli},
  {Morishima}, and {Walsh}}}]{raymond14}
{Raymond} S.~N., {Kokubo} E., {Morbidelli} A., {Morishima} R., and {Walsh}
  K.~J. (2014) \emph{{Terrestrial Planet Formation at Home and Abroad}},
  \emph{Protostars and Planets VI}, pp. 595--618.

\bibitem[{\emph{{Raymond} et~al.}(2006{\natexlab{a}})\emph{{Raymond},
  {Mandell}, and {Sigurdsson}}}]{raymond06c}
{Raymond} S.~N., {Mandell} A.~M., and {Sigurdsson} S. (2006{\natexlab{a}})
  \emph{{Exotic Earths: Forming Habitable Worlds with Giant Planet Migration}},
  \emph{Science}, \emph{313}, 1413--1416.

\bibitem[{\emph{{Raymond} and {Morbidelli}}(2014)}]{raymond14c}
{Raymond} S.~N. and {Morbidelli} A. (2014) in \emph{Complex Planetary Systems,
  Proceedings of the International Astronomical Union}, vol. 310 of \emph{IAU
  Symposium}, pp. 194--203.

\bibitem[{\emph{{Raymond} et~al.}(2009)\emph{{Raymond}, {O'Brien},
  {Morbidelli}, and {Kaib}}}]{raymond09c}
{Raymond} S.~N., {O'Brien} D.~P., {Morbidelli} A., and {Kaib} N.~A. (2009)
  \emph{{Building the terrestrial planets: Constrained accretion in the inner
  Solar System}}, \emph{Icarus}, \emph{203}, 644--662.

\bibitem[{\emph{{Raymond} et~al.}(2004)\emph{{Raymond}, {Quinn}, and
  {Lunine}}}]{raymond04}
{Raymond} S.~N., {Quinn} T., and {Lunine} J.~I. (2004) \emph{{Making other
  earths: dynamical simulations of terrestrial planet formation and water
  delivery}}, \emph{Icarus}, \emph{168}, 1--17.

\bibitem[{\emph{{Raymond} et~al.}(2006{\natexlab{b}})\emph{{Raymond}, {Quinn},
  and {Lunine}}}]{raymond06b}
{Raymond} S.~N., {Quinn} T., and {Lunine} J.~I. (2006{\natexlab{b}})
  \emph{{High-resolution simulations of the final assembly of Earth-like
  planets I. Terrestrial accretion and dynamics}}, \emph{Icarus}, \emph{183},
  265--282.

\bibitem[{\emph{{Raymond} et~al.}(2007{\natexlab{a}})\emph{{Raymond}, {Quinn},
  and {Lunine}}}]{raymond07a}
{Raymond} S.~N., {Quinn} T., and {Lunine} J.~I. (2007{\natexlab{a}})
  \emph{{High-Resolution Simulations of The Final Assembly of Earth-Like
  Planets. 2. Water Delivery And Planetary Habitability}}, \emph{Astrobiology},
  \emph{7}, 66--84.

\bibitem[{\emph{{Raymond} et~al.}(2007{\natexlab{b}})\emph{{Raymond}, {Scalo},
  and {Meadows}}}]{raymond07b}
{Raymond} S.~N., {Scalo} J., and {Meadows} V.~S. (2007{\natexlab{b}}) \emph{{A
  Decreased Probability of Habitable Planet Formation around Low-Mass Stars}},
  \emph{\apj}, \emph{669}, 606--614.

\bibitem[{\emph{{Ricci} et~al.}(2018)\emph{{Ricci}, {Isella}, {Liu}, and
  {Li}}}]{Ricci2018}
{Ricci} L., {Isella} A., {Liu} S., and {Li} H. (2018) in \emph{Science with a
  Next Generation Very Large Array} (E.~{Murphy}, ed.), vol. 517 of
  \emph{Astronomical Society of the Pacific Conference Series}, p. 147.

\bibitem[{\emph{{Rivkin} et~al.}(2014)\emph{{Rivkin}, {Asphaug}, and
  {Bottke}}}]{Rivkin2014}
{Rivkin} A.~S., {Asphaug} E., and {Bottke} W.~F. (2014) \emph{{The case of the
  missing Ceres family}}, \emph{\icarus}, \emph{243}, 429--439.

\bibitem[{\emph{{Rivkin} and {Emery}}(2010)}]{Rivkin2010}
{Rivkin} A.~S. and {Emery} J.~P. (2010) \emph{{Detection of ice and organics on
  an asteroidal surface}}, \emph{\nat}, \emph{464}, 1322--1323.

\bibitem[{\emph{{Rogers}}(2015)}]{rogers15}
{Rogers} L.~A. (2015) \emph{{Most 1.6 Earth-radius Planets are Not Rocky}},
  \emph{\apj}, \emph{801}, 41.

\bibitem[{\emph{{Ronnet} et~al.}(2018)\emph{{Ronnet}, {Mousis}, {Vernazza},
  {Lunine}, and {Crida}}}]{ronnet18}
{Ronnet} T., {Mousis} O., {Vernazza} P., {Lunine} J.~I., and {Crida} A. (2018)
  \emph{{Saturn{\textquoteright}s Formation and Early Evolution at the Origin
  of Jupiter{\textquoteright}s Massive Moons}}, \emph{\aj}, \emph{155}, 224.

\bibitem[{\emph{{Rubie} et~al.}(2007)\emph{{Rubie}, {Nimmo}, and
  {Melosh}}}]{Rubie2007}
{Rubie} D.~C., {Nimmo} F., and {Melosh} H.~J. (2007) \emph{{Formation of
  Earth's Core}}, p. 6054.

\bibitem[{\emph{{Rubin} et~al.}(2018)\emph{{Rubin}, {Altwegg}, {Balsiger},
  {Bar-Nun}, {Berthelier}, {Briois}, {Calmonte}, {Combi}, {De Keyser}, and
  {Fiethe}}}]{Rubin2018}
{Rubin} M., {Altwegg} K., {Balsiger} H., {Bar-Nun} A., {Berthelier} J.-J.,
  {Briois} C., {Calmonte} U., {Combi} M., {De Keyser} J., and {Fiethe} B.
  (2018) \emph{{Krypton isotopes and noble gas abundances in the coma of comet
  67P/Churyumov-Gerasimenko}}, \emph{Science Advances}, \emph{4}, eaar6297.

\bibitem[{\emph{{Santerne} et~al.}(2018)\emph{{Santerne}, {Brugger},
  {Armstrong}, {Adibekyan}, {Lillo-Box}, {Gosselin}, {Aguichine}, {Almenara},
  {Barrado}, and {Barros}}}]{santerne18}
{Santerne} A., {Brugger} B., {Armstrong} D.~J., {Adibekyan} V., {Lillo-Box} J.,
  {Gosselin} H., {Aguichine} A., {Almenara} J.~M., {Barrado} D., and {Barros}
  S.~C.~C. (2018) \emph{{An Earth-sized exoplanet with a Mercury-like
  composition}}, \emph{Nature Astronomy}, \emph{2}, 393--400.

\bibitem[{\emph{{Sasselov} and {Lecar}}(2000)}]{sasselov00}
{Sasselov} D.~D. and {Lecar} M. (2000) \emph{{On the Snow Line in Dusty
  Protoplanetary Disks}}, \emph{\apj}, \emph{528}, 995--998.

\bibitem[{\emph{{Sato} et~al.}(2016)\emph{{Sato}, {Okuzumi}, and
  {Ida}}}]{sato16}
{Sato} T., {Okuzumi} S., and {Ida} S. (2016) \emph{{On the water delivery to
  terrestrial embryos by ice pebble accretion}}, \emph{\aap}, \emph{589}, A15.

\bibitem[{\emph{{Schlichting} et~al.}(2015)\emph{{Schlichting}, {Sari}, and
  {Yalinewich}}}]{schlichting15}
{Schlichting} H.~E., {Sari} R., and {Yalinewich} A. (2015) \emph{{Atmospheric
  mass loss during planet formation: The importance of planetesimal impacts}},
  \emph{\icarus}, \emph{247}, 81--94.

\bibitem[{\emph{{Schorghofer}}(2008)}]{Schorghofer2008}
{Schorghofer} N. (2008) \emph{{The Lifetime of Ice on Main Belt Asteroids}},
  \emph{\apj}, \emph{682}, 697--705.

\bibitem[{\emph{{Scott} and {Krot}}(2014)}]{Scott2014}
{Scott} E.~R.~D. and {Krot} A.~N. (2014) \emph{{Chondrites and Their
  Components}}, vol.~1, pp. 65--137.

\bibitem[{\emph{{Selsis} et~al.}(2007)\emph{{Selsis}, {Chazelas}, {Bord{\'e}},
  {Ollivier}, {Brachet}, {Decaudin}, {Bouchy}, {Ehrenreich}, {Grie{\ss}meier},
  {Lammer}, {Sotin}, {Grasset}, {Moutou}, {Barge}, {Deleuil}, {Mawet},
  {Despois}, {Kasting}, and {L{\'e}ger}}}]{selsis07b}
{Selsis} F., {Chazelas} B., {Bord{\'e}} P., {Ollivier} M., {Brachet} F.,
  {Decaudin} M., {Bouchy} F., {Ehrenreich} D., {Grie{\ss}meier} J.-M., {Lammer}
  H., {Sotin} C., {Grasset} O., {Moutou} C., {Barge} P., {Deleuil} M., {Mawet}
  D., {Despois} D., {Kasting} J.~F., and {L{\'e}ger} A. (2007) \emph{{Could we
  identify hot ocean-planets with CoRoT, Kepler and Doppler velocimetry?}},
  \emph{Icarus}, \emph{191}, 453--468.

\bibitem[{\emph{{Sharp}}(2017)}]{sharp17}
{Sharp} Z.~D. (2017) \emph{{Nebular ingassing as a source of volatiles to the
  Terrestrial planets}}, \emph{Chemical Geology}, \emph{448}, 137--150.

\bibitem[{\emph{{Simon} et~al.}(2016)\emph{{Simon}, {Armitage}, {Li}, and
  {Youdin}}}]{simon16}
{Simon} J.~B., {Armitage} P.~J., {Li} R., and {Youdin} A.~N. (2016) \emph{{The
  Mass and Size Distribution of Planetesimals Formed by the Streaming
  Instability. I. The Role of Self-gravity}}, \emph{\apj}, \emph{822}, 55.

\bibitem[{\emph{{Snodgrass} et~al.}(2017)\emph{{Snodgrass}, {Agarwal}, {Combi},
  {Fitzsimmons}, {Guilbert-Lepoutre}, {Hsieh}, {Hui}, {Jehin}, {Kelley}, and
  {Knight}}}]{Snodgrass2017}
{Snodgrass} C., {Agarwal} J., {Combi} M., {Fitzsimmons} A., {Guilbert-Lepoutre}
  A., {Hsieh} H.~H., {Hui} M.-T., {Jehin} E., {Kelley} M. S.~P., and {Knight}
  M.~M. (2017) \emph{{The Main Belt Comets and ice in the Solar System}},
  \emph{\aapr}, \emph{25}, 5.

\bibitem[{\emph{{Spergel} et~al.}(2003)\emph{{Spergel}, {Verde}, {Peiris},
  {Komatsu}, {Nolta}, {Bennett}, {Halpern}, {Hinshaw}, {Jarosik}, {Kogut},
  {Limon}, {Meyer}, {Page}, {Tucker}, {Weiland}, {Wollack}, and
  {Wright}}}]{Spergel2003}
{Spergel} D.~N., {Verde} L., {Peiris} H.~V., {Komatsu} E., {Nolta} M.~R.,
  {Bennett} C.~L., {Halpern} M., {Hinshaw} G., {Jarosik} N., {Kogut} A.,
  {Limon} M., {Meyer} S.~S., {Page} L., {Tucker} G.~S., {Weiland} J.~L.,
  {Wollack} E., and {Wright} E.~L. (2003) \emph{{First-Year Wilkinson Microwave
  Anisotropy Probe (WMAP) Observations: Determination of Cosmological
  Parameters}}, \emph{\apjs}, \emph{148}, 175--194.

\bibitem[{\emph{{Starkey} et~al.}(2009)\emph{{Starkey}, {Stuart}, {Ellam},
  {Fitton}, {Basu}, and {Larsen}}}]{starkey2009}
{Starkey} N.~A., {Stuart} F.~M., {Ellam} R.~M., {Fitton} J.~G., {Basu} S., and
  {Larsen} L.~M. (2009) \emph{{Helium isotopes in early Iceland plume picrites:
  Constraints on the composition of high $^{3}$He/ $^{4}$He mantle}},
  \emph{Earth and Planetary Science Letters}, \emph{277}, 91--100.

\bibitem[{\emph{{Stevenson} and {Lunine}}(1988)}]{Stevenson1988}
{Stevenson} D.~J. and {Lunine} J.~I. (1988) \emph{{Rapid formation of Jupiter
  by diffusive redistribution of water vapor in the solar nebula}},
  \emph{\icarus}, \emph{75}, 146--155.

\bibitem[{\emph{{Stimpfl} et~al.}(2006)\emph{{Stimpfl}, {Walker}, {Drake}, {de
  Leeuw}, and {Deymier}}}]{stimpfl06}
{Stimpfl} M., {Walker} A.~M., {Drake} M.~J., {de Leeuw} N.~H., and {Deymier} P.
  (2006) \emph{{An {\r{a}}ngstr{\"o}m-sized window on the origin of water in
  the inner solar system: Atomistic simulation of adsorption of water on
  olivine}}, \emph{Journal of Crystal Growth}, \emph{294}, 83--95.

\bibitem[{\emph{{Stuart} et~al.}(2003)\emph{{Stuart}, {Lass-Evans}, {Godfrey
  Fitton}, and {Ellam}}}]{stuart2003}
{Stuart} F.~M., {Lass-Evans} S., {Godfrey Fitton} J., and {Ellam} R.~M. (2003)
  \emph{{High $^{3}$He/$^{4}$He ratios in picritic basalts from Baffin Island
  and the role of a mixed reservoir in mantle plumes}}, \emph{\nat},
  \emph{424}, 57--59.

\bibitem[{\emph{{Sugiura} and {Fujiya}}(2014)}]{Sugiura2014}
{Sugiura} N. and {Fujiya} W. (2014) \emph{{Correlated accretion ages and
  {\ensuremath{\in}}$^{54}$Cr of meteorite parent bodies and the evolution of
  the solar nebula}}, \emph{Meteoritics and Planetary Science}, \emph{49},
  772--787.

\bibitem[{\emph{{Svetsov}}(2007)}]{svetsov07}
{Svetsov} V.~V. (2007) \emph{{Atmospheric erosion and replenishment induced by
  impacts of cosmic bodies upon the Earth and Mars}}, \emph{Solar System
  Research}, \emph{41}, 28--41.

\bibitem[{\emph{{Tera} et~al.}(1974)\emph{{Tera}, {Papanastassiou}, and
  {Wasserburg}}}]{tera74}
{Tera} F., {Papanastassiou} D.~A., and {Wasserburg} G.~J. (1974)
  \emph{{Isotopic evidence for a terminal lunar cataclysm}}, \emph{Earth and
  Planetary Science Letters}, \emph{22}, 1.

\bibitem[{\emph{{Terquem} and {Papaloizou}}(2007)}]{terquem07}
{Terquem} C. and {Papaloizou} J.~C.~B. (2007) \emph{{Migration and the
  Formation of Systems of Hot Super-Earths and Neptunes}}, \emph{\apj},
  \emph{654}, 1110--1120.

\bibitem[{\emph{{Tielens} and {Hagen}}(1982)}]{Tielens1982}
{Tielens} A.~G.~G.~M. and {Hagen} W. (1982) \emph{{Model calculations of the
  molecular composition of interstellar grain mantles}}, \emph{\aap},
  \emph{114}, 245--260.

\bibitem[{\emph{{Tsiganis} et~al.}(2005)\emph{{Tsiganis}, {Gomes},
  {Morbidelli}, and {Levison}}}]{tsiganis05}
{Tsiganis} K., {Gomes} R., {Morbidelli} A., and {Levison} H.~F. (2005)
  \emph{{Origin of the orbital architecture of the giant planets of the Solar
  System}}, \emph{\nat}, \emph{435}, 459--461.

\bibitem[{\emph{{Turner} et~al.}(2014)\emph{{Turner}, {Fromang}, {Gammie},
  {Klahr}, {Lesur}, {Wardle}, and {Bai}}}]{turner14}
{Turner} N.~J., {Fromang} S., {Gammie} C., {Klahr} H., {Lesur} G., {Wardle} M.,
  and {Bai} X.-N. (2014) \emph{{Transport and Accretion in Planet-Forming
  Disks}}, \emph{Protostars and Planets VI}, pp. 411--432.

\bibitem[{\emph{{Udry} et~al.}(2007)\emph{{Udry}, {Bonfils}, {Delfosse},
  {Forveille}, {Mayor}, {Perrier}, {Bouchy}, {Lovis}, {Pepe}, {Queloz}, and
  {Bertaux}}}]{udry07}
{Udry} S., {Bonfils} X., {Delfosse} X., {Forveille} T., {Mayor} M., {Perrier}
  C., {Bouchy} F., {Lovis} C., {Pepe} F., {Queloz} D., and {Bertaux} J. (2007)
  \emph{{The HARPS search for southern extra-solar planets. XI. Super-Earths (5
  and 8 M$_\oplus$) in a 3-planet system}}, \emph{\aap}, \emph{469}, L43--L47.

\bibitem[{\emph{{Vaghi}}(1973)}]{Vaghi1973}
{Vaghi} S. (1973) \emph{{Orbital Evolution of Comets and Dynamical
  Characteristics of Jupiter`s Family}}, \emph{\aap}, \emph{29}, 85.

\bibitem[{\emph{{van Dishoeck} et~al.}(2014)\emph{{van Dishoeck}, {Bergin},
  {Lis}, and {Lunine}}}]{vanDischoeck2014}
{van Dishoeck} E.~F., {Bergin} E.~A., {Lis} D.~C., and {Lunine} J.~I. (2014) in
  \emph{Protostars and Planets VI} (H.~{Beuther}, R.~S. {Klessen}, C.~P.
  {Dullemond}, and T.~{Henning}, eds.), p. 835.

\bibitem[{\emph{{van Dishoeck} et~al.}(2013)\emph{{van Dishoeck}, {Herbst}, and
  {Neufeld}}}]{vandishoeck13}
{van Dishoeck} E.~F., {Herbst} E., and {Neufeld} D.~A. (2013)
  \emph{{Interstellar Water Chemistry: From Laboratory to Observations}},
  \emph{Chemical Reviews}, \emph{113}, 9043--9085.

\bibitem[{\emph{{Veras} and {Armitage}}(2006)}]{veras06}
{Veras} D. and {Armitage} P.~J. (2006) \emph{{Predictions for the Correlation
  between Giant and Terrestrial Extrasolar Planets in Dynamically Evolved
  Systems}}, \emph{\apj}, \emph{645}, 1509--1515.

\bibitem[{\emph{{Waite} et~al.}(2009)\emph{{Waite}, {Lewis}, {Magee}, {Lunine},
  {McKinnon}, {Glein}, {Mousis}, {Young}, {Brockwell}, and
  {Westlake}}}]{Waite2009}
{Waite} J., J.~H., {Lewis} W.~S., {Magee} B.~A., {Lunine} J.~I., {McKinnon}
  W.~B., {Glein} C.~R., {Mousis} O., {Young} D.~T., {Brockwell} T., and
  {Westlake} J. (2009) \emph{{Liquid water on Enceladus from observations of
  ammonia and $^{40}$Ar in the plume}}, \emph{\nat}, \emph{460}, 1164.

\bibitem[{\emph{{Walker}}(2009)}]{walker09}
{Walker} R.~J. (2009) \emph{{Highly siderophile elements in the Earth, Moon and
  Mars: Update and implications for planetary accretion and differentiation}},
  \emph{Chemie der Erde / Geochemistry}, \emph{69}, 101--125.

\bibitem[{\emph{{Walker} et~al.}(2015)\emph{{Walker}, {Bermingham}, {Liu},
  {Puchtel}, {Touboul}, and {Worsham}}}]{walker15}
{Walker} R.~J., {Bermingham} K., {Liu} J., {Puchtel} I.~S., {Touboul} M., and
  {Worsham} E.~A. (2015) \emph{{In search of late-stage planetary building
  blocks}}, \emph{Chemical Geology}, \emph{411}, 125--142.

\bibitem[{\emph{{Walsh} et~al.}(2011{\natexlab{a}})\emph{{Walsh}, {Morbidelli},
  {Raymond}, {O'Brien}, and {Mandell}}}]{walsh11}
{Walsh} K.~J., {Morbidelli} A., {Raymond} S.~N., {O'Brien} D.~P., and {Mandell}
  A.~M. (2011{\natexlab{a}}) \emph{{A low mass for Mars from Jupiter's early
  gas-driven migration}}, \emph{\nat}, \emph{475}, 206--209.

\bibitem[{\emph{{Walsh} et~al.}(2011{\natexlab{b}})\emph{{Walsh}, {Morbidelli},
  {Raymond}, {O'Brien}, and {Mandell}}}]{Walsh2011}
{Walsh} K.~J., {Morbidelli} A., {Raymond} S.~N., {O'Brien} D.~P., and {Mandell}
  A.~M. (2011{\natexlab{b}}) \emph{{A low mass for Mars from Jupiter's early
  gas-driven migration}}, \emph{\nat}, \emph{475}, 206--209.

\bibitem[{\emph{{Walsh} et~al.}(2012)\emph{{Walsh}, {Morbidelli}, {Raymond},
  {O'Brien}, and {Mandell}}}]{walsh12}
{Walsh} K.~J., {Morbidelli} A., {Raymond} S.~N., {O'Brien} D.~P., and {Mandell}
  A.~M. (2012) \emph{{Populating the asteroid belt from two parent source
  regions due to the migration of giant planets--``The Grand Tack''}},
  \emph{Meteoritics and Planetary Science}, \emph{47}, 1941--1947.

\bibitem[{\emph{{Ward}}(1997)}]{ward97}
{Ward} W.~R. (1997) \emph{{Protoplanet Migration by Nebula Tides}},
  \emph{Icarus}, \emph{126}, 261--281.

\bibitem[{\emph{{Warren}}(2011)}]{warren11}
{Warren} P.~H. (2011) \emph{{Stable-isotopic anomalies and the accretionary
  assemblage of the Earth and Mars: A subordinate role for carbonaceous
  chondrites}}, \emph{Earth and Planetary Science Letters}, \emph{311},
  93--100.

\bibitem[{\emph{{Weidenschilling}}(1977)}]{weidenschilling77b}
{Weidenschilling} S.~J. (1977) \emph{{Aerodynamics of solid bodies in the solar
  nebula}}, \emph{\mnras}, \emph{180}, 57--70.

\bibitem[{\emph{{Wetherill}}(1991)}]{wetherill91}
{Wetherill} G.~W. (1991) in \emph{Lunar and Planetary Institute Science
  Conference Abstracts}, vol.~22 of \emph{Lunar and Planetary Inst. Technical
  Report}, p. 1495.

\bibitem[{\emph{{Wetherill}}(1992)}]{wetherill92}
{Wetherill} G.~W. (1992) \emph{{An alternative model for the formation of the
  asteroids}}, \emph{\icarus}, \emph{100}, 307--325.

\bibitem[{\emph{{Wetherill}}(1996)}]{wetherill96}
{Wetherill} G.~W. (1996) \emph{{The Formation and Habitability of Extra-Solar
  Planets}}, \emph{Icarus}, \emph{119}, 219--238.

\bibitem[{\emph{{Wilde} et~al.}(2001)\emph{{Wilde}, {Valley}, {Peck}, and
  {Graham}}}]{wilde01}
{Wilde} S.~A., {Valley} J.~W., {Peck} W.~H., and {Graham} C.~M. (2001)
  \emph{{Evidence from detrital zircons for the existence of continental crust
  and oceans on the Earth 4.4Gyr ago}}, \emph{\nat}, \emph{409}, 175--178.

\bibitem[{\emph{{Willacy} and {Woods}}(2009)}]{Willacy2009}
{Willacy} K. and {Woods} P.~M. (2009) \emph{{Deuterium Chemistry in
  Protoplanetary Disks. II. The Inner 30 AU}}, \emph{\apj}, \emph{703},
  479--499.

\bibitem[{\emph{{Williams} and {Mukhopadhyay}}(2019)}]{williams19}
{Williams} C.~D. and {Mukhopadhyay} S. (2019) \emph{{Capture of nebular gases
  during Earth's accretion is preserved in deep-mantle neon}}, \emph{\nat},
  \emph{565}, 78--81.

\bibitem[{\emph{{Wolfgang} et~al.}(2016)\emph{{Wolfgang}, {Rogers}, and
  {Ford}}}]{wolfgang16}
{Wolfgang} A., {Rogers} L.~A., and {Ford} E.~B. (2016) \emph{{Probabilistic
  Mass-Radius Relationship for Sub-Neptune-Sized Planets}}, \emph{\apj},
  \emph{825}, 19.

\bibitem[{\emph{{Wood} et~al.}(2010)\emph{{Wood}, {Halliday}, and
  {Rehk{\"a}mper}}}]{wood10}
{Wood} B.~J., {Halliday} A.~N., and {Rehk{\"a}mper} M. (2010) \emph{{Volatile
  accretion history of the Earth}}, \emph{\nat}, \emph{467}, E6.

\bibitem[{\emph{{Yang} et~al.}(2017)\emph{{Yang}, {Johansen}, and
  {Carrera}}}]{yang17}
{Yang} C.-C., {Johansen} A., and {Carrera} D. (2017) \emph{{Concentrating small
  particles in protoplanetary disks through the streaming instability}},
  \emph{\aap}, \emph{606}, A80.

\bibitem[{\emph{{Yang} et~al.}(2013)\emph{{Yang}, {Ciesla}, and
  {Alexander}}}]{Yang2013}
{Yang} L., {Ciesla} F.~J., and {Alexander} C. M.~O.~D. (2013) \emph{{The D/H
  ratio of water in the solar nebula during its formation and evolution}},
  \emph{\icarus}, \emph{226}, 256--267.

\bibitem[{\emph{{Yokochi} et~al.}(2012)\emph{{Yokochi}, {Marboeuf}, {Quirico},
  and {Schmitt}}}]{Yokochi2012}
{Yokochi} R., {Marboeuf} U., {Quirico} E., and {Schmitt} B. (2012)
  \emph{{Pressure dependent trace gas trapping in amorphous water ice at 77 K:
  Implications for determining conditions of comet formation}}, \emph{\icarus},
  \emph{218}, 760--770.

\bibitem[{\emph{{Youdin} and {Goodman}}(2005)}]{youdin05}
{Youdin} A.~N. and {Goodman} J. (2005) \emph{{Streaming Instabilities in
  Protoplanetary Disks}}, \emph{\apj}, \emph{620}, 459--469.

\bibitem[{\emph{{Yurimoto} et~al.}(2008)\emph{{Yurimoto}, {Krot}, {Choi},
  {Aleon}, {Kunihiro}, and {Brearley}}}]{Yurimoto2008}
{Yurimoto} H., {Krot} A.~N., {Choi} B.~G., {Aleon} J., {Kunihiro} T., and
  {Brearley} A.~J. (2008) \emph{{Oxygen Isotopes of Chondritic Components}},
  \emph{Reviews in Mineralogy and Geochemistry}, \emph{68}, 141--186.

\bibitem[{\emph{{Yurimoto} and {Kuramoto}}(2004)}]{Yurimoto2004}
{Yurimoto} H. and {Kuramoto} K. (2004) \emph{{Molecular Cloud Origin for the
  Oxygen Isotope Heterogeneity in the Solar System}}, \emph{Science},
  \emph{305}, 1763--1766.

\bibitem[{\emph{{Zellner}}(2017)}]{zellner17}
{Zellner} N.~E.~B. (2017) \emph{{Cataclysm No More: New Views on the Timing and
  Delivery of Lunar Impactors}}, \emph{Origins of Life and Evolution of the
  Biosphere}, \emph{47}, 261--280.

\bibitem[{\emph{{Zeng} et~al.}(2019)\emph{{Zeng}, {Jacobsen}, {Sasselov},
  {Petaev}, {Vanderburg}, {Lopez-Morales}, {Perez-Mercader}, {Mattsson}, {Li},
  and {Heising}}}]{zeng19}
{Zeng} L., {Jacobsen} S.~B., {Sasselov} D.~D., {Petaev} M.~I., {Vanderburg} A.,
  {Lopez-Morales} M., {Perez-Mercader} J., {Mattsson} T.~R., {Li} G., and
  {Heising} M.~Z. (2019) \emph{{Growth Model Interpretation of Planet Size
  Distribution}}, \emph{arXiv e-prints}, arXiv:1906.04253.

\bibitem[{\emph{{Zhang} and {Zhou}}(2010)}]{zhang10}
{Zhang} H. and {Zhou} J.-L. (2010) \emph{{On the Orbital Evolution of a Giant
  Planet Pair Embedded in a Gaseous Disk. I. Jupiter-Saturn Configuration}},
  \emph{\apj}, \emph{714}, 532--548.

\bibitem[{\emph{{Zhang} et~al.}(2013)\emph{{Zhang}, {Pontoppidan}, {Salyk}, and
  {Blake}}}]{Zhang2013}
{Zhang} K., {Pontoppidan} K.~M., {Salyk} C., and {Blake} G.~A. (2013)
  \emph{{Evidence for a Snow Line beyond the Transitional Radius in the TW Hya
  Protoplanetary Disk}}, \emph{\apj}, \emph{766}, 82.

\end{thebibliography}

\end{document}